\documentclass[%
    aps,
    prb,
    reprint,
    superscriptaddress
    amsfonts,
    amsmath,
    amssymb,
    eprint,
    longbibliography
    twocolumn,preprintnumbers,unsortedaddress
    ]{revtex4-2}

\usepackage{graphicx}
\usepackage{dcolumn}
\usepackage{amsmath,amsbsy,amssymb,scalefnt}
\usepackage{float} 
\usepackage{lipsum} 
\usepackage{verbatim}
\usepackage{bm}
\usepackage{epstopdf}
\usepackage{amsfonts}
\usepackage{textcomp}
\usepackage{soul}
\usepackage{mathtools}
\usepackage{hyperref}
\usepackage{hyperref}
\usepackage{comment}
\usepackage{fancyhdr}
\usepackage{mathtools}
\usepackage[normalem]{ulem}
\usepackage[x11names]{xcolor}
\hypersetup{
 pdfnewwindow=true, colorlinks=true,
 linkcolor=blue, anchorcolor=blue,
 citecolor=blue, filecolor=blue,
 menucolor=blue, urlcolor=blue
}

\begin{document}


\title{Enhanced Thermoelectricity in Nanowires with inhomogeneous Helical states}

\author{Zahra Aslani}
\email{zahraslani@gmail.com}
\affiliation{NEST, Istituto Nanoscienze-CNR and Scuola Normale Superiore, I-56126 Pisa, Italy}
\affiliation{ICTP,
Strada Costiera 11, 34151 Trieste, Italy}
\author{Fabio Taddei}
\affiliation{NEST, Istituto Nanoscienze-CNR and Scuola Normale Superiore, I-56126 Pisa, Italy}
\author{Fabrizio Dolcini}
\affiliation{Dipartimento di Scienza Applicata e Tecnologia del Politecnico di Torino, I-10129 Torino, Italy}
\author{Alessandro Braggio}
\email{alessandro.braggio@nano.cnr.it}
\affiliation{NEST, Istituto Nanoscienze-CNR and Scuola Normale Superiore, I-56126 Pisa, Italy}


\date{\today}

\begin{abstract}
Semiconductor nanowires (NWs) with strong Rashba spin-orbit coupling (RSOC), when exposed to a suitably applied Zeeman field, exhibit one-dimensional helical channels with a spin orientation locked to the propagation direction within the magnetic energy gap.  Here, by adopting a scattering-matrix approach applied to a tight-binding model of the NW, we demonstrate that the thermoelectric (TE) properties can be widely controlled by tuning the misalignment angle $\phi$ between the spin-orbit directions of two NW segments.
In particular, when the RSOC vectors are antiparallel (Dirac paradox configuration) we predict a significant violation of the Wiedemann-Franz law, and a strong enhancement of the Seebeck coefficient and the $ZT$ figure of merit. We also show that the Zeeman gap determines the optimal energy window for doping and temperatures. These results suggest that controlling the spin-orbit field direction, which can be achieved with suitably applied wrap gates, is a promising alternative for tuning and optimizing the TE response in quantum-coherent semiconducting NW devices.

\end{abstract}

\pacs{73.22.-f,71.70.Ej,73.63.-b}
                              
\maketitle

\section{\label{introduction}Introduction}
Developing devices with enhanced thermoelectric (TE) performance that optimize the conversion of wasted heat into work is crucial for advancing energy-harvesting technologies in the face of climate change~\cite{Petsagkourakis2018,Singh2024}.
It is also a key challenge for quantum technologies, such as quantum computers, where heat management and heat harvesting are crucial for large-scale integration at ultralow temperatures~\cite{Chen2019, Ishibe2018, Ma2021, chen2019thermoelectrics}.
Various strategies have been proposed to leverage TE properties.
Quantum confinement, for instance, implemented by reducing the device size in comparison to the electron wavelength, strongly affects the device energy filtering capabilities~\cite{Hicks1993a, Hicks1993b, Mao2016}. The contribution of the spin degree of freedom~\cite{Bauer2012,Yang2023}, also in the presence of spin-orbit effect, has been discussed for TE materials~\cite{Yuan2018,Hong2020,Tian2021}.
Also, the ongoing discoveries on topological materials have spurred the interest in exploiting topology in TE response~\cite{yang2025,Xu2017,Gooth2015ThermoelectricTI,Toriyama2025}, leading to the prediction of unconventional effects in topological Josephson junctions and revealing the sensitivity of thermoelectric effects to the underlying transport regime, distinguishing ballistic and dissipative behaviors~\cite{Blasi2020, Blasi2021,Arrachea2025, Mukhopadhyay2021, Alisultanov2025}. Furthermore, nanomaterials characterized by van der Waals layered structure or spin-orbit coupling have been proposed for various thermoelectric and optoelectronic applications~\cite{Liu2025, Hochbaum2008, Huber_2012, Karwacki2018, Manya2022, Gumbs2010_SOI}.

In this context, nanowires (NWs) appear particularly promising as a platform for TE devices for several reasons. Indeed, lateral confinement gives rise to discrete energy subbands with a modified density of states (DOS) or even van-Hove divergencies, which are expected to strongly enhance TE performances~\cite{Yang2004, Latronico2024,Roddaro2013}.
Also, in comparison with bulk materials, NWs are characterized by low thermal capacity and high sensitivity to heat flow~\cite{ElSachat2021, Sergej2020}, which makes them ideal for bolometric and quantum sensing technologies, where the variations in electronic temperature are a key issue~\cite{Chalopin2008,Sawtelle2019}.   
Furthermore, various studies have pointed out that inhomogeneity in the spin-orbit coupling of NWs can lead to a strongly energy-dependent transmission coefficient~\cite{Sanchez2006,Sanchez2008,Sadreev2013,Cayao2015,Gogin_2022_1,Gogin_2022_2}, which ultimately is determined by a unique mismatching spin-effect. This novel resource candidate is an intriguing addition to the more conventional quantum-confinement approach~\cite{Hicks1993a} for enhancing thermoelectric figures of merit.
Finally, the fabrication of clean ballistic NWs, which allows quantum coherent transport~\cite{Li2016, EstradaSaldana2018, Shani_2024}, with a Rashba spin-orbit coupling (RSOC) that is electrically tunable over a wide range of values~\cite{Scherubl2016, Takase2017, Liang2012,Sasaki2013, micolich_2015, Bindel2016,Takase2021,Pan2025} offers a dramatic improvement in terms of controllability~\cite{DattaDas1990, Bauer2012,Liang2012}.

Over the last two decades,  NWs such as InAs and InSb have attracted a lot of attention given the possibility to combine their strong RSOC with a magnetic field to effectively generate helical states similar to those of 2D topological insulators (TIs),  where the electron propagation direction is locked to the spin orientation~\cite{Streda2003,Yuval2010,Roman2010,Kloeffel2011,Kammhuber2017,vanWeperen2013,Quay2010}.
Notably, the helical states in TIs are edge states of a 2D bulk, and uncontrolled edge-bulk coupling may restrict their use in thermoelectric applications~\cite{Yong2014}. However, in NWs, the helical states are protected by the magnetic gap, which can even be externally controlled. Moreover, by applying local gates to different NW segments, one can generate inhomogeneous RSOC profiles~\cite{Sanchez2006, Sanchez2008,Sadreev2013,Szumniak2024, Klinovaja2015,Cayao2015,Klinovaja2018, Dolcini2018,Rossi2020,Gogin_2022_1,Gogin_2022_2} and thereby control independently the helicity of the electron states in such segments, a feature that would be quite hard to realize in the edge states of a TI. Finally, networks of NWs can be individually connected to reservoirs through Ohmic contacts~\cite{Roddaro2011,Aseev2019,Fadaly2017}. So far, most studies on the helical states of NWs have focused on the proximity-induced effects of superconductors to realize Majorana bound states~\cite{Roman2010, Yuval2010, Mourik_2012, Heiblum_2012, Marcus_2016,Prada_2020}. However,  the potential of NW helical states for TE properties remains largely unexplored.

In the present work, inspired by the promising features and versatility of NWs, we demonstrate that a suitable engineering of {\it inhomogeneous} RSOCs can dramatically enhance the TE properties of a NW. This result suggests new routes for electrically manipulating TE features in nanostructures of broader applicability.

\begin{figure}[t]
\centering
\includegraphics[width=0.47\textwidth]{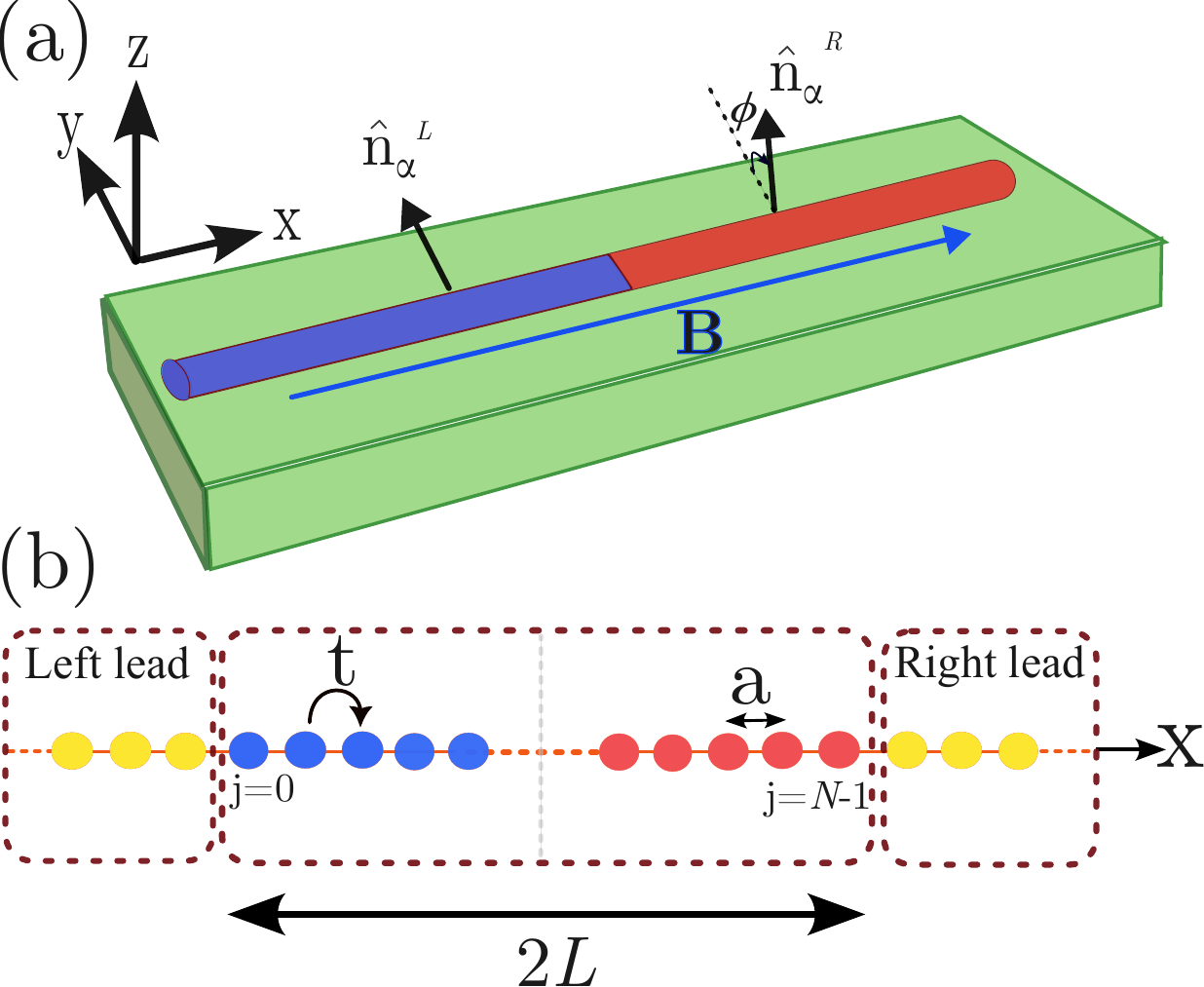}
\caption{Sketch of the setup: (a) Rashba NW oriented along the \(x\) direction and deposited on an insulating substrate (green). An external magnetic field is applied along the direction of $x$. The Rashba spin-orbit interaction is oriented along a unit vector $\hat{\mathbf{n}}_\alpha^{L(R)}$, which differs between the left and right regions of the NW. In the left region, $\hat{\mathbf{n}}_\alpha^{L} = (0,1,0)$, pointing along the \(y\)-direction. In the right region, the vector lies in the \(yz\)-plane, i.e., $\hat{\mathbf{n}}_\alpha^{R} = (0, \cos\phi, \sin\phi)$, forming an angle \(\phi\) with the $\hat{\mathbf{n}}_\alpha^{L}$. (b) 1D chain with TB parameters as described by Eqs.~\ref{H0-def}--\ref{HZ-def} and the number of sites $N$ in the chain sets the NW length $2L = (N-1)a$.}
\label{fig:System}
\end{figure}

The structure of the paper is as follows. In section~\ref{Sec2}, we present our model of the NW, based on a Tight-Binding (TB) Hamiltonian.
We discuss the general conditions for obtaining helical behavior in a Rashba-dominated system, 
including the inhomogeneous case. After we adopt the scattering-matrix approach to compute thermoelectric transport in those quasi-1D systems. To efficiently solve the inhomogeneous case for a range of parameters, such as the misalignment angle or the chemical potential, we use the Python package KWANT~\cite{Groth2014}. In section~\ref{Sec3}, we consider an NW device consisting of two segments with different RSOC directions, connected to two semi-infinite metallic leads. In particular, we discuss zero-temperature transport and investigate the effect of the RSOC direction on the transport. In section~\ref{Sec4}, we study the TE properties of the nonuniform RSOC NW in the presence of a uniform magnetic field. We focus on the role of the RSOC angle difference and the different length scales relative to magnetic and spin-orbit lengths. We also discuss doping and temperature dependence, identifying optimal regimes to achieve maximum performance.
 Finally, in section~\ref{Sec5}, we summarize our results and conclusions.

\section{\label{Sec2}Model and Formalism}
\subsection{System Hamiltonian}
We shall consider an NW deposited on a substrate and denote by $x$ the longitudinal axis direction, and by $z$ the direction perpendicular to the substrate, as illustrated in Fig.~\ref{fig:System}(a). 
In our analysis, we will assume that transversal confinement of the NW  is sufficiently strong to allow us to focus on the TE properties of a single spin-split energy band, which is well separated from higher-energy subbands. We shall therefore deal with a single-channel physics, where the electron motion is effectively 1D~\footnote{The reader should be aware that the results, including the TE properties, can be generalized if other independent bands are also present~\cite{May2012,Agrawal2024}}.

The structural inversion asymmetry arising from the substrate generates an effective built-in electric field along $z$, which in turn corresponds to an RSOC field directed along $y$~\cite{Kammhuber2017}. However, the magnitude of the RSOC can also be controlled by applying a gate voltage between the NW and the back gate of the insulating substrate. Moreover, the direction $\hat{\mathbf{n}}_\alpha$ of the RSOC field can be tuned by introducing additional side gates~\cite{Roddaro2011} and/or by means of substrate inhomogeneities~\cite{Gan2019, Takase2021,Klinovaja2015}. Therefore, in the following, we shall assume that $\hat{\mathbf{n}}_\alpha$ always lies in the $y$-$z$ plane. However,  we shall consider that such a direction may be inhomogeneous along the NW axis, i.e. $\hat{\mathbf{n}}_\alpha=\hat{\mathbf{n}}_\alpha(x)$. In particular, we shall analyze the case where two NW segments, depicted in blue and red in Fig.~\ref{fig:System}, are characterized by two different directions $\hat{\mathbf{n}}_\alpha^L$ and $\hat{\mathbf{n}}_\alpha^R$, differing by an angle~$\phi$. 
Additionally, we consider a magnetic field uniformly directed along $x$, i.e., perpendicular to the RSOC field direction $\hat{\mathbf{n}}_\alpha$. \footnote{It is easy to generalize the results also to the cases where the magnetic field is not oriented along $x$. However, especially in the inhomogeneous case, one needs to specify the two angles between the magnetic fields and the RSOC, making the analysis slightly more involved.} 
As is well known, under such conditions, a magnetic gap in the spin-split bands effectively generates helical states~\cite{Streda2003,Yuval2010,Roman2010,Kloeffel2011,Kammhuber2017,vanWeperen2013,Quay2010}, if the RSOC is sufficiently strong.

To investigate the transport properties, we will describe the nonuniform RSOC NW using a standard TB approach with a Hamiltonian $
\mathcal{H}_{\mathrm{NW}} = \mathcal{H}_{0} + \mathcal{H}_{\mathrm{SOC}} + \mathcal{H}_\mathrm{Z}$. Here,
\(\mathcal{H}_{0}\) represents the simple 1D-TB Hamiltonian with electron hopping $t$ between the nearest neighboring sites and $\epsilon_{\mathrm{NW}}$ the Fermi energy of the NW. It is~\cite{Szumniak2024, PhysRevB.110.075415}
\begin{equation}
\begin{aligned}\label{H0-def}
\mathcal{H}_{0} = & -t \sum_j\left( \Psi_{j+1}^\dagger\sigma_0 \Psi_j + \text{H.c.}\right) \\
 &+ (2t-\epsilon_{\mathrm{NW}}) \sum_j \Psi_j^ \dagger \sigma_{0} \Psi_j,
\end{aligned}
\end{equation}
where $\Psi_{j}^\dagger = (c_{j\uparrow}^\dagger,c_{j\downarrow}^\dagger)$ and $c_{j\beta}^\dagger ~(c_{j\beta})$ is the creation (annihilation) operator is acting on an electron with spin $ \beta \in \{\uparrow, \downarrow\}$ located at a site $j$ in a 1D lattice of $N$ sites. In Eq.~(\ref{H0-def}), $t$ is the hopping amplitude parameter [see Fig.~\ref{fig:System}(b)] and $\sigma_0$ denotes the identity matrix for the spin degree of freedom. The spectrum of Eq.~(\ref{H0-def}) is doubly degenerate in spin and near its band bottom at $k=0$, is well approximated by a parabola with an effective mass $m^*$ connected to the hopping amplitude parameter through $t = {\hbar^2}/{2m^* a^2}$, where $a$ is the lattice spacing.

The \(\mathcal{H}_{\mathrm{SOC}}\) represents the RSOC term 
\begin{equation}\label{HSOC-def}
\mathcal{H}_{\mathrm{SOC}} = - i \tilde{\alpha}\sum_j\Psi_{j+1}^\dagger (\hat{\mathbf{n}}^j_{\alpha}\cdot\boldsymbol{\sigma})\Psi_j + \text{H.c.},
\end{equation}
where $\tilde{\alpha}$  
is the spin-flip hopping amplitude arising from the RSOC~\cite{Szumniak2024, Klinovaja2015,Klinovaja2018,Cayao2015}. For the sake of clarity, we mention that the relation between $\tilde{\alpha}$ and the RSOC $\alpha$ customarily used in NW continuum models is 
$\tilde{\alpha}=  \alpha/2a$. In Eq.~(\ref{HSOC-def}), 
$\boldsymbol{\sigma}=(0,\sigma_y,\sigma_z)$ are Pauli matrices, while the unit vector \(\hat{\mathbf{n}}^j_{\alpha}=(0,n^j_{\alpha,y},n^j_{\alpha,z})\) identifies the RSOC field direction in the $y$-$z$ plane at the site $j$, as illustrated in Fig.~\ref{fig:System}(a).
Its possible dependence on $j$ corresponds to the case where the inhomogeneity is along the NW axis.

The Zeeman energy contribution arising from the coupling with the external magnetic field along the $x$-axis is \(\mathcal{H}_{\mathrm{Z}}\) and it will be written as
\begin{equation}
\label{HZ-def}
\mathcal{H}_{\mathrm{Z}} = {\frac{\Delta_{\mathrm{Z}}}{2}} \sum_j \Psi_j^\dagger \sigma_x  \Psi_j,
\end{equation}
with $\Delta_{\mathrm{Z}}=g^* \mu_B B_x$, where $B_x$ is the strength of the external magnetic field directed along $x$,  $g^*$ is the effective Landé g-factor, and \( \mu_B \) is the Bohr magneton.

\begin{figure*}[ht]
\centering
\includegraphics[width=0.85\linewidth]{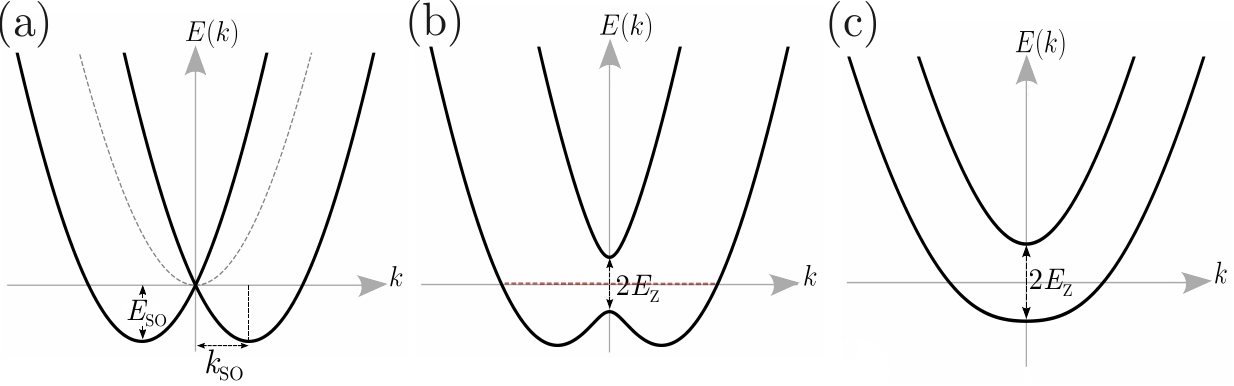}
    \caption{The electronic band structure of a 1D NW  with a constant RSOC profile, for different values of the Zeeman field. (a) The Zeeman field is absent, and the RSOC lifts the spin degeneracy of the parabolic band (dashed gray line). This results in two spin-split bands, each shifted by $\pm k_{\mathrm{SO}}$	
  in momentum and lowered in energy by $E_{\mathrm{SO}}$.
 (b) A finite Zeeman field along the $x$-axis opens a gap at $k = 0$, creating a single helical state when the chemical potential lies within the gap. Spins align with the Zeeman field at low $k$ and tilt toward the RSOC direction at higher $k$. (c) Zeeman-dominated regime resulting in almost complete spin polarization of the bands along the field direction.}
    \label{fig:BS}
\end{figure*}

\subsection{Homogeneous NW}
For a homogeneous NW, the vector $\mathbf{n}_{\alpha}^j$ is spatially uniform ($\mathbf{n}_{\alpha}^j \equiv \mathbf{n}_{\alpha}\,\, \forall j$), so that the whole system is translationally invariant and the Hamiltonian $\mathcal{H}_{\mathrm{NW}}$ can be rewritten in the momentum base. Indeed, by applying the Fourier transform $\Psi_{{j}}  = \frac{1}{\sqrt{N}} \sum_{k} e^{ik(ja)} \Psi_k$, with $N$ denoting the number of lattice sites with periodic boundary conditions, the TB Hamiltonian becomes diagonal in \(k\)-space and can be written~as
\begin{equation}\label{H-tot}
\mathcal{H}_{\mathrm{NW}} =  \sum_{k} \Psi_{k}^\dagger H(k)\, \Psi_{k},
\end{equation}
where
\begin{equation}
\begin{aligned}
H(k) & =  \big(2t\left[1 - \cos(ka)\right]-\epsilon_\mathrm{NW}\big)\sigma_{0} \\
& - 2\tilde{\alpha}\, \sin(ka)(\hat{\mathbf{n}}_{\alpha}\cdot\boldsymbol{\sigma})  + (\Delta_\mathrm{Z}/2)\, \sigma_{x}. \\
\end{aligned}
\end{equation}
In such a case, the dispersion relation consists of two bands 
\begin{equation}\label{spectrum}
\begin{aligned}
E_\pm(k) & = \big[2t\big(1-\cos (ka)\big)-\epsilon_\mathrm{NW}\big] \\
 & \pm\sqrt{4\tilde{\alpha}^2\sin^2(ka)+(\Delta_\mathrm{Z}/2)^2}\, .
\end{aligned}
\end{equation}
Notably, the spectrum (\ref{spectrum})  is independent of the direction $\hat{\mathbf{n}}_\alpha$ of the RSOC, and leads one to identify two characteristic energy scales, namely the RSOC energy \begin{equation}\label{ESO-def}
E_{\mathrm{SO}} = \tilde{\alpha}^2/t,
\end{equation} 
which reflects the strength of the spin–orbit interaction, and the Zeeman energy 
\begin{equation}\label{EZ-def}
E_\mathrm{Z} = |\Delta_\mathrm{Z}|/2 =|g^* \mu_B B|/2 \quad.
\end{equation} 
Depending on the value of $E_\mathrm{Z}/E_\mathrm{SO}$, the two bands in Eq.~(\ref{spectrum}) exhibit qualitatively different  behaviors, as summarized in Fig.~\ref{fig:BS} for $|k|\ll \pi/a$.

When the magnetic field is absent, i.e. $E_\mathrm{Z}=0$  [see Fig.~\ref{fig:BS}(a)], the two bands  exhibit a horizontal spin splitting, and their local minima are located at $k=\pm k_{SO}$, where 
\begin{equation}\label{kSO-def}
    k_{SO} = \frac{1}{a}\arctan\left(\frac{\tilde{\alpha}}{t}\right)
\end{equation}
denotes the spin-orbit wavevector. Note that, for $\tilde{\alpha}  \ll t$,   Eq.(\ref{kSO-def})  reduces to the continuum limit expression  $k_{SO}\approx |\alpha|m^*/\hbar^2=\sqrt{2 m^* E_{SO}}/\hbar$.

When a finite Zeeman field $E_\mathrm{Z} \neq 0$ is further applied, an energy gap   $\Delta_Z=2E_\mathrm{Z}$ opens up at \(k = 0\) and two possible regimes can be realized: Rashba- or Zeeman-dominated.  

In the Rashba-dominated regime, where $E_{\mathrm{Z}} < 2E_{\mathrm{SO}}$,  the lower band exhibits a local maximum at $k = 0$ and two degenerate minima at finite momenta 
$k = \pm k_{\mathrm{SO}} \sqrt{[1 - ({E_{\mathrm{Z}}}/{2E_{\mathrm{SO}})^2}}]$ as shown in Fig.~\ref{fig:BS}(b).
Importantly, in the deep Rashba-dominated regime, where \( E_{\mathrm{Z}} \ll 2E_{\mathrm{SO}} \), for energies in the magnetic gap, the lower band features effective helical states, 
with a locking between the propagation direction and the spin orientation~\cite{Streda2003,Yuval2010,Roman2010,Kloeffel2011,Kammhuber2017,vanWeperen2013,Quay2010}. More precisely, inside the Zeeman gap (\( |E| \ll E_{\mathrm{Z}} \)), the spin of right-moving electrons is mainly directed along the direction $+\hat{\mathbf{n}}_\alpha$, while the left-moving electrons have their spins locked along $-\hat{\mathbf{n}}_\alpha$. This emulates the physics of helical edge states of a 2DTI~\cite{Qi2011, Brune2012, Konig2007}. However, differently from that case, here their spin-momentum locking (helicity) is completely tunable by the direction of the Rashba vector $\hat{\mathbf{n}}_\alpha$.

Finally, in the Zeeman-dominated regime, when $E_\mathrm{Z} > 2E_{\mathrm{SO}}$, both spin‐split bands exhibit their minima at $k = 0$ with energies $\pm E_\mathrm{Z}$ reflecting the dispersion of a  Zeeman‐split system, as illustrated in Fig.~\ref{fig:BS}(c). Similarly, the wavevector associated with the Zeeman energy $E_Z$ is 

\begin{equation}\label{kZ-def}
    k_Z=\frac{2}{a}  \arcsin\left(\sqrt{\frac{E_Z}{4t}}\right), 
\end{equation}
which, for $|\Delta_Z| \ll t$, reduces to the continuum limit expression  $k_Z\approx\sqrt{2 m^* E_Z}/\hbar$. 

\subsection{Inhomogeneous NW}
\label{ino}
Let us now consider a spatially inhomogeneous NW. Specifically, we shall focus on the case depicted in Fig.~\ref{fig:System}(a), where two NW segments are characterized by two different RSOC directions \(\hat{\mathbf{n}}^\mathrm{L}_\alpha\) and \(\hat{\mathbf{n}}^\mathrm{R}_\alpha\) forming an angle $\phi$.  
For simplicity, we shall assume that the strength $\tilde{\alpha}$  is uniform everywhere and that the two segments have equal length $L$  each. This situation is modelled by adopting in Eq.~(\ref{HSOC-def})  the profile
$\hat{\mathbf{n}}_{\alpha}^j=\hat{\mathbf{n}}_{\alpha}^\mathrm{L} g(j-j^*)+\hat{\mathbf{n}}_{\alpha}^\mathrm{R} [1-g(j^*-j+1)]$, where  $g(j)$ is a crossover function  such  that $0\leq g(j)\leq 1$,  $g(\infty)=0$ and $g(-\infty)=1$, and by setting $j^* a=L$. We shall mainly focus on the situation where such a crossover is sharp relative to the other involved length scales. This will enable us to assume a  
piecewise-constant profile $g(j)=\theta(-j)$, with $\theta(x)$ denoting the Heaviside function. We have verified that the results obtained for such a sharp profile are only marginally affected when considering a smoother profile, provided that the crossover length $\ell$ is still shorter than the natural length scale associated with RSOC strength, i.e.~the spin-orbit length defined as $L_{SO}=2\pi/k_{SO}$~\cite{Gogin_2022_2}. 

Since $\phi$ identifies the relative orientation between $\hat{\mathbf{n}}^L_\alpha$ and $\hat{\mathbf{n}}^R_\alpha$, for symmetry reasons it will be sufficient to consider the range $\phi \in[0,\pi]$, with the extremal value  \(\phi=0\) (\(\phi=\pi\)) corresponding to the homogeneous (antiparallel) case. As we shall see in detail in Sec.~\ref {Sec3}, for $\phi \neq 0$, the different spin orientations of the RSOC field strongly affect both the charge transport and the TE properties of the systems.  

A comment is in order for the case $\phi=\pi$, which corresponds to the antiparallel orientation. In this case,  the helical states inside the magnetic gap of the two NW segments are characterized by exactly opposite helicities, realizing the  'Dirac paradox' configuration discussed in Refs.~\onlinecite{Rossi2020,Gogin_2022_1,Gogin_2022_2}. This paradox consists in the seeming impossibility for a quasiparticle to either propagate, due to the opposite helicity of the state across the interface, or to be backscattered, due to the opposite helicity of the reflected state on the same side. The solution to such a paradox lies in the existence of evanescent modes localized at the interface between two regions with opposite RSOC directions. Although these localized modes do not carry current directly, their spin textures affect wavefunction matching at the interface, thereby indirectly determining the transmission properties of the propagating modes.
The evanescent modes operate as a sort of magnetic impurity, whose effects on the helical states depend on the crossover length $\ell$ between the regions of opposite RSOC directions. In particular, if the crossover length $\ell$ is much shorter than the spin-orbit length $L_{\mathrm{SO}}$,  a strong suppression of the charge transport occurs in the magnetic gap~\cite{Gogin_2022_1}. Of course, for energies $|E|>E_\mathrm{Z}$ outside the magnetic gap, the evanescent modes become propagating [see Fig.~\ref{fig:BS}], giving rise to a strong energy-dependent transmission at the magnetic band-gap edge.

\subsection{Scattering matrix approach}
We consider a two-terminal setup where the two reservoirs are kept at different temperatures ($T_\mathrm{R}=T$, on the right, and $T_\mathrm{L}=T+\Delta T$, on the left) and electrochemical potentials ($\mu_\mathrm{R} = \mu$, on the right, and $\mu_\mathrm{L} = \mu +e\Delta V$, on the left).
Using the Landauer-B\"uttiker formalism~\cite{Blanter2000, Buttiker1985, Lv2012} and the scattering matrix theory, the steady state electric $J_e$ and heat $J_h$ currents flowing in the right lead toward the NW can be computed as~\cite{Oji1984, Benenti2017}
\begin{equation}
\binom{J_e}{J_h} = \frac{1}{h}\int_{-\infty}^{+\infty}\!\!\!\!\mathrm{d}E\; \binom{e}{E-\mu} \, \mathcal{T}(E)\, [f_\mathrm{R}(E)-f_\mathrm{L}(E)].
\label{I}  
\end{equation}
In this equation, $\mathcal{T}(E)$ denotes the transmission coefficient from the right to the left lead. The Fermi distribution functions of the reservoirs for a given electrochemical potential $\mu_i$ and temperature $T_i$ are $f_i (E)=1/[e^{(E-\mu_i)/k_BT_i}+1]$, where $k_B$ denotes the Boltzmann constant.
In the linear response regime, $\lvert\Delta T \rvert \ll T $ and $e\lvert \Delta V \rvert \ll k_B T $, the electric and heat currents can be written as~\cite{Benenti2017,Groot2013} 
\begin{equation}
     \binom{J_e}{J_h} =\frac{1}{T}  \begin{pmatrix}
           L_{ee} & L_{eh}  \\
          L_{he} & L_{hh}    \\
 \end{pmatrix}\binom{\Delta V}{\Delta T/T},
\end{equation}
where $L_{ab}$ ($a,b = e,h$) are the Onsager coefficients.
The latter can be expressed in terms of energy integrals of the transmission coefficient weighted by different powers of $(E-\mu)$ and by the equilibrium Fermi distribution derivative ${\partial f_0(E)}/{\partial E}$ (see Ref.~\cite{Benenti2017}).
The linear transport coefficients such as conductances, electrical $G$ and thermal $K$, and thermopower $S$ can all be expressed in terms of the Onsager coefficients as~\cite{Benenti2017}
\begin{equation}
G=\frac{L_{ee}}{T}, \,\, K=\frac{1}{T^2}[L_{hh}-\frac{L_{eh}^2}{L_{ee}}], \,\, S=\frac{1}{T}\frac{L_{eh}}{L_{ee}}.
\label{eq:G_K_S}
\end{equation}

Thus, all electrical and TE properties for the two-terminal device are determined by the transmission coefficient $\mathcal{T}(E)$, at energy $E$, which we have evaluated numerically by using the Python package KWANT~\cite{Groth2014}.
To this purpose,  the NW, divided into the two segments with different RSOC vectors [see Fig.~\ref{fig:System}(b)], is connected to two normal leads, which are described by a TB Hamiltonian ${H}_\mathrm{lead}(k) =  \{ 2t\left[1 - \cos(ka)\right]-\epsilon_\mathrm{N}\} \sigma_{0}$, where the hopping amplitude parameter \(t\) is assumed to be the same as in the NW and the interfaces, while the RSOC and the Zeeman field terms are negligible. Moreover, we consider the Fermi energies in the normal leads to be much higher than in the semiconducting NW, that is, $\epsilon_\mathrm{N}\gg \epsilon_{\mathrm{NW}}$, ensuring that the bottom of the lead band lies much lower than that of the NW.\\

\subsection{Experimental implementation}

To clearly evaluate the transport and thermoelectric signatures of the discussed Dirac paradox, it is convenient to consider a realistic experimental example. This provides us with a quantitative estimate of the reported effects.
However, the general mechanism described hereafter is completely general and can be reproduced in other systems, after the appropriate rescaling.

A possible realization of the proposed NW setup makes use of InSb ballistic NWs having an effective electron mass of $m^* = 0.015\,m_e$ and a g-factor $g^* \approx 50$~\cite{Lekwongderm2019,Fan2015,Gogin_2022_1}. In these systems,
RSOC can be sufficiently strong to access the deep Rashba-dominated regime.
Indeed, one finds that \(E_{\mathrm{SO}} \sim 1\)~meV, which is much larger than the Zeeman splitting energy corresponding to a magnetic field strength $B \sim 70 $ mT, for which $E_\mathrm{Z} \sim 0.1$~meV. This implies that $L_{\mathrm{SO}}=    L_\mathrm{Z}/\sqrt{10}\,  \simeq 0.3L_{\mathrm{Z}}$.
Cryogenic operating temperatures, however, are needed.

\section{Zero-temperature transport}
\label{Sec3}
\begin{figure}[t]
    \centering
\includegraphics[width=0.98
    \linewidth]{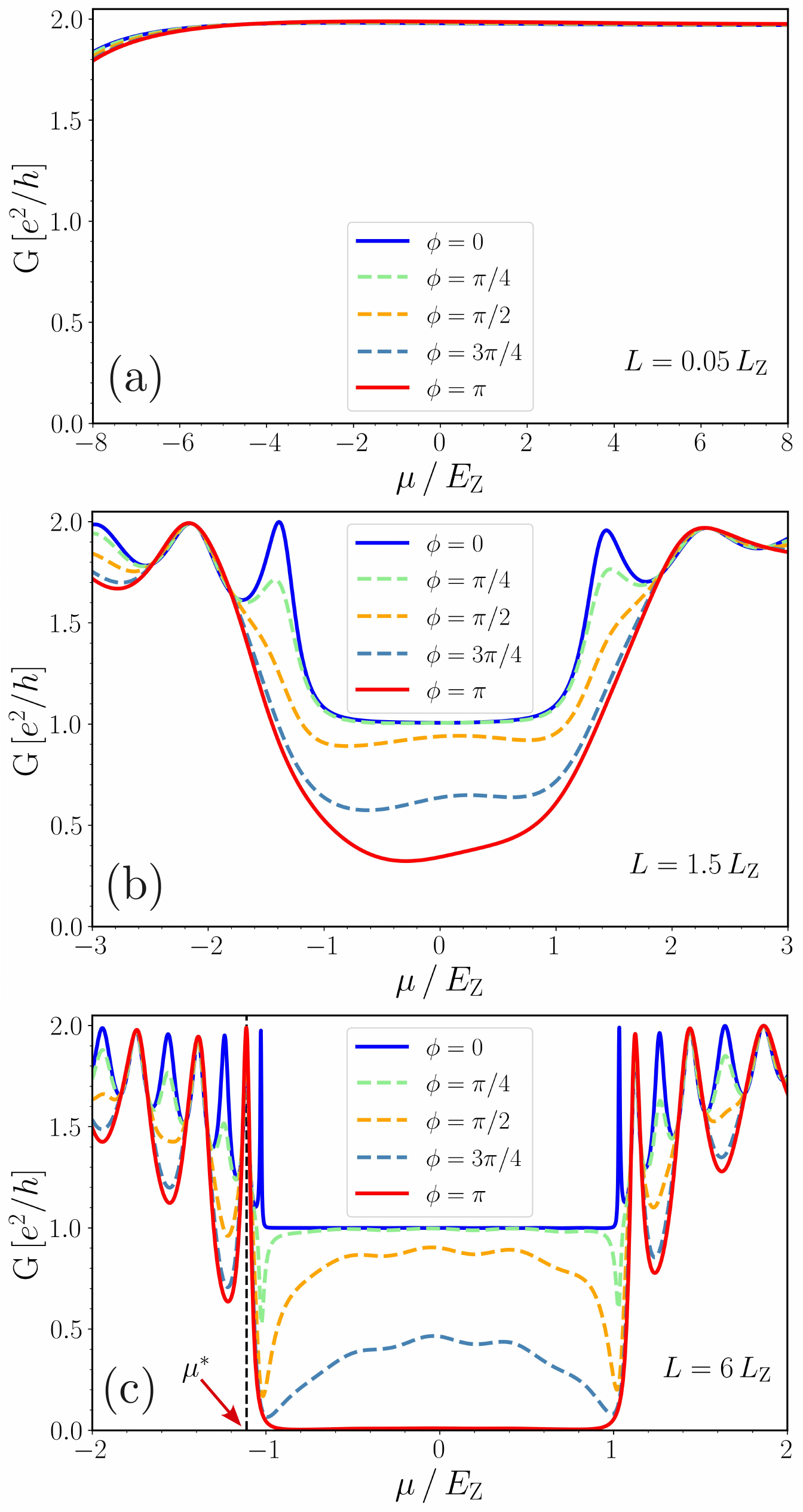}
    \caption{ The electrical zero-temperature conductance $G$ of an inhomogeneous RSOC NW, shown in units of the conductance quantum $G_0=e^2/h$, is plotted as a function of the normalized chemical potential $\mu/E_\mathrm{Z}$ for different misalignment angles $\phi$, and for various lengths: (a)  
    $L= 0.05L_\mathrm{Z}$, (b) 
    $L= 1.5L_\mathrm{Z}$,  
    and (c) 
    $L= 6L_\mathrm{Z}$. We set $E_{\mathrm{SO}}/E_\mathrm{Z} = 10$, i.e.~we are in the deep Rashba-dominated regime.}
    \label{fig:G_0}
\end{figure}

Various studies have predicted that an inhomogeneous  RSOC in NWs leads to a strongly energy-dependent transmission coefficient~\cite{Sanchez2006,Sanchez2008,Sadreev2013,Cayao2015,Dolcini2018,Gogin_2022_1,Gogin_2022_2,Rossi2020}, which is a particularly interesting feature in the context of TE properties.
However, these studies are mainly limited to the analysis of inhomogeneities in the magnitude of the RSOC~\footnote{The effect of inhomogeneities in the RSOC direction has been mainly considered on the Andreev bound levels and Josephson current in the case of NWs proximitized by superconductors~\cite{Szumniak2024, Klinovaja2015,Klinovaja2018,Cayao2015}.}.

In this section, we show that the inhomogeneities of RSOC {\it direction} have a strong impact on the transport.  Specifically, we analyze how the zero-temperature conductance $G=G_0{\cal T}(\mu)$, where $G_0 = e^2/h$ is the conductance quantum, depends on the angle  $\phi$ between the RSOC field directions $\hat{\mathbf{n}}_\alpha^L$ and $\hat{\mathbf{n}}_\alpha^R$ in the two NW segments of length $L$ depicted in Fig.~\ref{fig:System}.
For definiteness, we assume that both NW segments are in the deep Rashba-dominated regime, i.e.~$E_{\mathrm{Z}} \ll 2E_{\mathrm{SO}}$.

The results are shown in Fig.~\ref{fig:G_0}, which displays $G$ as a function of the normalized chemical potential $\mu/E_\mathrm{Z}$. The three panels refer to three different values of length $L$, compared to the magnetic length $L_\mathrm{Z} = 2\pi/k_\mathrm{Z}$, with $k_Z$ given by Eq.(\ref{kZ-def}). 
Note that, in the deep Rashba-dominated regime,  the magnetic length is larger than the spin-orbit length, i.e. $L_{\mathrm{SO}} < L_\mathrm{Z}$. 
In each panel, the solid blue curve shows the conductance for the case of identical RSOC directions ($\phi = 0$) corresponding to the homogeneous NW, while the solid red curve showcases the antiparallel case ($\phi = \pi$), where the  Dirac-paradox configuration is realized. The dashed curves correspond to intermediate misalignment cases with angles $\phi = \pi/4,\, \pi/2,\, 3\pi/4$ (see caption).\\
  
{\it Short NW.} For very short NW lengths $(L \ll L_{\mathrm{SO}}, L_\mathrm{Z})$,  the conductance is very close to $2G_0$, as for two ideal channels, for any value of the relative angle $\phi$ [see Fig.~\ref{fig:G_0}(a)]. This behavior originates from the fact that this is the ballistic regime, where the NW behaves as a weak localized impurity whose internal inhomogeneities are irrelevant, even for values of the chemical potential in the range $|\mu|<E_{\mathrm{Z}}$ (i.e. in the magnetic gap). At very low chemical potential values, where the NW band bottom is depleted, backscattering increases due to momentum mismatch with the normal-metal leads, leading to a small reduction in conductance.

{\it Intermediate-length NW.} By increasing the NW length to $L\sim L_{\mathrm{Z}}$, the evanescent modes are then localized at the interface between two regions (when the chemical potential lies within the magnetic gap). Since their localization length is $\sim L_Z$, they can no longer fully mediate transport (see discussion in Sec.~\ref {ino}). A clear signature of this is the drop of conductance in the range $|\mu|<E_\mathrm{Z}$ for all $\phi$-values, as shown in Fig.~\ref{fig:G_0}(b). In particular, for the homogeneous case ($\phi=0$), $G$ is roughly equal to $G_0$ in the whole magnetic gap, implying that charge transport is carried by the almost perfectly transmitted helical states.  
A non-zero misalignment angle ($\phi \neq 0$) between the two NW segments causes a further reduction in conductance. Such a reduction is smallest when the segments are antiparallel ($\phi=\pi$), though it remains finite. 

For energies outside the Zeeman gap ($|\mu|>E_\mathrm{Z}$), both NW bands [see Fig.~\ref{fig:BS}(b)] contribute to transport, and one observes an oscillatory behavior of the conductance as a function of $\mu$. 
This is due to a resonant Fabry–P\'erot-like interference effect, where electron back-scattering occurs at the interfaces with the leads and, in the inhomogeneous case where $\phi\neq 0$, also at the interface between the two NW segments. 
The backscattering at the NW-lead interfaces arises from the momenta mismatch, i.e. the discontinuities in the Fermi energies, but also from the spin degree of freedom, a consequence of the RSOC inhomogeneity at the interfaces~\cite{Cayao2015}.
In the inhomogeneous case, $\phi\neq 0$, however, an additional backscattering occurs at the interface between the two NW segments and it is \emph{only} due to spin texture mismatch. 

The period of the conductance oscillations in the chemical potential $\mu$ is determined by the resonant condition that is related to the effective length of the Fabry-P\'erot cavity with respect to the Fermi wavelength. 
In the inhomogeneous case, this period 
is roughly twice that of the homogeneous one ($\phi=0$), because the effective length of the Fabry-P\'erot cavity is $L$ rather than $2L$. However, the specific shape of the conductance oscillations also depends on the relative angle $\phi$ between the RSOC fields in the two NW segments, as this angle determines the strength of the back-scattering at their interface.\\

{\it Long NW.} In the regime of a sufficiently long NW, $L> L_{\mathrm{Z}}$ [see Fig.~\ref{fig:G_0}(c)], two interesting features emerge. 
First, in the antiparallel configuration $\phi=\pi$,   a strong suppression of the conductance ($G\approx 0$) is observed over almost the entire magnetic gap range $|\mu|<E_{\mathrm{Z}}$.
Indeed, the spin misalignment of the propagating modes is so strong that it cannot be compensated by the spin texture of the evanescent modes at the interface.  
This effect is a consequence of the sharp interface regime assumed here, where the crossover length between the two segments $\ell$ of opposite helicities is the shortest lengthscale ($\ell\ll L_{\mathrm{SO}}$)~\cite{Gogin_2022_1} (see also the discussion in Sec.~\ref{ino}). We tested (not shown) that for a smoother interface, the conductance suppression progressively disappears. 

The second noteworthy effect in Fig.~\ref{fig:G_0}(c) occurs at intermediate values of the angle $\phi$ (dashed curves). While the conductance remains finite in the middle of the magnetic gap $|\mu| \ll E_\mathrm{Z}$, 
conductance anti-resonances appear at $|\mu|\sim E_\mathrm{Z}$.
This result generalizes the finding of Ref.~\cite{Gogin_2022_2}, showing that anti-resonances occur not only from inhomogeneities in the RSOC magnitude, but also from inhomogeneities in the RSOC field direction.
Importantly, this effect resembles the Fano-Rashba antiresonances that appear when a bound state of an NW couples to the continuum states of the leads ~\cite{Sanchez2008,Cayao2015}. 
In our case, the antiresonances are mainly pinned at the magnetic-gap boundaries, $|\mu|\lesssim E_\mathrm{Z}$. They are due to a Fano-like mechanism involving the states propagating along the NW and the states localized at the central interface when the two NW segments exhibit a different spin-orbit orientation ($\phi\neq0$). This result extends to the case of inhomogeneities in the spin-orbit direction the effect found in Ref.~\cite{Gogin_2022_2}  in the presence of inhomogeneities in the RSOC magnitude. The role of the bound states localized at the central interface becomes more pronounced as the misalignment angle $\phi$ increases, leading to a wider and deeper Fano-Rashba antiresonance. In particular, for the antiparallel case  ($\phi=\pi$), the spin texture of these bound states hinders the transmission of the propagating modes across the junction. 

Finally, in the homogeneous case, at $|\mu|\sim E_\mathrm{Z}$, the chemical potential is fixed at the opening of the bottom (top) of the top (bottom) band, as shown in Fig.~\ref{fig:BS}(b).
For such a case, we observe a strong resonant peak in the zero-temperature conductance at $|\mu|\approx E_\mathrm{Z}$. For the antiparallel configuration $\phi=\pi$, such a peak is shifted to slightly high energies $|\mu|=\mu^*>E_Z$ (see red arrow in Fig.~\ref{fig:G_0}(c))   due to the additional Fabry-P\'erot resonant contribution. 
As we will see below, these resonances may affect TE properties because they are strong, energy-dependent features.  

In the following, we will mainly focus on the case of long NWs, $L\gtrsim L_Z$, since in this regime transport properties exhibit the most pronounced energy dependence. \\

\section{Thermoelectric properties}
\label{Sec4}
We now turn to the thermoelectric properties of the NW, comparing the homogeneous case with inhomogeneous RSOC profiles at different misalignment angles~$\phi$. 

\subsection{Wiedemann-Franz ratio and TE Onsager coefficient}
As anticipated, the inhomogeneous NW configurations with a misalignment angle $\phi \neq 0$ in the RSOC direction exhibit energy-dependent transport properties, resulting in significant thermoelectric effects. Here, we shall show that such an effect can be so strong that it also causes the violation of the Wiedemann-Franz (WF) law.
We recall that, according to such law, the WF ratio  $\mathcal{L}=K/(G\,T)$, where $K$ is the thermal conductance and $G$ the electrical conductance introduced in Sec.~\ref{Sec3}, reaches a universal value
\begin{equation}
\mathcal{L}_0=\frac{\pi^2}{3}\left(\frac{k_B}{e}\right)^2\approx 2.44\times10^{-8}\ \mathrm{W\Omega K^{-2}}
\end{equation}
in the low temperature limit, 
where $\mathcal{L}_0$ is the Lorenz factor. 
This is because the carriers responsible for the entropic flow are also the ones that carry the charge. While the system considered here does reach this universal value in the limit $T\to 0$, at finite temperature a different scenario emerges. We start by focusing on the finite but small temperature regime, where the thermal energy is smaller than the magnetic gap, i.e., $ k_BT \ll E_\mathrm{Z}$.
We consider an NW of length $L = 6\,L_\mathrm{Z}$ to be in the long NW configuration to ensure maximal contrast between the homogeneous and the Dirac-paradox configuration [see Fig.\ref{fig:G_0}(c)].

\begin{figure}[t]
\centering
\includegraphics[width=0.47\textwidth]{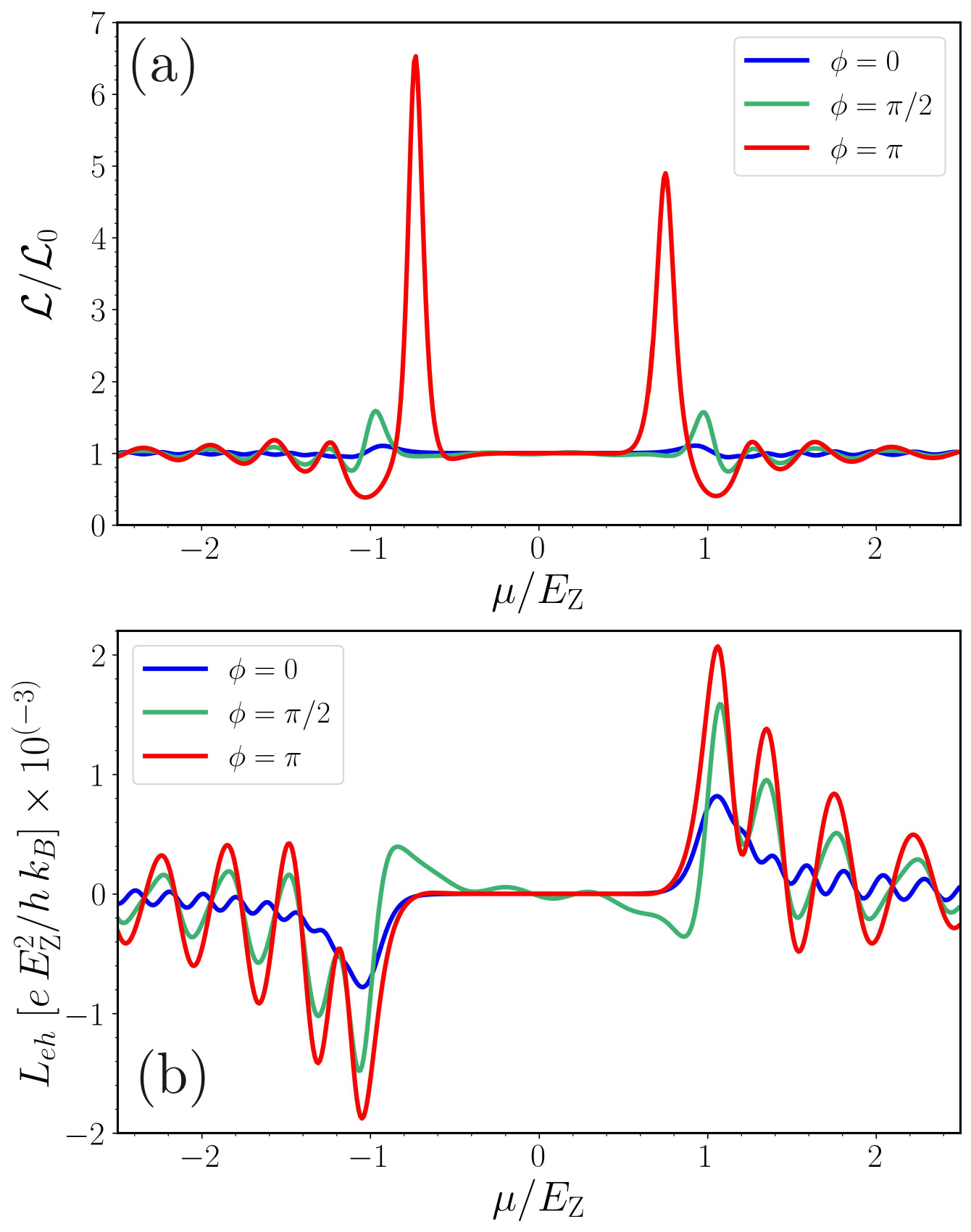}\caption{(a) Wiedemann–Franz ratio $\mathcal{L}/\mathcal{L}_0={K}/{(G\,T\mathcal{L}_0)},
$ as a function of the normalized chemical potential \(\mu/E_\mathrm{Z}\) for three different RSOC misalignment angles (see label) where RSOC NW segments have a length \(L=6\,L_\mathrm{Z}\). (b) Thermoelectric Onsager coefficient $L_{eh}$ for the same  $\phi$ values as in panel (b). The temperature is set such that $k_B T=0.05 E_\mathrm{Z}$. Other parameters as in Fig.~\ref{fig:G_0}.}
\label{fig:L_eh_Widdman}
\end{figure}

 \begin{figure*}[t!]
\includegraphics[width=\linewidth]{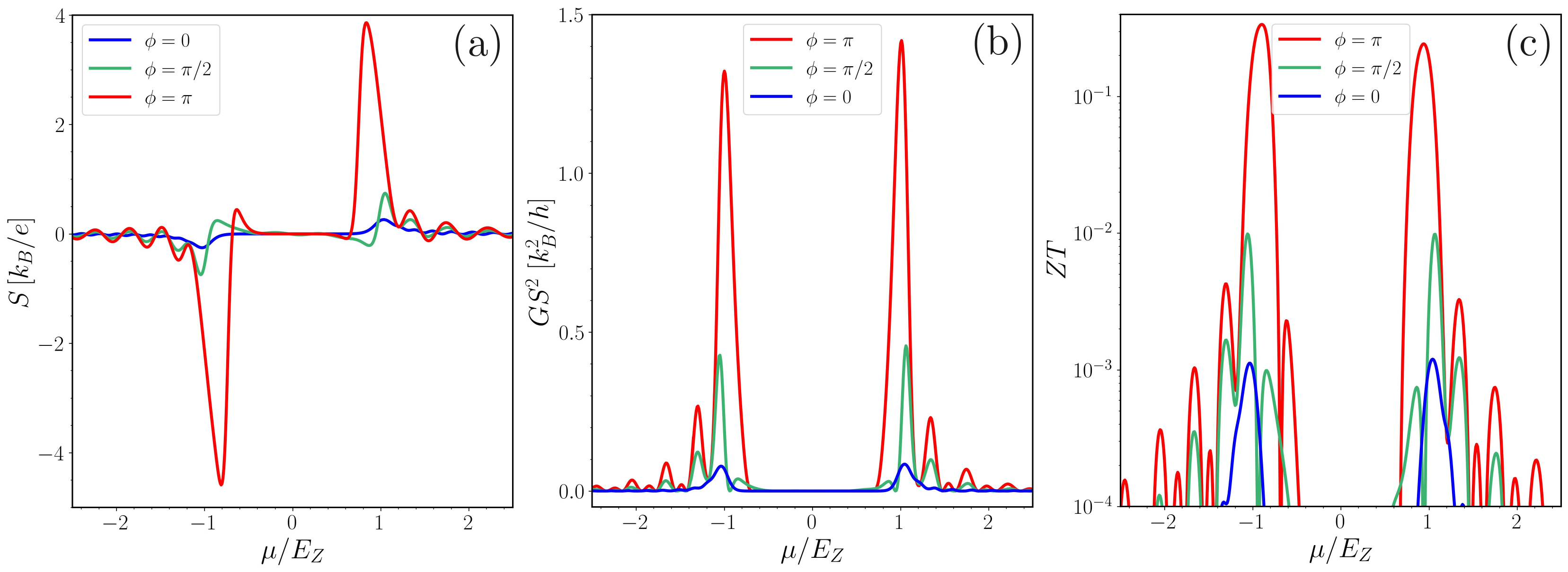} 
\caption{Thermoelectric transport properties of a RSOC NW with segments of length 
\(L=6\,L_\mathrm{Z}\) as functions of the normalized chemical potential \(\mu/E_\mathrm{Z}\) for different RSOC misalignement angle $\phi$ (see label). (a) Seebeck coefficient \(S\), (b) thermoelectric power factor $GS^2$, and (c) figure of merit \(ZT\) in logarithmic scale. Temperature is fixed such that $k_B T=0.05\,E_\mathrm{Z}$. Other parameters as in Fig.~\ref{fig:G_0}.}
\label{fig:Thermal_T}
\end{figure*}
In Fig.~\ref{fig:L_eh_Widdman}(a) we show the dimensionless WF ratio \(\mathcal{L}/\mathcal{L}_0\) as a function of \(\mu / E_\mathrm{Z}\), for various values of the misalignment angle $\phi$. 
For the homogeneous case (blue curve), one sees small deviations of the WF ratio from the universal value ($\mathcal{L}/\mathcal{L}_0=1$) at the magnetic gap boundary. This is essentially due to the opening of propagating channels related to the bottom or the top bands, which are indeed separated by the magnetic gap (see Fig.~\ref{fig:BS}). 
However, still near the magnetic gap boundaries, it is just in the antiparallel configuration $\phi=\pi$ that the WF ratio strongly deviates from the universal value. Indeed, such
large deviations correspond exactly to the chemical potential values where the transmission coefficient ${\cal T}(E)$ exhibits a strong energy dependence [see red curve in Fig.~\ref{fig:G_0}(c)]. Indeed, for $\phi \simeq \pi$, the transmission near the magnetic gap $|\mu| \lesssim E_Z$ is strongly suppressed by the  RSOC misalignment between the two NW regions.

For values of chemical potential well below the magnetic gap, $|\mu| \ll E_\mathrm{Z}$, the WF ratio approaches its universal value and the same occurs far above the magnetic gap, for $|\mu|\gg E_\mathrm{Z}$, up to a residual oscillatory behavior due to Fabry-P\'erot-like resonances. Those are detectable as long as their characteristic energy scales are bigger than the thermal energy.\\ 

{\it Onsager coefficient \(L_{eh}\).} This quantity, directly associated with the thermoelectrical response,  is shown in Fig.~\ref{fig:L_eh_Widdman}(b)    as a function of the normalized chemical potential, for various $\phi$.
As one can see,  near the gap edges \( L_{eh}\) exhibits pronounced peaks,  for all $\phi$ values. The maximal peak height is observed for the antiparallel configuration ($\phi=\pi$), in agreement with the strong energy dependence of the transmission function in this configuration, which also leads to a strong violation of the WF law.

Furthermore, for the two extreme cases, $\phi=0,\,\pi$ (blue and red curves, respectively), the sign of such \(L_{eh}\) peaks is controlled by the sign of the chemical potential. This is because when $\mu\approx- E_\mathrm{Z}$ (\(\mu\approx + E_\mathrm{Z}\)),  \(L_{eh}\) is mainly dominated by the opening of the hole (electron) band [see Fig.~\ref{fig:BS}(b)] at small momentum. However, in the intermediate misalignment case, $\phi=\pi/2$ (green curve), one can also observe the appearance of 
shoulders near the boundaries of the magnetic gap, $|\mu| \lesssim E_Z$. Such 
overshooting shoulders are characterized by a sign change of~$L_{eh}$ across the magnetic gap boundaries $\mu \approx\pm E_Z$.
This peculiar behavior stems from the energy dependence of the transmission function ${\cal T}(E)$, which features Fano-Rashba antiresonances within the magnetic gap. Indeed, according to the Mott formula, at sufficiently low temperatures, the thermoelectric coefficient, \( L_{eh}\), is expected to be proportional to the energy derivative of transmission: $L_{eh}(\mu)\sim\partial_\mu \mathcal{T}(\mu)$~\cite{Benenti2017}. 
For a similar reason, when the chemical potential is well within the magnetic gap ($|\mu|\ll E_\mathrm{Z}$), the thermoelectric coefficient $L_{eh}$ is strongly suppressed in both the homogeneous and antiparallel configurations. This suppression occurs because the transmission function is nearly constant over this energy range.

\subsection{Figures of merit}
Let us now analyze the behavior of various linear thermoelectric figures of merit, which quantify the ability of a TE material to convert heat into electricity.

\begin{figure*}[ht]
\includegraphics[width=\linewidth]{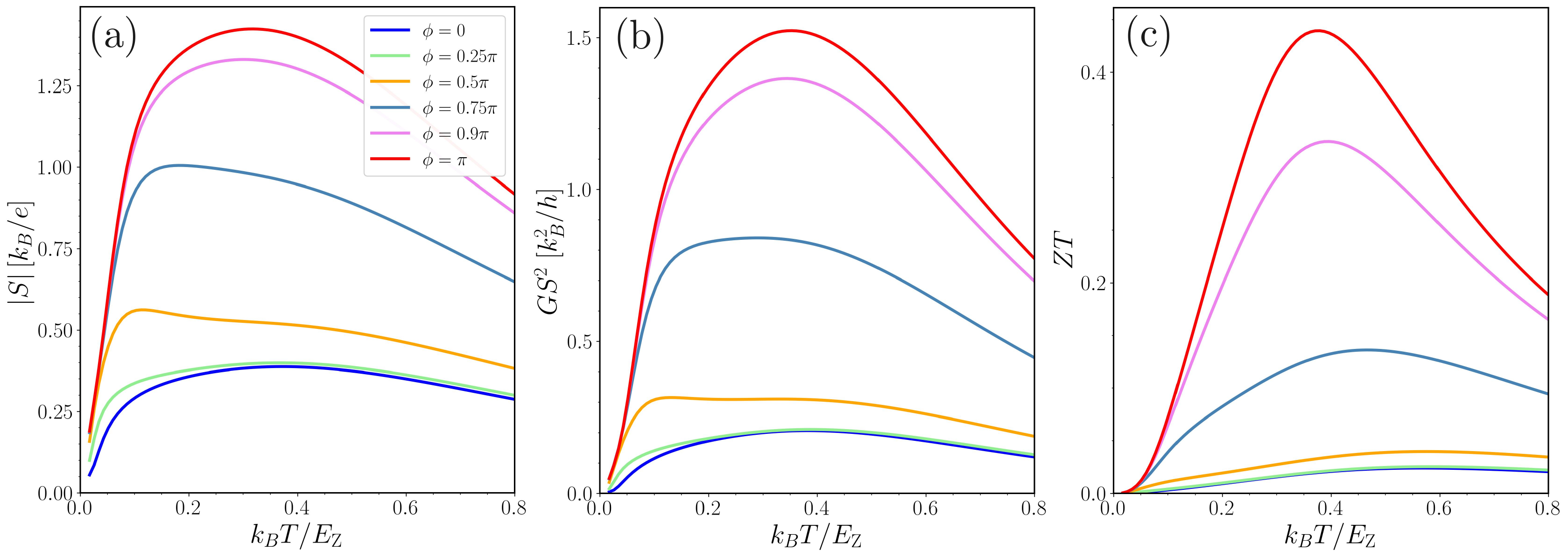}
    \caption{Temperature dependence of the TE properties for different misalignment angles \(\phi\) and for a fixed chemical potential \( \mu = \mu^* \). The Seebeck coefficient \( |S| \), panel (a), the power factor \( G S^2 \), panel (b), and the figure of merit \( ZT \), panel (c), are plotted as functions of the normalized temperature \( k_BT / E_\mathrm{Z} \).  Other parameters as in Fig.~\ref{fig:G_0}. }
    \label{fig:S_GS2_log(ZT)_T}
\end{figure*}

\subsubsection{Seebeck coefficient}
We first consider the Seebeck coefficient $S$, defined in Eq.~(\ref{eq:G_K_S}) and shown in Fig.~\ref{fig:Thermal_T}(a) as a funtion of \(\mu / E_\mathrm{Z}\). We note that, for a homogeneous NW (blue curve), $S$ is strongly suppressed within the magnetic gap and reaches at most a value of about \(\sim 0.3\,k_B/e\), at the magnetic-gap boundaries. However, for the antiparallel configuration $\phi=\pi$ (red curve), $S$ is significantly enhanced with respect to the homogeneous case,  by more than a factor $15$, reaching a maximum value of \(\sim 4.6\,k_B/e\) at $\mu\approx -E_\mathrm{Z}$.
It is worth emphasizing that such an enhancement is mainly due to  the strong suppression of the coefficient $L_{ee}$ appearing in the denominator of Eq.~(\ref{eq:G_K_S}), and partly also to a moderate increase of $L_{eh}$ at the denominator of  Eq.~(\ref{eq:G_K_S}). The former effect is seen in the behavior of the conductance $G=TL_{ee}$ displayed in Fig.~\ref{fig:G_0}(c), whereas the latter increase of $L_{eh}$ by a factor $\sim 2.5$ is shown in  Fig.~\ref{fig:L_eh_Widdman}(b).
We also observe that, for a misalignment angle $\phi=\pi/2$ (green curve), $S$ decreases with respect to the antiparallel configuration, suggesting that the misalignment angle can be harnessed as a knob to tune the thermoelectric effects.

Finally, we wish to comment on the lack of anti-symmetry of $S$ as a function of $\mu$, which is particularly evident for the peaks located around \(\mu=\pm\, E_\mathrm{Z}\). This feature reflects the inherent particle–hole asymmetry of the dispersive spectrum at the bottom of the band. 

\subsubsection{Power factor and thermoelectric efficiency}
The analysis of the power factor $GS^2$ and of the thermoelectric efficiency figure of merit~\cite{Benenti2017}
\begin{equation}
ZT = \frac{G S^2}{K} T
\label{eq:ZT}
\end{equation}
further clarifies the crucial role of the RSOC inhomogeneity.
On the one hand, the power factor, shown in Fig.~\ref{fig:Thermal_T}(b), exhibits an enhancement by a factor of \(\sim 18\) in the antiparallel configuration (red curve) with respect to the homogeneous case (blue curve), an enhancement comparable to that of the Seebeck coefficient.
On the other hand, the figure of merit $ZT$ in the antiparallel configuration ($\phi=\pi$) is   {\it two orders of magnitude} larger than in the homogeneous case ($\phi=0$), as shown in the log-scale plot of Fig.~\ref{fig:Thermal_T}(c).
Indeed, for $\phi=\pi$ we find that $ZT$ reaches values up to \(\sim 0.34\) at \(\mu \approx -E_Z\) and \(\sim 0.24\) at \(\mu \approx +E_Z\).
Although such \(ZT\) values are moderate, it is worth stressing that such a significant increase by a factor of $\sim 340$ implies that the spin-orbit direction inhomogeneities can be regarded as a knob to turn a poor TE system into a good one. Moreover, as we shall discuss below, the \(ZT\) values can be further increased by increasing the working temperature.

Importantly, we have verified that the large enhancement of $ZT$ in the antiparallel configuration compared to the homogeneous configuration is
mainly due to the significant reduction in the thermal conductance~$K$, appearing at the denominator of Eq.~(\ref{eq:ZT}). This effect is much stronger than the corresponding reduction of the conductance $G$ at the numerator of Eq.~(\ref{eq:ZT}). Additionally, the increase of the Seebeck $S$, which appears quadratically at the numerator of Eq.~(\ref{eq:ZT}) also favors the increase of $ZT$.

The figure also clearly demonstrates the strong tunability of the TE properties to the misalignment angle $\phi$ (see Fig.~\ref{fig:Thermal_T}), which can be regarded as a promising strategy to control and substantially enhance the thermoelectric performance of NWs.

Finally, we stress that the plots in Fig.~\ref{fig:Thermal_T} refer to a low temperature, specifically $k_BT=0.05\,E_\mathrm{Z}$. With increasing temperature, thermoelectrical figures of merit are typically expected to improve. Indeed, as we shall discuss below, the optimal temperature for maximizing $GS^2$ and $ZT$ is significantly higher.
Nevertheless, we have verified that the behavior of $S$, $GS^2$, and $ZT$ as functions of the chemical potential, as discussed above, remains qualitatively valid also at higher temperatures, provided that $k_BT \lesssim E_\mathrm{Z}$.

\begin{figure*}[ht]
\includegraphics[width=\linewidth]{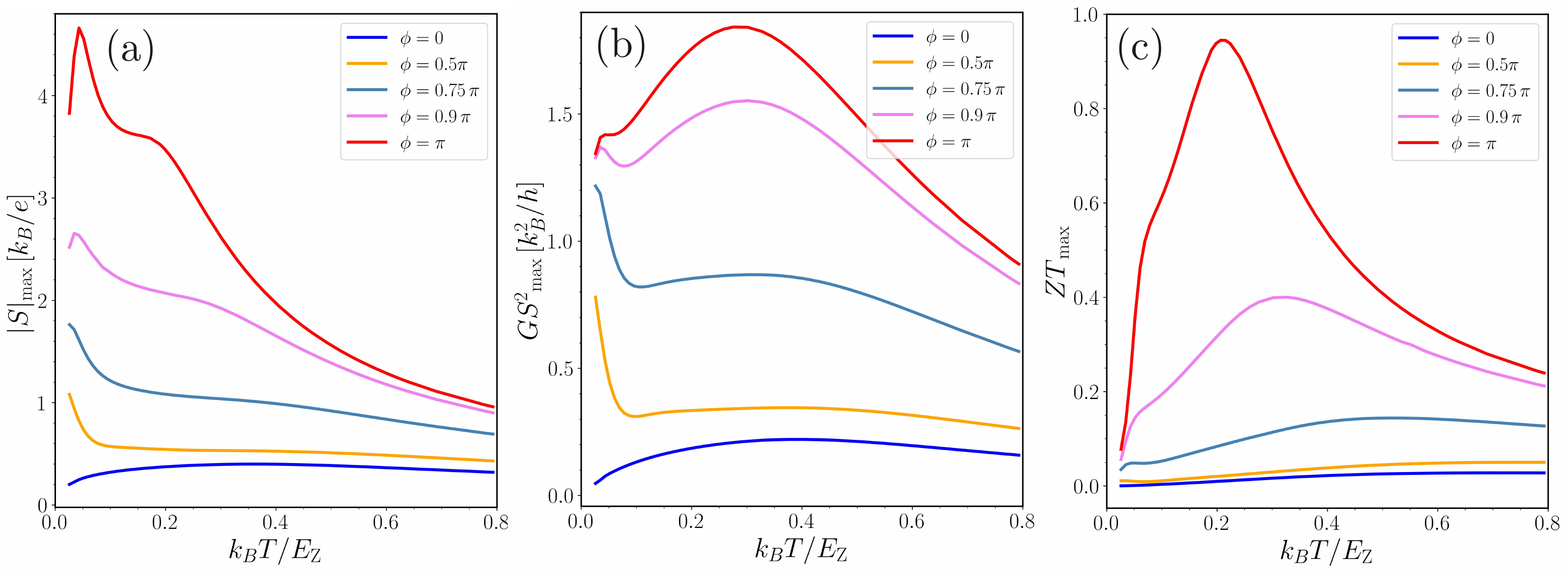}
\caption{Temperature dependence of the TE properties at the optimal chemical potential. The optimal value is calculated over the chemical potential values, for the modulus of the Seebeck coefficient \( \big|S\big|_\mathrm{max} \) (a), of the the power factor \( G S^2_\mathrm{max} \) (b), and of the the figure of merit \( ZT_\mathrm{max} \) (c), as functions of the normalized temperature \( k_B T / E_\mathrm{Z} \).  Other parameters as in Fig.~\ref{fig:G_0}.}
    \label{fig:Smax}
\end{figure*}

\subsection{Temperature dependence of the TE properties}

We now turn to a detailed investigation of how the operating temperature influences the results presented so far. 
We start by outlining the expected temperature dependence of the various TE figures of merit. On the one hand, in the low-temperature limit, $k_B T \ll E_Z$, one typically expects properties such as the Seebeck coefficient to increase with temperature $T$~\cite{Benenti2017}. On the other hand, for temperatures much larger than the magnetic gap ($k_B T \gg 2 E_Z$), thermal broadening is expected to cause effective averaging over the entire energy range, thereby washing out the energy-dependent features due to the magnetic gap and the helical states characterizing the NW. In such a high temperature limit, one expects to recover the conventional TE properties described by a quadratic dispersion relation, where any difference between the homogeneous and inhomogeneous cases disappears. As a consequence,  optimal TE performances should occur in the intermediate temperature range. We anticipate that, for realistic parameter values, the optimal temperature window indeed lies in the range $0.01 < k_B T/E_\mathrm{Z} < 1$.

{\it Fixed doping}.  
Let us first study the temperature dependence of the TE figures of merit for a fixed chemical potential $\mu$.
We choose a value $\mu^*$ that corresponds to a common maximum of the Fabry-P\'erot-like resonance in the zero-temperature conductance, occurring for both the homogeneous and inhomogeneous configurations (see the resonance marked by a red arrow and the vertical dashed line in Fig.~\ref{fig:G_0}).
This choice allows a meaningful comparison of the temperature dependence in the two configurations, without being significantly influenced by the specific Fabry-Perot resonant pattern.

 Figure~\ref{fig:S_GS2_log(ZT)_T}, illustrates how the Seebeck coefficient $S$, the power factor $GS^2$, and the $ZT$ figure of merit depend on the temperature, at such a chemical potential value $\mu^*$. The results are plotted in panels (a), (b), and (c), respectively, for different misalignment angles $\phi$ in different colors.
All curves exhibit maxima at intermediate temperatures for all misalignment angles, with values approaching zero at very low temperatures.
A quite strong dependence on the misalignment angle $\phi$ can be observed, especially in the range $\pi/2 < \phi < \pi$. In particular, all TE figures of merit are higher in the antiparallel configuration compared to the homogeneous case, in full agreement with the discussion in the previous section.

Let us focus e.g. on the Seebeck coefficient [Fig.\ref{fig:S_GS2_log(ZT)_T}(a)]: starting from the homogeneous case [blue line in Fig.~\ref{fig:S_GS2_log(ZT)_T}(a)], one observes a maximum of $\sim  0.3\,k_B/e$ for $k_B T\sim 0.4\,E_\mathrm{Z}$ and a smooth reduction by increasing temperature. This is consistent with the expectation that, for very high temperatures $k_BT\gg E_\mathrm{Z}$,   the homogeneous and inhomogeneous cases should return results identical to a spin-degenerate open channel.
For the antiparallel case (red curve),  $S$ is strongly enhanced, up to $\sim 1.4 k_B/e$ at $k_BT \sim 0.3 E_\mathrm{Z}$. 
As far as the power factor and $ZT$ figure of merit, shown in Figs.~\ref{fig:S_GS2_log(ZT)_T}(b) and (c), respectively, the optimal performance in the antiparallel case is also reached at nearly the same temperature ($k_BT \sim 0.3 E_\mathrm{Z}$).
Notably, \emph{all} three figures of merit are maximized at roughly the same temperature.

These results show that, in general, the magnetic gap energy scale $E_{\mathrm{Z}}$ sets the temperature scale for the maximum TE performance.
An inhomogeneity in the RSOC direction profile characterizing the helical states could indeed provide a strong TE enhancement at a temperature comparable to roughly 30-40\% of the magnetic gap (i.e.~$k_BT\approx 0.3 -0.4E_\mathrm{Z}$).

{\it Optimal doping}.
It is now instructive to examine the temperature dependence of the TE figures of merit evaluated at the optimal chemical potential — that is, the value of chemical potential which maximizes the respective TE property.

The Fig.~\ref{fig:Smax} reports the maximum Seebeck coefficient [panel (a)], power factor [panel (b)], and $ZT$ figure of merit [panel (c)] as functions of temperature.
The result 
confirms that the TE quantities are very sensitive to the misalignment angle $\phi$ and that they improve as $\phi$ increases, with the best performance observed in the antiparallel configuration ($\phi=\pi$).

However, the comparison between Fig.~\ref{fig:Smax} and Fig.~\ref{fig:S_GS2_log(ZT)_T} also features some remarkable differences.
Regarding the Seebeck coefficient, for instance, Fig.~\ref{fig:Smax}(a) showcases, for the antiparallel configuration,   an additional peak located at low temperatures, where $k_BT\sim 0.02 E_\mathrm{Z}$.
This is  due to the Fabry-P\'erot resonances, which occur on a similar energy scale ($\sim(\mu^*-E_Z)$), see Fig.~\ref{fig:G_0}(c).
This phenomenon is similar to the behavior observed in resonant systems, such as quantum dots, which exhibit a single resonance in the zero-temperature conductance.
In such cases, the Seebeck coefficient $S$ increases as the temperature $T$ decreases, provided that $k_B T$ remains larger than the intrinsic resonance linewidth $\gamma$~\cite{Svensson2012,Erdman2017}.
This explains the peak-like behavior of $S$ at low temperatures, including its suppression as $T$ approaches zero. Indeed, in the zero-temperature limit, $S$ must vanish for temperatures satisfying $k_B T \ll \gamma$.

For the other inhomogeneous cases, depicted by the pink, cyan, and yellow curves, the peak shifts to even lower temperatures, in some cases below the minimum temperature considered in the calculation. As a result, only the descending part of the peak is visible for the cyan and yellow curves.
Note that in the homogeneous case, no peak is observed at low temperatures.

Let us now turn to the power factor [Fig.~\ref{fig:Smax}(b)]. In this case, the influence of the resonances is less pronounced than in $S$, at least for strong misalignment $\phi \sim \pi$ (red and pink curves). For such angles, the global maximum occurs at intermediate temperatures, where a higher broader peak is observed. Such a broad peak, is a feature common to all the curves, as in Fig.~\ref{fig:S_GS2_log(ZT)_T}(b), and it is a consequence of the magnetic gap not being affected by the Fabry-Perot-like resonances, which are averaged out at such temperatures. 
     
Finally, for the figure of merit  $ZT$, which expresses the TE thermodynamic efficiency, the optimal value reaches even 0.9 for the antiparallel configuration, around $0.2 E_Z$ [see   Fig.~\ref{fig:Smax}(c)]. This represents a doubling in comparison with Fig.~\ref{fig:S_GS2_log(ZT)_T}(c).
We also report a high sensitivity to the misalignment angle $\phi$, as even a small change from $\phi=\pi$ to $\phi=0.9$ can halve the value of $ZT_{\mathrm{max}}$.

\subsection{Effects of NW length}
Before concluding, we shall focus on the antiparallel configuration and analyze the effects of the NW length on the optimal TE performance.

\begin{figure}[t!]
    \centering
    \includegraphics[width=1.02
    \linewidth]{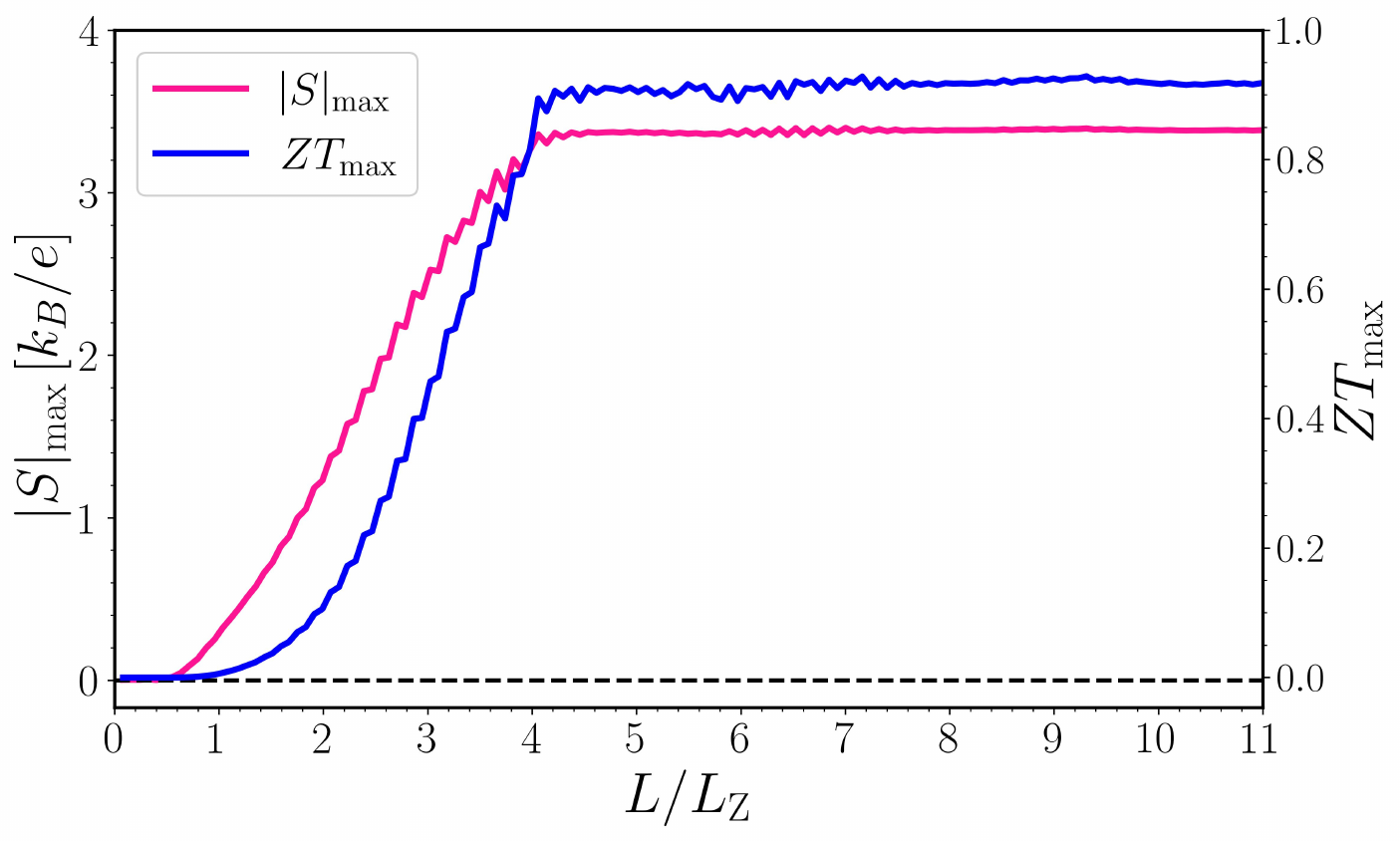}
    \caption{Maximum absolute value of the Seebeck coefficient $\big|S
    \big|_{\mathrm{max}}$ and the maximum of $ZT$
    as a function of the dimensionless length \(L/L_\mathrm{Z}\) for the inhomogeneous RSOC configuration with misalignment angle \(\phi=\pi\). The temperature is fixed such that $k_BT\approx 0.21 E_Z$ to maximize the magnitude of the $ZT_{max}$ coefficient in agreement with the TE properties shown in Fig.~\ref{fig:Smax}(c).  Other parameters as in Fig.~\ref{fig:G_0}.}
    \label{fig:S_L}
\end{figure} 
Figure~\ref{fig:S_L} shows the maximum absolute value of the Seebeck coefficient \(|S|_ \mathrm{max}\) and the maximum TE efficiency $ZT_{\mathrm{max}}$, both computed by optimizing the chemical potential $\mu$, as functions of the dimensionless length $L/L_\mathrm{Z}$. 
We set the temperature to $k_BT=0.21 E_Z$, which corresponds to the value that precisely maximizes $ZT$ [see Fig.~\ref{fig:Smax}(c)]. 
The curves show that both \(\lvert S \rvert_\mathrm{{max}}\) and  $ZT_{\mathrm{max}}$ increase with NW length, eventually saturating to a constant value when $L/L_\mathrm{Z}\gtrsim 4$.
Thus, increasing the length beyond a few magnetic lengths yields no further advantage.
The small irregularities, due to resonant Fabry-P\'erot-like oscillations, do not significantly affect the results.

\section{\label{Sec5}SUMMARY AND CONCLUSION}
In this paper, we have shown that the helical states emerging in the magnetic gap of NWs with strong RSOC and a Zeeman field applied orthogonally to the spin-orbit direction can be exploited to control and enhance the TE properties.
Specifically, we have considered a NW setup consisting of two segments in the Rashba-dominated regime, separated by a sharp interface, that exhibits a misalignment angle $\phi$ between the spin-orbit fields lying in the $y$-$z$ plane  [see Fig.~\ref{fig:System}].

Starting from the zero-temperature limit, we first analyzed electron transport along the NW. When the length $L$ of each NW segment becomes comparable to the magnetic length $L_Z$, the electron conductance $G$ acquires a strong dependence on the chemical potential $\mu$. In particular, for values of $\mu$ outside the magnetic gap $G$ displays a Fabry-P\'erot oscillation pattern due to the quantum interference caused by the backscattering at the interfaces, an effect that is present already for a homogeneous NW ($\phi=0$) and that, for $\phi \neq 0$, is only modified by the presence of the additional backscattering at the sharp interface separating the two segments. However, for values of $\mu$  inside the magnetic gap, the value of $G$ is strongly affected by the RSOC inhomogeneity, and the role of the misalignment angle $\phi$ becomes significant. In particular, for the antiparallel configuration $\phi=\pi$, which realizes the recently studied Dirac-paradox configuration, $G$ gets strongly suppressed as compared to the homogeneous case ($\phi=0$), due to the helicity mismatch in the helical states characterizing the two NW segments [see Fig.~\ref{fig:G_0}]. This analysis has enabled us to identify the long NW regime ($L > L_Z$) as the most suitable for TE properties, given the strongly energy-dependent transmission of the NW.

Then, focusing on the finite but small temperature regime $k_B T \ll E_Z$, we have demonstrated that, when the antiparallel configuration $\phi=\pi$ is realized, and the chemical potential is set near the boundaries of the magnetic gap, $\mu \simeq \pm E_Z$, 
the TE properties are strongly enhanced with respect to a homogeneous NW. In particular, we have shown that these conditions lead to a strong violation of the Wiedemann-Franz law [see Fig.~\ref{fig:L_eh_Widdman}(a)] and to the increase of the thermoelectric Onsager coefficient [see Fig.~\ref{fig:L_eh_Widdman}(b)]. In particular, the latter effect, combined with the strong suppression of the conductance, results in a strong enhancement of the
Seebeck coefficient $S$ in Eq.~(\ref{eq:G_K_S}),  by more than an order of magnitude, as displayed in Fig.~\ref{fig:Thermal_T}. In particular, we predict up to $S \sim 3.5 k_B/e$ and a power factor of $GS^2\sim 1.8\,k_B^2/h$ for temperatures in $T\sim 0.24\,$K. 
\\
One of the most relevant result of our analysis is that the $ZT$ figure of merit, which can reach the value $ZT\sim 0.9$ for the antiparallel configuration, is increased by more than {\it two orders} of magnitude with respect to a homogeneous NW [see  Fig.~\ref{fig:Thermal_T}], an effect that is mainly caused by the significant reduction of the thermal conductance $K$.

We have identified that the optimal TE performance is obtained for temperatures spanning $k_B T/E_Z\sim 0.2\div 0.4 E_Z$, highlighting the roles of the magnetic gap and the inhomogeneous RSOC profile [see Figs.~\ref{fig:S_GS2_log(ZT)_T} and \ref{fig:Smax}].
Indeed, it is the Dirac-paradox configuration, realized with antiparallel alignment ($\phi=\pi$), that yields the highest TE efficiencies at moderate temperatures. We have further investigated how the NW length may affect the TE properties, finding that once the NW length is at least a few magnetic lengths enhancement effect should be appreciable.\\

Most of the figures concerning the TE properties have been obtained using the parameters for a  ballistic InSb NW. Indeed, such NWs are characterized by low electron density, a very small effective electron mass, and a strong $g$-factor. The deep Rashba-dominated regime is accessible at low magnetic field strengths, while RSOC orientations can be controlled using wrap-gate geometries without significantly altering the NW electron density. 
We emphasize, however, that the predicted tunability of TE properties in terms of the spin-orbit orientation is quite general, and not limited to the NW-based, low-temperature platforms.

Our findings suggest that the RSOC‐engineered system can serve as a promising, versatile platform for reversible control of TE in a coherent quantum transport regime, paving the way for novel applications in nanodevices for high‑performance energy harvesting and temperature sensing. Finally, although we have not specifically investigated the spin-dependent transport properties of the effect discussed here, our proposal also has potential applicability in spintronics and spin-thermoelectric nanodevices~\cite{Yang2023}.

\section{Acknowledgements}
Z.A and A.B. acknowledge the TRIL Program by ICTP and CNR project QTHERMONANO. F.T., F.D., and A.B. acknowledge the 
MUR-PRIN 2022—Grant No. 2022B9P8LN-(PE3)-Project NEThEQS “Non-equilibrium coherent thermal effects in quantum systems” in PNRR Mission 4-Component 2-Investment 1.1 “Fondo per il Programma Nazionale di Ricerca e Progetti di Rilevante Interesse Nazionale (PRIN)” funded by the European Union-Next Generation E. 

\section{Author contributions}
Conceptualization, Z. Aslani and A. Braggio; Funding acquisition, A. Braggio, F. Taddei and F. Dolcini; Formal analysis, all authors; Software,Z. Aslani;  Methodology: all authors; Writing-original draft,  Z. Aslani and A. Braggio; Writing-review \& editing, all authors; Supervision, A. Braggio.

\section{Declaration of interests}
The authors declare no competing interests.

\nocite{*}
\bibliography{name}

\begin{thebibliography}{101}%
\makeatletter
\providecommand \@ifxundefined [1]{%
 \@ifx{#1\undefined}
}%
\providecommand \@ifnum [1]{%
 \ifnum #1\expandafter \@firstoftwo
 \else \expandafter \@secondoftwo
 \fi
}%
\providecommand \@ifx [1]{%
 \ifx #1\expandafter \@firstoftwo
 \else \expandafter \@secondoftwo
 \fi
}%
\providecommand \natexlab [1]{#1}%
\providecommand \enquote  [1]{``#1''}%
\providecommand \bibnamefont  [1]{#1}%
\providecommand \bibfnamefont [1]{#1}%
\providecommand \citenamefont [1]{#1}%
\providecommand \href@noop [0]{\@secondoftwo}%
\providecommand \href [0]{\begingroup \@sanitize@url \@href}%
\providecommand \@href[1]{\@@startlink{#1}\@@href}%
\providecommand \@@href[1]{\endgroup#1\@@endlink}%
\providecommand \@sanitize@url [0]{\catcode `\\12\catcode `\$12\catcode `\&12\catcode `\#12\catcode `\^12\catcode `\_12\catcode `\%12\relax}%
\providecommand \@@startlink[1]{}%
\providecommand \@@endlink[0]{}%
\providecommand \url  [0]{\begingroup\@sanitize@url \@url }%
\providecommand \@url [1]{\endgroup\@href {#1}{\urlprefix }}%
\providecommand \urlprefix  [0]{URL }%
\providecommand \Eprint [0]{\href }%
\providecommand \doibase [0]{https://doi.org/}%
\providecommand \selectlanguage [0]{\@gobble}%
\providecommand \bibinfo  [0]{\@secondoftwo}%
\providecommand \bibfield  [0]{\@secondoftwo}%
\providecommand \translation [1]{[#1]}%
\providecommand \BibitemOpen [0]{}%
\providecommand \bibitemStop [0]{}%
\providecommand \bibitemNoStop [0]{.\EOS\space}%
\providecommand \EOS [0]{\spacefactor3000\relax}%
\providecommand \BibitemShut  [1]{\csname bibitem#1\endcsname}%
\let\auto@bib@innerbib\@empty
\bibitem [{\citenamefont {Petsagkourakis}\ \emph {et~al.}(2018)\citenamefont {Petsagkourakis}, \citenamefont {Tybrandt}, \citenamefont {Crispin}, \citenamefont {Ohkubo}, \citenamefont {Satoh},\ and\ \citenamefont {Mori}}]{Petsagkourakis2018}%
  \BibitemOpen
  \bibfield  {author} {\bibinfo {author} {\bibfnamefont {I.}~\bibnamefont {Petsagkourakis}}, \bibinfo {author} {\bibfnamefont {K.}~\bibnamefont {Tybrandt}}, \bibinfo {author} {\bibfnamefont {X.}~\bibnamefont {Crispin}}, \bibinfo {author} {\bibfnamefont {I.}~\bibnamefont {Ohkubo}}, \bibinfo {author} {\bibfnamefont {N.}~\bibnamefont {Satoh}},\ and\ \bibinfo {author} {\bibfnamefont {T.}~\bibnamefont {Mori}},\ }\bibfield  {title} {\bibinfo {title} {Thermoelectric materials and applications for energy harvesting power generation},\ }\href {https://doi.org/10.1080/14686996.2018.1530938} {\bibfield  {journal} {\bibinfo  {journal} {Sci. Technol. Adv. Mater.}\ }\textbf {\bibinfo {volume} {19}},\ \bibinfo {pages} {836} (\bibinfo {year} {2018})}\BibitemShut {NoStop}%
\bibitem [{\citenamefont {Singh}\ \emph {et~al.}(2024)\citenamefont {Singh}, \citenamefont {Dogra}, \citenamefont {Dixit}, \citenamefont {Vatin}, \citenamefont {Bhardwaj}, \citenamefont {Sundramoorthy}, \citenamefont {Perera}, \citenamefont {Patole}, \citenamefont {Mishra},\ and\ \citenamefont {Arya}}]{Singh2024}%
  \BibitemOpen
  \bibfield  {author} {\bibinfo {author} {\bibfnamefont {R.}~\bibnamefont {Singh}}, \bibinfo {author} {\bibfnamefont {S.}~\bibnamefont {Dogra}}, \bibinfo {author} {\bibfnamefont {S.}~\bibnamefont {Dixit}}, \bibinfo {author} {\bibfnamefont {N.~I.}\ \bibnamefont {Vatin}}, \bibinfo {author} {\bibfnamefont {R.}~\bibnamefont {Bhardwaj}}, \bibinfo {author} {\bibfnamefont {A.~K.}\ \bibnamefont {Sundramoorthy}}, \bibinfo {author} {\bibfnamefont {H.~C.~S.}\ \bibnamefont {Perera}}, \bibinfo {author} {\bibfnamefont {S.~P.}\ \bibnamefont {Patole}}, \bibinfo {author} {\bibfnamefont {R.~K.}\ \bibnamefont {Mishra}},\ and\ \bibinfo {author} {\bibfnamefont {S.}~\bibnamefont {Arya}},\ }\bibfield  {title} {\bibinfo {title} {Advancements in thermoelectric materials for efficient waste heat recovery and renewable energy generation},\ }\href {https://doi.org/10.1016/j.hybadv.2024.100176} {\bibfield  {journal} {\bibinfo  {journal} {Hybrid Adv.}\ }\textbf {\bibinfo {volume} {5}},\ \bibinfo {pages} {100176} (\bibinfo {year}
  {2024})}\BibitemShut {NoStop}%
\bibitem [{\citenamefont {Chen}\ \emph {et~al.}(2019{\natexlab{a}})\citenamefont {Chen}, \citenamefont {Lee}, \citenamefont {Lee},\ and\ \citenamefont {Li}}]{Chen2019}%
  \BibitemOpen
  \bibfield  {author} {\bibinfo {author} {\bibfnamefont {R.}~\bibnamefont {Chen}}, \bibinfo {author} {\bibfnamefont {J.}~\bibnamefont {Lee}}, \bibinfo {author} {\bibfnamefont {W.}~\bibnamefont {Lee}},\ and\ \bibinfo {author} {\bibfnamefont {D.}~\bibnamefont {Li}},\ }\bibfield  {title} {\bibinfo {title} {Thermoelectrics of nanowires},\ }\href {https://doi.org/10.1021/acs.chemrev.8b00627} {\bibfield  {journal} {\bibinfo  {journal} {Chem. Rev.}\ }\textbf {\bibinfo {volume} {119}},\ \bibinfo {pages} {9260} (\bibinfo {year} {2019}{\natexlab{a}})}\BibitemShut {NoStop}%
\bibitem [{\citenamefont {Ishibe}\ \emph {et~al.}(2018)\citenamefont {Ishibe}, \citenamefont {Tomeda}, \citenamefont {Watanabe}, \citenamefont {Kamakura}, \citenamefont {Mori}, \citenamefont {Naruse}, \citenamefont {Mera}, \citenamefont {Yamashita},\ and\ \citenamefont {Nakamura}}]{Ishibe2018}%
  \BibitemOpen
  \bibfield  {author} {\bibinfo {author} {\bibfnamefont {T.}~\bibnamefont {Ishibe}}, \bibinfo {author} {\bibfnamefont {A.}~\bibnamefont {Tomeda}}, \bibinfo {author} {\bibfnamefont {K.}~\bibnamefont {Watanabe}}, \bibinfo {author} {\bibfnamefont {Y.}~\bibnamefont {Kamakura}}, \bibinfo {author} {\bibfnamefont {N.}~\bibnamefont {Mori}}, \bibinfo {author} {\bibfnamefont {N.}~\bibnamefont {Naruse}}, \bibinfo {author} {\bibfnamefont {Y.}~\bibnamefont {Mera}}, \bibinfo {author} {\bibfnamefont {Y.}~\bibnamefont {Yamashita}},\ and\ \bibinfo {author} {\bibfnamefont {Y.}~\bibnamefont {Nakamura}},\ }\bibfield  {title} {\bibinfo {title} {Methodology of thermoelectric power factor enhancement by controlling nanowire interface},\ }\href {https://doi.org/10.1021/acsami.8b13528} {\bibfield  {journal} {\bibinfo  {journal} {ACS Appl. Mater. Interfaces}\ }\textbf {\bibinfo {volume} {10}},\ \bibinfo {pages} {37709} (\bibinfo {year} {2018})}\BibitemShut {NoStop}%
\bibitem [{Ma2()}]{Ma2021}%
  \BibitemOpen
  \bibfield  {title} {\bibinfo {title} {{Review of experimental approaches for improving zT of thermoelectric materials}},\ }\href@noop {} {\bibinfo  {journal} {Mater. Sci. Semicond. Process.}\ }\BibitemShut {NoStop}%
\bibitem [{\citenamefont {Chen}\ \emph {et~al.}(2019{\natexlab{b}})\citenamefont {Chen}, \citenamefont {Lee}, \citenamefont {Lee},\ and\ \citenamefont {Li}}]{chen2019thermoelectrics}%
  \BibitemOpen
\bibfield  {journal} {  }\bibfield  {author} {\bibinfo {author} {\bibfnamefont {R.}~\bibnamefont {Chen}}, \bibinfo {author} {\bibfnamefont {J.}~\bibnamefont {Lee}}, \bibinfo {author} {\bibfnamefont {W.}~\bibnamefont {Lee}},\ and\ \bibinfo {author} {\bibfnamefont {D.}~\bibnamefont {Li}},\ }\bibfield  {title} {\bibinfo {title} {Thermoelectrics of nanowires},\ }\href {https://doi.org/10.1021/acs.chemrev.8b00627} {\bibfield  {journal} {\bibinfo  {journal} {Chem. Rev.}\ }\textbf {\bibinfo {volume} {119}},\ \bibinfo {pages} {9260} (\bibinfo {year} {2019}{\natexlab{b}})}\BibitemShut {NoStop}%
\bibitem [{\citenamefont {Hicks}\ and\ \citenamefont {Dresselhaus}(1993{\natexlab{a}})}]{Hicks1993a}%
  \BibitemOpen
  \bibfield  {author} {\bibinfo {author} {\bibfnamefont {L.~D.}\ \bibnamefont {Hicks}}\ and\ \bibinfo {author} {\bibfnamefont {M.~S.}\ \bibnamefont {Dresselhaus}},\ }\bibfield  {title} {\bibinfo {title} {Effect of quantum-well structures on the thermoelectric figure of merit},\ }\href {https://doi.org/10.1103/PhysRevB.47.12727} {\bibfield  {journal} {\bibinfo  {journal} {Phys. Rev. B}\ }\textbf {\bibinfo {volume} {47}},\ \bibinfo {pages} {12727} (\bibinfo {year} {1993}{\natexlab{a}})}\BibitemShut {NoStop}%
\bibitem [{\citenamefont {Hicks}\ and\ \citenamefont {Dresselhaus}(1993{\natexlab{b}})}]{Hicks1993b}%
  \BibitemOpen
  \bibfield  {author} {\bibinfo {author} {\bibfnamefont {L.~D.}\ \bibnamefont {Hicks}}\ and\ \bibinfo {author} {\bibfnamefont {M.~S.}\ \bibnamefont {Dresselhaus}},\ }\bibfield  {title} {\bibinfo {title} {Thermoelectric figure of merit of a one-dimensional conductor},\ }\href {https://doi.org/10.1103/PhysRevB.47.16631} {\bibfield  {journal} {\bibinfo  {journal} {Phys. Rev. B}\ }\textbf {\bibinfo {volume} {47}},\ \bibinfo {pages} {16631} (\bibinfo {year} {1993}{\natexlab{b}})}\BibitemShut {NoStop}%
\bibitem [{\citenamefont {Mao}\ \emph {et~al.}(2016)\citenamefont {Mao}, \citenamefont {Liu},\ and\ \citenamefont {Ren}}]{Mao2016}%
  \BibitemOpen
  \bibfield  {author} {\bibinfo {author} {\bibfnamefont {J.}~\bibnamefont {Mao}}, \bibinfo {author} {\bibfnamefont {Z.}~\bibnamefont {Liu}},\ and\ \bibinfo {author} {\bibfnamefont {Z.}~\bibnamefont {Ren}},\ }\bibfield  {title} {\bibinfo {title} {Size effect in thermoelectric materials},\ }\href {https://doi.org/10.1038/npjquantmats.2016.28} {\bibfield  {journal} {\bibinfo  {journal} {npj Quantum Mater.}\ }\textbf {\bibinfo {volume} {1}},\ \bibinfo {pages} {16028} (\bibinfo {year} {2016})}\BibitemShut {NoStop}%
\bibitem [{\citenamefont {Bauer}\ \emph {et~al.}(2012)\citenamefont {Bauer}, \citenamefont {Saitoh},\ and\ \citenamefont {van Wees}}]{Bauer2012}%
  \BibitemOpen
  \bibfield  {author} {\bibinfo {author} {\bibfnamefont {G.~E.~W.}\ \bibnamefont {Bauer}}, \bibinfo {author} {\bibfnamefont {E.}~\bibnamefont {Saitoh}},\ and\ \bibinfo {author} {\bibfnamefont {B.~J.}\ \bibnamefont {van Wees}},\ }\bibfield  {title} {\bibinfo {title} {Spin caloritronics},\ }\href {https://doi.org/10.1038/nmat3301} {\bibfield  {journal} {\bibinfo  {journal} {Nat. Mater.}\ }\textbf {\bibinfo {volume} {11}},\ \bibinfo {pages} {391} (\bibinfo {year} {2012})}\BibitemShut {NoStop}%
\bibitem [{\citenamefont {Yang}\ \emph {et~al.}(2023)\citenamefont {Yang}, \citenamefont {Sang}, \citenamefont {Zhang}, \citenamefont {Ye}, \citenamefont {Hamilton}, \citenamefont {Fuhrer},\ and\ \citenamefont {Wang}}]{Yang2023}%
  \BibitemOpen
  \bibfield  {author} {\bibinfo {author} {\bibfnamefont {G.}~\bibnamefont {Yang}}, \bibinfo {author} {\bibfnamefont {L.}~\bibnamefont {Sang}}, \bibinfo {author} {\bibfnamefont {C.}~\bibnamefont {Zhang}}, \bibinfo {author} {\bibfnamefont {N.}~\bibnamefont {Ye}}, \bibinfo {author} {\bibfnamefont {A.}~\bibnamefont {Hamilton}}, \bibinfo {author} {\bibfnamefont {M.~S.}\ \bibnamefont {Fuhrer}},\ and\ \bibinfo {author} {\bibfnamefont {X.}~\bibnamefont {Wang}},\ }\bibfield  {title} {\bibinfo {title} {The role of spin in thermoelectricity},\ }\href {https://doi.org/10.1038/s42254-023-00604-0} {\bibfield  {journal} {\bibinfo  {journal} {Nat. Rev. Phys.}\ }\textbf {\bibinfo {volume} {5}},\ \bibinfo {pages} {466} (\bibinfo {year} {2023})}\BibitemShut {NoStop}%
\bibitem [{\citenamefont {Yuan}\ \emph {et~al.}(2018)\citenamefont {Yuan}, \citenamefont {Cai}, \citenamefont {Shen}, \citenamefont {Xiao}, \citenamefont {Ren}, \citenamefont {Wang}, \citenamefont {Feng},\ and\ \citenamefont {Yan}}]{Yuan2018}%
  \BibitemOpen
  \bibfield  {author} {\bibinfo {author} {\bibfnamefont {J.}~\bibnamefont {Yuan}}, \bibinfo {author} {\bibfnamefont {Y.}~\bibnamefont {Cai}}, \bibinfo {author} {\bibfnamefont {L.}~\bibnamefont {Shen}}, \bibinfo {author} {\bibfnamefont {Y.}~\bibnamefont {Xiao}}, \bibinfo {author} {\bibfnamefont {J.-C.}\ \bibnamefont {Ren}}, \bibinfo {author} {\bibfnamefont {A.}~\bibnamefont {Wang}}, \bibinfo {author} {\bibfnamefont {Y.~P.}\ \bibnamefont {Feng}},\ and\ \bibinfo {author} {\bibfnamefont {X.}~\bibnamefont {Yan}},\ }\bibfield  {title} {\bibinfo {title} {One-dimensional thermoelectrics induced by {Rashba} spin-orbit coupling in two-dimensional bisb monolayer},\ }\href {https://doi.org/10.1016/j.nanoen.2018.07.041} {\bibfield  {journal} {\bibinfo  {journal} {Nano Energy}\ }\textbf {\bibinfo {volume} {52}},\ \bibinfo {pages} {163} (\bibinfo {year} {2018})}\BibitemShut {NoStop}%
\bibitem [{\citenamefont {Hong}\ \emph {et~al.}(2020)\citenamefont {Hong}, \citenamefont {Lyv}, \citenamefont {Li}, \citenamefont {Xu}, \citenamefont {Sun}, \citenamefont {Zou},\ and\ \citenamefont {Chen}}]{Hong2020}%
  \BibitemOpen
  \bibfield  {author} {\bibinfo {author} {\bibfnamefont {M.}~\bibnamefont {Hong}}, \bibinfo {author} {\bibfnamefont {W.}~\bibnamefont {Lyv}}, \bibinfo {author} {\bibfnamefont {M.}~\bibnamefont {Li}}, \bibinfo {author} {\bibfnamefont {S.}~\bibnamefont {Xu}}, \bibinfo {author} {\bibfnamefont {Q.}~\bibnamefont {Sun}}, \bibinfo {author} {\bibfnamefont {J.}~\bibnamefont {Zou}},\ and\ \bibinfo {author} {\bibfnamefont {Z.-G.}\ \bibnamefont {Chen}},\ }\bibfield  {title} {\bibinfo {title} {Rashba effect maximizes thermoelectric performance of gete derivatives},\ }\href {https://doi.org/10.1016/j.joule.2020.07.021} {\bibfield  {journal} {\bibinfo  {journal} {Joule}\ }\textbf {\bibinfo {volume} {4}},\ \bibinfo {pages} {2030} (\bibinfo {year} {2020})}\BibitemShut {NoStop}%
\bibitem [{\citenamefont {Tian}\ \emph {et~al.}(2021)\citenamefont {Tian}, \citenamefont {Zhang}, \citenamefont {Qin},\ and\ \citenamefont {Qin}}]{Tian2021}%
  \BibitemOpen
  \bibfield  {author} {\bibinfo {author} {\bibfnamefont {Q.}~\bibnamefont {Tian}}, \bibinfo {author} {\bibfnamefont {W.}~\bibnamefont {Zhang}}, \bibinfo {author} {\bibfnamefont {Z.}~\bibnamefont {Qin}},\ and\ \bibinfo {author} {\bibfnamefont {G.}~\bibnamefont {Qin}},\ }\bibfield  {title} {\bibinfo {title} {Novel optimization perspectives for thermoelectric properties based on {Rashba} spin splitting: a mini review},\ }\href {https://doi.org/10.1039/D1NR04323D} {\bibfield  {journal} {\bibinfo  {journal} {Nanoscale}\ }\textbf {\bibinfo {volume} {13}},\ \bibinfo {pages} {18032} (\bibinfo {year} {2021})}\BibitemShut {NoStop}%
\bibitem [{\citenamefont {Yang}\ \emph {et~al.}(2025)\citenamefont {Yang}, \citenamefont {Sang}, \citenamefont {Zhang}, \citenamefont {See}, \citenamefont {Hamilton}, \citenamefont {Fuhrer}, \citenamefont {Ye}, \citenamefont {Snyder},\ and\ \citenamefont {Wang}}]{yang2025}%
  \BibitemOpen
  \bibfield  {author} {\bibinfo {author} {\bibfnamefont {G.}~\bibnamefont {Yang}}, \bibinfo {author} {\bibfnamefont {L.}~\bibnamefont {Sang}}, \bibinfo {author} {\bibfnamefont {C.}~\bibnamefont {Zhang}}, \bibinfo {author} {\bibfnamefont {K.}~\bibnamefont {See}}, \bibinfo {author} {\bibfnamefont {A.}~\bibnamefont {Hamilton}}, \bibinfo {author} {\bibfnamefont {M.}~\bibnamefont {Fuhrer}}, \bibinfo {author} {\bibfnamefont {N.}~\bibnamefont {Ye}}, \bibinfo {author} {\bibfnamefont {G.~J.}\ \bibnamefont {Snyder}},\ and\ \bibinfo {author} {\bibfnamefont {X.}~\bibnamefont {Wang}},\ }\bibfield  {title} {\bibinfo {title} {Advances in topological thermoelectrics: Harnessing quantum materials for energy applications},\ }\href {https://doi.org/https://doi.org/10.1002/adma.202506417} {\bibfield  {journal} {\bibinfo  {journal} {Adv. Mater.}\ }\textbf {\bibinfo {volume} {37}},\ \bibinfo {pages} {2506417} (\bibinfo {year} {2025})}\BibitemShut {NoStop}%
\bibitem [{\citenamefont {Xu}\ \emph {et~al.}(2017)\citenamefont {Xu}, \citenamefont {Xu},\ and\ \citenamefont {Zhu}}]{Xu2017}%
  \BibitemOpen
  \bibfield  {author} {\bibinfo {author} {\bibfnamefont {N.}~\bibnamefont {Xu}}, \bibinfo {author} {\bibfnamefont {Y.}~\bibnamefont {Xu}},\ and\ \bibinfo {author} {\bibfnamefont {J.}~\bibnamefont {Zhu}},\ }\bibfield  {title} {\bibinfo {title} {Topological insulators for thermoelectrics},\ }\href {https://doi.org/10.1038/s41535-017-0054-3} {\bibfield  {journal} {\bibinfo  {journal} {npj Quantum Mater.}\ }\textbf {\bibinfo {volume} {2}},\ \bibinfo {pages} {51} (\bibinfo {year} {2017})}\BibitemShut {NoStop}%
\bibitem [{\citenamefont {Gooth}\ \emph {et~al.}(2015)\citenamefont {Gooth}, \citenamefont {Gluschke}, \citenamefont {Zierold}, \citenamefont {Leijnse}, \citenamefont {Linke},\ and\ \citenamefont {Nielsch}}]{Gooth2015ThermoelectricTI}%
  \BibitemOpen
  \bibfield  {author} {\bibinfo {author} {\bibfnamefont {J.}~\bibnamefont {Gooth}}, \bibinfo {author} {\bibfnamefont {J.~G.}\ \bibnamefont {Gluschke}}, \bibinfo {author} {\bibfnamefont {R.}~\bibnamefont {Zierold}}, \bibinfo {author} {\bibfnamefont {M.}~\bibnamefont {Leijnse}}, \bibinfo {author} {\bibfnamefont {H.}~\bibnamefont {Linke}},\ and\ \bibinfo {author} {\bibfnamefont {K.}~\bibnamefont {Nielsch}},\ }\bibfield  {title} {\bibinfo {title} {Thermoelectric performance of classical topological insulator nanowires},\ }\href {https://doi.org/10.1088/0268-1242/30/1/015015} {\bibfield  {journal} {\bibinfo  {journal} {Semicond. Sci. Technol.}\ }\textbf {\bibinfo {volume} {30}},\ \bibinfo {pages} {015015} (\bibinfo {year} {2015})}\BibitemShut {NoStop}%
\bibitem [{\citenamefont {Toriyama}\ and\ \citenamefont {Snyder}(2025)}]{Toriyama2025}%
  \BibitemOpen
  \bibfield  {author} {\bibinfo {author} {\bibfnamefont {M.~Y.}\ \bibnamefont {Toriyama}}\ and\ \bibinfo {author} {\bibfnamefont {G.~J.}\ \bibnamefont {Snyder}},\ }\bibfield  {title} {\bibinfo {title} {Topological insulators for thermoelectrics: A perspective from beneath the surface},\ }\href {https://doi.org/10.1016/j.xinn.2024.100782} {\bibfield  {journal} {\bibinfo  {journal} {Innov.}\ }\textbf {\bibinfo {volume} {6}} (\bibinfo {year} {2025})}\BibitemShut {NoStop}%
\bibitem [{\citenamefont {Blasi}\ \emph {et~al.}(2020)\citenamefont {Blasi}, \citenamefont {Taddei}, \citenamefont {Arrachea}, \citenamefont {Carrega},\ and\ \citenamefont {Braggio}}]{Blasi2020}%
  \BibitemOpen
  \bibfield  {author} {\bibinfo {author} {\bibfnamefont {G.}~\bibnamefont {Blasi}}, \bibinfo {author} {\bibfnamefont {F.}~\bibnamefont {Taddei}}, \bibinfo {author} {\bibfnamefont {L.}~\bibnamefont {Arrachea}}, \bibinfo {author} {\bibfnamefont {M.}~\bibnamefont {Carrega}},\ and\ \bibinfo {author} {\bibfnamefont {A.}~\bibnamefont {Braggio}},\ }\bibfield  {title} {\bibinfo {title} {Nonlocal thermoelectricity in a superconductor–topological-insulator–superconductor junction in contact with a normal-metal probe: Evidence for helical edge states},\ }\href {https://doi.org/10.1103/PhysRevLett.124.227701} {\bibfield  {journal} {\bibinfo  {journal} {Phys. Rev. Lett.}\ }\textbf {\bibinfo {volume} {124}},\ \bibinfo {pages} {227701} (\bibinfo {year} {2020})}\BibitemShut {NoStop}%
\bibitem [{\citenamefont {Blasi}\ \emph {et~al.}(2021)\citenamefont {Blasi}, \citenamefont {Taddei}, \citenamefont {Arrachea}, \citenamefont {Carrega},\ and\ \citenamefont {Braggio}}]{Blasi2021}%
  \BibitemOpen
  \bibfield  {author} {\bibinfo {author} {\bibfnamefont {G.}~\bibnamefont {Blasi}}, \bibinfo {author} {\bibfnamefont {F.}~\bibnamefont {Taddei}}, \bibinfo {author} {\bibfnamefont {L.}~\bibnamefont {Arrachea}}, \bibinfo {author} {\bibfnamefont {M.}~\bibnamefont {Carrega}},\ and\ \bibinfo {author} {\bibfnamefont {A.}~\bibnamefont {Braggio}},\ }\bibfield  {title} {\bibinfo {title} {Nonlocal thermoelectric engines in hybrid topological josephson junctions},\ }\href {https://doi.org/10.1103/PhysRevB.103.235434} {\bibfield  {journal} {\bibinfo  {journal} {Phys. Rev. B}\ }\textbf {\bibinfo {volume} {103}},\ \bibinfo {pages} {235434} (\bibinfo {year} {2021})}\BibitemShut {NoStop}%
\bibitem [{\citenamefont {Arrachea}\ \emph {et~al.}(2025{\natexlab{a}})\citenamefont {Arrachea}, \citenamefont {Braggio}, \citenamefont {Burset}, \citenamefont {Lee}, \citenamefont {Yeyati},\ and\ \citenamefont {Sánchez}}]{Arrachea2025}%
  \BibitemOpen
  \bibfield  {author} {\bibinfo {author} {\bibfnamefont {L.}~\bibnamefont {Arrachea}}, \bibinfo {author} {\bibfnamefont {A.}~\bibnamefont {Braggio}}, \bibinfo {author} {\bibfnamefont {P.}~\bibnamefont {Burset}}, \bibinfo {author} {\bibfnamefont {E.~J.~H.}\ \bibnamefont {Lee}}, \bibinfo {author} {\bibfnamefont {A.~L.}\ \bibnamefont {Yeyati}},\ and\ \bibinfo {author} {\bibfnamefont {R.}~\bibnamefont {Sánchez}},\ }\bibfield  {title} {\bibinfo {title} {Thermoelectric processes of quantum normal-superconductor interfaces},\ }\href {https://doi.org/10.1002/andp.202500197} {\bibfield  {journal} {\bibinfo  {journal} {Ann. Phys. (Leipzig)}\ ,\ \bibinfo {pages} {e00197}} (\bibinfo {year} {2025}{\natexlab{a}})}\BibitemShut {NoStop}%
\bibitem [{\citenamefont {Mukhopadhyay}\ and\ \citenamefont {Das}(2021)}]{Mukhopadhyay2021}%
  \BibitemOpen
  \bibfield  {author} {\bibinfo {author} {\bibfnamefont {A.}~\bibnamefont {Mukhopadhyay}}\ and\ \bibinfo {author} {\bibfnamefont {S.}~\bibnamefont {Das}},\ }\bibfield  {title} {\bibinfo {title} {Thermal signature of the majorana fermion in a josephson junction},\ }\href {https://doi.org/10.1103/PhysRevB.103.144502} {\bibfield  {journal} {\bibinfo  {journal} {Phys. Rev. B}\ }\textbf {\bibinfo {volume} {103}},\ \bibinfo {pages} {144502} (\bibinfo {year} {2021})}\BibitemShut {NoStop}%
\bibitem [{\citenamefont {Alisultanov}\ \emph {et~al.}(2025)\citenamefont {Alisultanov}, \citenamefont {Idrisov},\ and\ \citenamefont {Kavokin}}]{Alisultanov2025}%
  \BibitemOpen
  \bibfield  {author} {\bibinfo {author} {\bibfnamefont {Z.~Z.}\ \bibnamefont {Alisultanov}}, \bibinfo {author} {\bibfnamefont {E.~G.}\ \bibnamefont {Idrisov}},\ and\ \bibinfo {author} {\bibfnamefont {A.~V.}\ \bibnamefont {Kavokin}},\ }\bibfield  {title} {\bibinfo {title} {Thermoelectric effects in two-dimensional topological insulators},\ }\href {https://doi.org/10.1103/PhysRevB.111.155430} {\bibfield  {journal} {\bibinfo  {journal} {Phys. Rev. B}\ }\textbf {\bibinfo {volume} {111}},\ \bibinfo {pages} {155430} (\bibinfo {year} {2025})}\BibitemShut {NoStop}%
\bibitem [{\citenamefont {Liu}\ \emph {et~al.}(2025)\citenamefont {Liu}, \citenamefont {Jin}, \citenamefont {Pandey}, \citenamefont {Sun}, \citenamefont {Liu}, \citenamefont {Li}, \citenamefont {Rodriguez}, \citenamefont {Wang}, \citenamefont {Zhou}, \citenamefont {Chen}, \citenamefont {Sun}, \citenamefont {Yang}, \citenamefont {Chrzan}, \citenamefont {Lindsay}, \citenamefont {Wu},\ and\ \citenamefont {Yao}}]{Liu2025}%
  \BibitemOpen
  \bibfield  {author} {\bibinfo {author} {\bibfnamefont {Y.}~\bibnamefont {Liu}}, \bibinfo {author} {\bibfnamefont {L.}~\bibnamefont {Jin}}, \bibinfo {author} {\bibfnamefont {T.}~\bibnamefont {Pandey}}, \bibinfo {author} {\bibfnamefont {H.}~\bibnamefont {Sun}}, \bibinfo {author} {\bibfnamefont {Y.}~\bibnamefont {Liu}}, \bibinfo {author} {\bibfnamefont {X.}~\bibnamefont {Li}}, \bibinfo {author} {\bibfnamefont {A.}~\bibnamefont {Rodriguez}}, \bibinfo {author} {\bibfnamefont {Y.}~\bibnamefont {Wang}}, \bibinfo {author} {\bibfnamefont {T.}~\bibnamefont {Zhou}}, \bibinfo {author} {\bibfnamefont {R.}~\bibnamefont {Chen}}, \bibinfo {author} {\bibfnamefont {Y.}~\bibnamefont {Sun}}, \bibinfo {author} {\bibfnamefont {Y.}~\bibnamefont {Yang}}, \bibinfo {author} {\bibfnamefont {D.~C.}\ \bibnamefont {Chrzan}}, \bibinfo {author} {\bibfnamefont {L.}~\bibnamefont {Lindsay}}, \bibinfo {author} {\bibfnamefont {J.}~\bibnamefont {Wu}},\ and\ \bibinfo {author} {\bibfnamefont {J.}~\bibnamefont {Yao}},\ }\bibfield  {title}
  {\bibinfo {title} {Anomalous thermal transport in eshelby twisted van der waals nanowires},\ }\href {https://doi.org/10.1038/s41563-024-02108-3} {\bibfield  {journal} {\bibinfo  {journal} {Nat. Mater.}\ }\textbf {\bibinfo {volume} {24}},\ \bibinfo {pages} {728} (\bibinfo {year} {2025})}\BibitemShut {NoStop}%
\bibitem [{\citenamefont {Hochbaum}\ \emph {et~al.}(2008)\citenamefont {Hochbaum}, \citenamefont {Chen}, \citenamefont {Delgado}, \citenamefont {Liang}, \citenamefont {Garnett}, \citenamefont {Najarian}, \citenamefont {Majumdar},\ and\ \citenamefont {Yang}}]{Hochbaum2008}%
  \BibitemOpen
  \bibfield  {author} {\bibinfo {author} {\bibfnamefont {A.~I.}\ \bibnamefont {Hochbaum}}, \bibinfo {author} {\bibfnamefont {R.}~\bibnamefont {Chen}}, \bibinfo {author} {\bibfnamefont {R.~D.}\ \bibnamefont {Delgado}}, \bibinfo {author} {\bibfnamefont {W.}~\bibnamefont {Liang}}, \bibinfo {author} {\bibfnamefont {E.~C.}\ \bibnamefont {Garnett}}, \bibinfo {author} {\bibfnamefont {M.}~\bibnamefont {Najarian}}, \bibinfo {author} {\bibfnamefont {A.}~\bibnamefont {Majumdar}},\ and\ \bibinfo {author} {\bibfnamefont {P.}~\bibnamefont {Yang}},\ }\bibfield  {title} {\bibinfo {title} {Enhanced thermoelectric performance of rough silicon nanowires},\ }\href {https://doi.org/10.1038/nature06381} {\bibfield  {journal} {\bibinfo  {journal} {Nature}\ }\textbf {\bibinfo {volume} {451}},\ \bibinfo {pages} {163} (\bibinfo {year} {2008})}\BibitemShut {NoStop}%
\bibitem [{\citenamefont {Huber}\ \emph {et~al.}(2012)\citenamefont {Huber}, \citenamefont {Owusu}, \citenamefont {Johnson}, \citenamefont {Nikolaeva}, \citenamefont {Konopko}, \citenamefont {Johnson},\ and\ \citenamefont {Graf}}]{Huber_2012}%
  \BibitemOpen
  \bibfield  {author} {\bibinfo {author} {\bibfnamefont {T.~E.}\ \bibnamefont {Huber}}, \bibinfo {author} {\bibfnamefont {K.}~\bibnamefont {Owusu}}, \bibinfo {author} {\bibfnamefont {S.}~\bibnamefont {Johnson}}, \bibinfo {author} {\bibfnamefont {A.}~\bibnamefont {Nikolaeva}}, \bibinfo {author} {\bibfnamefont {L.}~\bibnamefont {Konopko}}, \bibinfo {author} {\bibfnamefont {R.~C.}\ \bibnamefont {Johnson}},\ and\ \bibinfo {author} {\bibfnamefont {M.~J.}\ \bibnamefont {Graf}},\ }\bibfield  {title} {\bibinfo {title} {Thermoelectric prospects of nanomaterials with spin-orbit surface bands},\ }\href {https://doi.org/10.1063/1.3686206} {\bibfield  {journal} {\bibinfo  {journal} {J. Appl. Phys.}\ }\textbf {\bibinfo {volume} {111}},\ \bibinfo {pages} {043705} (\bibinfo {year} {2012})}\BibitemShut {NoStop}%
\bibitem [{\citenamefont {Karwacki}\ and\ \citenamefont {Barna\ifmmode~\acute{s}\else \'{s}\fi{}}(2018)}]{Karwacki2018}%
  \BibitemOpen
  \bibfield  {author} {\bibinfo {author} {\bibfnamefont {L.}~\bibnamefont {Karwacki}}\ and\ \bibinfo {author} {\bibfnamefont {J.}~\bibnamefont {Barna\ifmmode~\acute{s}\else \'{s}\fi{}}},\ }\bibfield  {title} {\bibinfo {title} {Thermoelectric properties of a quantum dot coupled to magnetic leads by {Rashba} spin-orbit interaction},\ }\href {https://doi.org/10.1103/PhysRevB.98.075413} {\bibfield  {journal} {\bibinfo  {journal} {Phys. Rev. B}\ }\textbf {\bibinfo {volume} {98}},\ \bibinfo {pages} {075413} (\bibinfo {year} {2018})}\BibitemShut {NoStop}%
\bibitem [{\citenamefont {Manya}\ \emph {et~al.}(2022)\citenamefont {Manya}, \citenamefont {Martins},\ and\ \citenamefont {Figueira}}]{Manya2022}%
  \BibitemOpen
  \bibfield  {author} {\bibinfo {author} {\bibfnamefont {M.~A.}\ \bibnamefont {Manya}}, \bibinfo {author} {\bibfnamefont {G.~B.}\ \bibnamefont {Martins}},\ and\ \bibinfo {author} {\bibfnamefont {M.~S.}\ \bibnamefont {Figueira}},\ }\bibfield  {title} {\bibinfo {title} {Spin-orbit coupling effects on thermoelectric transport properties in quantum dots},\ }\href {https://doi.org/10.1103/PhysRevB.105.165421} {\bibfield  {journal} {\bibinfo  {journal} {Phys. Rev. B}\ }\textbf {\bibinfo {volume} {105}},\ \bibinfo {pages} {165421} (\bibinfo {year} {2022})}\BibitemShut {NoStop}%
\bibitem [{\citenamefont {Gumbs}\ \emph {et~al.}(2010)\citenamefont {Gumbs}, \citenamefont {Balassis},\ and\ \citenamefont {Huang}}]{Gumbs2010_SOI}%
  \BibitemOpen
  \bibfield  {author} {\bibinfo {author} {\bibfnamefont {G.}~\bibnamefont {Gumbs}}, \bibinfo {author} {\bibfnamefont {A.}~\bibnamefont {Balassis}},\ and\ \bibinfo {author} {\bibfnamefont {D.}~\bibnamefont {Huang}},\ }\bibfield  {title} {\bibinfo {title} {Energy bands, conductance and thermoelectric power for ballistic electrons in a nanowire with spin–orbit interaction},\ }\href {https://doi.org/10.1063/1.3462060} {\bibfield  {journal} {\bibinfo  {journal} {J. Appl. Phys.}\ }\textbf {\bibinfo {volume} {108}},\ \bibinfo {pages} {093704} (\bibinfo {year} {2010})}\BibitemShut {NoStop}%
\bibitem [{\citenamefont {Yang}(2004)}]{Yang2004}%
  \BibitemOpen
  \bibfield  {author} {\bibinfo {author} {\bibfnamefont {J.}~\bibnamefont {Yang}},\ }\bibfield  {title} {\bibinfo {title} {Theory of thermal conductivity},\ }in\ \href@noop {} {\emph {\bibinfo {booktitle} {Thermal Conductivity: Theory, Properties, and Applications}}},\ \bibinfo {editor} {edited by\ \bibinfo {editor} {\bibfnamefont {T.~M.}\ \bibnamefont {Tritt}}}\ (\bibinfo  {publisher} {Springer},\ \bibinfo {year} {2004})\ pp.\ \bibinfo {pages} {1--20}\BibitemShut {NoStop}%
\bibitem [{\citenamefont {Latronico}\ \emph {et~al.}(2024)\citenamefont {Latronico}, \citenamefont {Asnaashari~Eivari}, \citenamefont {Mele},\ and\ \citenamefont {Assadi}}]{Latronico2024}%
  \BibitemOpen
  \bibfield  {author} {\bibinfo {author} {\bibfnamefont {G.}~\bibnamefont {Latronico}}, \bibinfo {author} {\bibfnamefont {H.}~\bibnamefont {Asnaashari~Eivari}}, \bibinfo {author} {\bibfnamefont {P.}~\bibnamefont {Mele}},\ and\ \bibinfo {author} {\bibfnamefont {M.~H.~N.}\ \bibnamefont {Assadi}},\ }\bibfield  {title} {\bibinfo {title} {Insights into one-dimensional thermoelectric materials: A concise review of nanowires and nanotubes},\ }\href {https://doi.org/10.3390/nano14151272} {\bibfield  {journal} {\bibinfo  {journal} {Nanomaterials}\ }\textbf {\bibinfo {volume} {14}},\ \bibinfo {pages} {1272} (\bibinfo {year} {2024})}\BibitemShut {NoStop}%
\bibitem [{\citenamefont {Roddaro}\ \emph {et~al.}(2013)\citenamefont {Roddaro}, \citenamefont {Ercolani}, \citenamefont {Safeen}, \citenamefont {Suomalainen}, \citenamefont {Rossella}, \citenamefont {Giazotto}, \citenamefont {Sorba},\ and\ \citenamefont {Beltram}}]{Roddaro2013}%
  \BibitemOpen
  \bibfield  {author} {\bibinfo {author} {\bibfnamefont {S.}~\bibnamefont {Roddaro}}, \bibinfo {author} {\bibfnamefont {D.}~\bibnamefont {Ercolani}}, \bibinfo {author} {\bibfnamefont {M.~A.}\ \bibnamefont {Safeen}}, \bibinfo {author} {\bibfnamefont {S.}~\bibnamefont {Suomalainen}}, \bibinfo {author} {\bibfnamefont {F.}~\bibnamefont {Rossella}}, \bibinfo {author} {\bibfnamefont {F.}~\bibnamefont {Giazotto}}, \bibinfo {author} {\bibfnamefont {L.}~\bibnamefont {Sorba}},\ and\ \bibinfo {author} {\bibfnamefont {F.}~\bibnamefont {Beltram}},\ }\bibfield  {title} {\bibinfo {title} {Giant thermovoltage in single {InAs} nanowire field-effect transistors},\ }\href {https://doi.org/10.1021/nl401482p} {\bibfield  {journal} {\bibinfo  {journal} {Nano Lett.}\ }\textbf {\bibinfo {volume} {13}},\ \bibinfo {pages} {3638} (\bibinfo {year} {2013})}\BibitemShut {NoStop}%
\bibitem [{\citenamefont {El~Sachat}\ \emph {et~al.}(2021)\citenamefont {El~Sachat}, \citenamefont {Alzina}, \citenamefont {Sotomayor~Torres},\ and\ \citenamefont {Chavez-Angel}}]{ElSachat2021}%
  \BibitemOpen
  \bibfield  {author} {\bibinfo {author} {\bibfnamefont {A.}~\bibnamefont {El~Sachat}}, \bibinfo {author} {\bibfnamefont {F.}~\bibnamefont {Alzina}}, \bibinfo {author} {\bibfnamefont {C.~M.}\ \bibnamefont {Sotomayor~Torres}},\ and\ \bibinfo {author} {\bibfnamefont {E.}~\bibnamefont {Chavez-Angel}},\ }\bibfield  {title} {\bibinfo {title} {Heat transport control and thermal characterization of low-dimensional materials: A review},\ }\href {https://doi.org/10.3390/nano11010175} {\bibfield  {journal} {\bibinfo  {journal} {Nanomaterials}\ }\textbf {\bibinfo {volume} {11}},\ \bibinfo {pages} {175} (\bibinfo {year} {2021})}\BibitemShut {NoStop}%
\bibitem [{\citenamefont {Fust}\ \emph {et~al.}(2020)\citenamefont {Fust}, \citenamefont {Faustmann}, \citenamefont {Carrad}, \citenamefont {Bissinger}, \citenamefont {Loitsch}, \citenamefont {Döblinger}, \citenamefont {Becker}, \citenamefont {Abstreiter}, \citenamefont {Finley},\ and\ \citenamefont {Koblmüller}}]{Sergej2020}%
  \BibitemOpen
  \bibfield  {author} {\bibinfo {author} {\bibfnamefont {S.}~\bibnamefont {Fust}}, \bibinfo {author} {\bibfnamefont {A.}~\bibnamefont {Faustmann}}, \bibinfo {author} {\bibfnamefont {D.~J.}\ \bibnamefont {Carrad}}, \bibinfo {author} {\bibfnamefont {J.}~\bibnamefont {Bissinger}}, \bibinfo {author} {\bibfnamefont {B.}~\bibnamefont {Loitsch}}, \bibinfo {author} {\bibfnamefont {M.}~\bibnamefont {Döblinger}}, \bibinfo {author} {\bibfnamefont {J.}~\bibnamefont {Becker}}, \bibinfo {author} {\bibfnamefont {G.}~\bibnamefont {Abstreiter}}, \bibinfo {author} {\bibfnamefont {J.~J.}\ \bibnamefont {Finley}},\ and\ \bibinfo {author} {\bibfnamefont {G.}~\bibnamefont {Koblmüller}},\ }\bibfield  {title} {\bibinfo {title} {Quantum-confinement-enhanced thermoelectric properties in modulation-doped {GaAs}--{AlGaAs} core--shell nanowires},\ }\href {https://doi.org/10.1002/adma.201905458} {\bibfield  {journal} {\bibinfo  {journal} {Adv. Mater.}\ }\textbf {\bibinfo {volume} {32}},\ \bibinfo {pages} {1905458} (\bibinfo {year}
  {2020})}\BibitemShut {NoStop}%
\bibitem [{\citenamefont {Chalopin}\ \emph {et~al.}(2008)\citenamefont {Chalopin}, \citenamefont {Gillet},\ and\ \citenamefont {Volz}}]{Chalopin2008}%
  \BibitemOpen
  \bibfield  {author} {\bibinfo {author} {\bibfnamefont {Y.}~\bibnamefont {Chalopin}}, \bibinfo {author} {\bibfnamefont {J.-N.}\ \bibnamefont {Gillet}},\ and\ \bibinfo {author} {\bibfnamefont {S.}~\bibnamefont {Volz}},\ }\bibfield  {title} {\bibinfo {title} {Predominance of thermal contact resistance in a silicon nanowire on a planar substrate},\ }\href {https://doi.org/10.1103/PhysRevB.77.233309} {\bibfield  {journal} {\bibinfo  {journal} {Phys. Rev. B}\ }\textbf {\bibinfo {volume} {77}},\ \bibinfo {pages} {233309} (\bibinfo {year} {2008})}\BibitemShut {NoStop}%
\bibitem [{\citenamefont {Sawtelle}\ and\ \citenamefont {Reed}(2019)}]{Sawtelle2019}%
  \BibitemOpen
  \bibfield  {author} {\bibinfo {author} {\bibfnamefont {S.~D.}\ \bibnamefont {Sawtelle}}\ and\ \bibinfo {author} {\bibfnamefont {M.~A.}\ \bibnamefont {Reed}},\ }\bibfield  {title} {\bibinfo {title} {Temperature-dependent thermal conductivity and suppressed lorenz number in ultrathin gold nanowires},\ }\href {https://doi.org/10.1103/PhysRevB.99.054304} {\bibfield  {journal} {\bibinfo  {journal} {Phys. Rev. B}\ }\textbf {\bibinfo {volume} {99}},\ \bibinfo {pages} {054304} (\bibinfo {year} {2019})}\BibitemShut {NoStop}%
\bibitem [{\citenamefont {S\'anchez}\ and\ \citenamefont {Serra}(2006)}]{Sanchez2006}%
  \BibitemOpen
  \bibfield  {author} {\bibinfo {author} {\bibfnamefont {D.}~\bibnamefont {S\'anchez}}\ and\ \bibinfo {author} {\bibfnamefont {L.~m.~c.}\ \bibnamefont {Serra}},\ }\bibfield  {title} {\bibinfo {title} {Fano-{R}ashba effect in a quantum wire},\ }\href {https://doi.org/10.1103/PhysRevB.74.153313} {\bibfield  {journal} {\bibinfo  {journal} {Phys. Rev. B}\ }\textbf {\bibinfo {volume} {74}},\ \bibinfo {pages} {153313} (\bibinfo {year} {2006})}\BibitemShut {NoStop}%
\bibitem [{\citenamefont {S\'anchez}\ \emph {et~al.}(2008)\citenamefont {S\'anchez}, \citenamefont {Serra},\ and\ \citenamefont {Choi}}]{Sanchez2008}%
  \BibitemOpen
  \bibfield  {author} {\bibinfo {author} {\bibfnamefont {D.}~\bibnamefont {S\'anchez}}, \bibinfo {author} {\bibfnamefont {L.~m.~c.}\ \bibnamefont {Serra}},\ and\ \bibinfo {author} {\bibfnamefont {M.-S.}\ \bibnamefont {Choi}},\ }\bibfield  {title} {\bibinfo {title} {Strongly modulated transmission of a spin-split quantum wire with local {R}ashba interaction},\ }\href {https://doi.org/10.1103/PhysRevB.77.035315} {\bibfield  {journal} {\bibinfo  {journal} {Phys. Rev. B}\ }\textbf {\bibinfo {volume} {77}},\ \bibinfo {pages} {035315} (\bibinfo {year} {2008})}\BibitemShut {NoStop}%
\bibitem [{\citenamefont {Sadreev}\ and\ \citenamefont {Sherman}(2013)}]{Sadreev2013}%
  \BibitemOpen
  \bibfield  {author} {\bibinfo {author} {\bibfnamefont {A.~F.}\ \bibnamefont {Sadreev}}\ and\ \bibinfo {author} {\bibfnamefont {E.~Y.}\ \bibnamefont {Sherman}},\ }\bibfield  {title} {\bibinfo {title} {Effect of gate-driven spin resonance on the conductance through a one-dimensional quantum wire},\ }\href {https://doi.org/10.1103/PhysRevB.88.115302} {\bibfield  {journal} {\bibinfo  {journal} {Phys. Rev. B}\ }\textbf {\bibinfo {volume} {88}},\ \bibinfo {pages} {115302} (\bibinfo {year} {2013})}\BibitemShut {NoStop}%
\bibitem [{\citenamefont {Cayao}\ \emph {et~al.}(2015)\citenamefont {Cayao}, \citenamefont {Prada}, \citenamefont {San-Jose},\ and\ \citenamefont {Aguado}}]{Cayao2015}%
  \BibitemOpen
  \bibfield  {author} {\bibinfo {author} {\bibfnamefont {J.}~\bibnamefont {Cayao}}, \bibinfo {author} {\bibfnamefont {E.}~\bibnamefont {Prada}}, \bibinfo {author} {\bibfnamefont {P.}~\bibnamefont {San-Jose}},\ and\ \bibinfo {author} {\bibfnamefont {R.}~\bibnamefont {Aguado}},\ }\bibfield  {title} {\bibinfo {title} {Sns junctions in nanowires with spin-orbit coupling: Role of confinement and helicity on the subgap spectrum},\ }\href {https://doi.org/10.1103/PhysRevB.91.024514} {\bibfield  {journal} {\bibinfo  {journal} {Phys. Rev. B}\ }\textbf {\bibinfo {volume} {91}},\ \bibinfo {pages} {024514} (\bibinfo {year} {2015})}\BibitemShut {NoStop}%
\bibitem [{\citenamefont {Gogin}\ \emph {et~al.}(2022{\natexlab{a}})\citenamefont {Gogin}, \citenamefont {Rossi}, \citenamefont {Rossi},\ and\ \citenamefont {Dolcini}}]{Gogin_2022_1}%
  \BibitemOpen
  \bibfield  {author} {\bibinfo {author} {\bibfnamefont {L.}~\bibnamefont {Gogin}}, \bibinfo {author} {\bibfnamefont {L.}~\bibnamefont {Rossi}}, \bibinfo {author} {\bibfnamefont {F.}~\bibnamefont {Rossi}},\ and\ \bibinfo {author} {\bibfnamefont {F.}~\bibnamefont {Dolcini}},\ }\bibfield  {title} {\bibinfo {title} {The {D}irac paradox in 1 + 1 dimensions and its realization with spin–orbit coupled nanowires},\ }\href {https://doi.org/10.1088/1367-2630/ac6cfe} {\bibfield  {journal} {\bibinfo  {journal} {New J. Phys.}\ }\textbf {\bibinfo {volume} {24}},\ \bibinfo {pages} {053045} (\bibinfo {year} {2022}{\natexlab{a}})}\BibitemShut {NoStop}%
\bibitem [{\citenamefont {Gogin}\ \emph {et~al.}(2022{\natexlab{b}})\citenamefont {Gogin}, \citenamefont {Rossi},\ and\ \citenamefont {Dolcini}}]{Gogin_2022_2}%
  \BibitemOpen
  \bibfield  {author} {\bibinfo {author} {\bibfnamefont {L.}~\bibnamefont {Gogin}}, \bibinfo {author} {\bibfnamefont {F.}~\bibnamefont {Rossi}},\ and\ \bibinfo {author} {\bibfnamefont {F.}~\bibnamefont {Dolcini}},\ }\bibfield  {title} {\bibinfo {title} {Electron transport in quantum channels with spin--orbit interaction: effects of the sign of the {Rashba} coupling and applications to nanowires},\ }\href {https://doi.org/10.1088/1367-2630/ac8f66} {\bibfield  {journal} {\bibinfo  {journal} {New J. Phys.}\ }\textbf {\bibinfo {volume} {24}},\ \bibinfo {pages} {093025} (\bibinfo {year} {2022}{\natexlab{b}})}\BibitemShut {NoStop}%
\bibitem [{\citenamefont {Li}\ \emph {et~al.}(2016)\citenamefont {Li}, \citenamefont {Kang}, \citenamefont {Fan}, \citenamefont {Wang}, \citenamefont {Huang}, \citenamefont {Caroff},\ and\ \citenamefont {Xu}}]{Li2016}%
  \BibitemOpen
  \bibfield  {author} {\bibinfo {author} {\bibfnamefont {S.}~\bibnamefont {Li}}, \bibinfo {author} {\bibfnamefont {N.}~\bibnamefont {Kang}}, \bibinfo {author} {\bibfnamefont {D.~X.}\ \bibnamefont {Fan}}, \bibinfo {author} {\bibfnamefont {L.~B.}\ \bibnamefont {Wang}}, \bibinfo {author} {\bibfnamefont {Y.~Q.}\ \bibnamefont {Huang}}, \bibinfo {author} {\bibfnamefont {P.}~\bibnamefont {Caroff}},\ and\ \bibinfo {author} {\bibfnamefont {H.~Q.}\ \bibnamefont {Xu}},\ }\bibfield  {title} {\bibinfo {title} {Coherent charge transport in ballistic {I}n{S}b nanowire josephson junctions},\ }\href {https://doi.org/10.1038/srep24822} {\bibfield  {journal} {\bibinfo  {journal} {Sci. Rep.}\ }\textbf {\bibinfo {volume} {6}},\ \bibinfo {pages} {24822} (\bibinfo {year} {2016})}\BibitemShut {NoStop}%
\bibitem [{\citenamefont {Estrada~Saldaña}\ \emph {et~al.}(2018)\citenamefont {Estrada~Saldaña}, \citenamefont {Niquet}, \citenamefont {Cleuziou}, \citenamefont {Lee}, \citenamefont {Car}, \citenamefont {Plissard}, \citenamefont {Bakkers},\ and\ \citenamefont {De~Franceschi}}]{EstradaSaldana2018}%
  \BibitemOpen
  \bibfield  {author} {\bibinfo {author} {\bibfnamefont {J.~C.}\ \bibnamefont {Estrada~Saldaña}}, \bibinfo {author} {\bibfnamefont {Y.-M.}\ \bibnamefont {Niquet}}, \bibinfo {author} {\bibfnamefont {J.-P.}\ \bibnamefont {Cleuziou}}, \bibinfo {author} {\bibfnamefont {E.~J.~H.}\ \bibnamefont {Lee}}, \bibinfo {author} {\bibfnamefont {D.}~\bibnamefont {Car}}, \bibinfo {author} {\bibfnamefont {S.~R.}\ \bibnamefont {Plissard}}, \bibinfo {author} {\bibfnamefont {E.~P. A.~M.}\ \bibnamefont {Bakkers}},\ and\ \bibinfo {author} {\bibfnamefont {S.}~\bibnamefont {De~Franceschi}},\ }\bibfield  {title} {\bibinfo {title} {Split-channel ballistic transport in an {InSb} nanowire},\ }\href {https://doi.org/10.1021/acs.nanolett.7b03854} {\bibfield  {journal} {\bibinfo  {journal} {Nano Lett.}\ }\textbf {\bibinfo {volume} {18}},\ \bibinfo {pages} {2282} (\bibinfo {year} {2018})}\BibitemShut {NoStop}%
\bibitem [{\citenamefont {Shani}\ \emph {et~al.}(2024)\citenamefont {Shani}, \citenamefont {Lueb}, \citenamefont {Menning}, \citenamefont {Gupta}, \citenamefont {Riggert}, \citenamefont {Littmann}, \citenamefont {Hackbarth}, \citenamefont {Rossi}, \citenamefont {Jung}, \citenamefont {Badawy}, \citenamefont {Verheijen}, \citenamefont {Crowell}, \citenamefont {Bakkers},\ and\ \citenamefont {Pribiag}}]{Shani_2024}%
  \BibitemOpen
  \bibfield  {author} {\bibinfo {author} {\bibfnamefont {L.}~\bibnamefont {Shani}}, \bibinfo {author} {\bibfnamefont {P.}~\bibnamefont {Lueb}}, \bibinfo {author} {\bibfnamefont {G.}~\bibnamefont {Menning}}, \bibinfo {author} {\bibfnamefont {M.}~\bibnamefont {Gupta}}, \bibinfo {author} {\bibfnamefont {C.}~\bibnamefont {Riggert}}, \bibinfo {author} {\bibfnamefont {T.}~\bibnamefont {Littmann}}, \bibinfo {author} {\bibfnamefont {F.}~\bibnamefont {Hackbarth}}, \bibinfo {author} {\bibfnamefont {M.}~\bibnamefont {Rossi}}, \bibinfo {author} {\bibfnamefont {J.}~\bibnamefont {Jung}}, \bibinfo {author} {\bibfnamefont {G.}~\bibnamefont {Badawy}}, \bibinfo {author} {\bibfnamefont {M.~A.}\ \bibnamefont {Verheijen}}, \bibinfo {author} {\bibfnamefont {P.~A.}\ \bibnamefont {Crowell}}, \bibinfo {author} {\bibfnamefont {E.~P. A.~M.}\ \bibnamefont {Bakkers}},\ and\ \bibinfo {author} {\bibfnamefont {V.~S.}\ \bibnamefont {Pribiag}},\ }\bibfield  {title} {\bibinfo {title} {Diffusive and ballistic transport in thin {I}n{S}b nanowire
  devices using a few-layer-graphene-{A}l{O}\textsubscript{x} gate},\ }\href {https://doi.org/10.1088/2633-4356/ad2d6b} {\bibfield  {journal} {\bibinfo  {journal} {Mater. Quantum Technol.}\ }\textbf {\bibinfo {volume} {4}},\ \bibinfo {pages} {015101} (\bibinfo {year} {2024})}\BibitemShut {NoStop}%
\bibitem [{\citenamefont {Scher\"ubl}\ \emph {et~al.}(2016)\citenamefont {Scher\"ubl}, \citenamefont {F\"ul\"op}, \citenamefont {Madsen}, \citenamefont {Nyg\aa{}rd},\ and\ \citenamefont {Csonka}}]{Scherubl2016}%
  \BibitemOpen
  \bibfield  {author} {\bibinfo {author} {\bibfnamefont {Z.}~\bibnamefont {Scher\"ubl}}, \bibinfo {author} {\bibfnamefont {G.~m.~H.}\ \bibnamefont {F\"ul\"op}}, \bibinfo {author} {\bibfnamefont {M.~H.}\ \bibnamefont {Madsen}}, \bibinfo {author} {\bibfnamefont {J.}~\bibnamefont {Nyg\aa{}rd}},\ and\ \bibinfo {author} {\bibfnamefont {S.}~\bibnamefont {Csonka}},\ }\bibfield  {title} {\bibinfo {title} {Electrical tuning of {R}ashba spin-orbit interaction in multigated {I}n{A}s nanowires},\ }\href {https://doi.org/10.1103/PhysRevB.94.035444} {\bibfield  {journal} {\bibinfo  {journal} {Phys. Rev. B}\ }\textbf {\bibinfo {volume} {94}},\ \bibinfo {pages} {035444} (\bibinfo {year} {2016})}\BibitemShut {NoStop}%
\bibitem [{\citenamefont {Takase}\ \emph {et~al.}(2017)\citenamefont {Takase}, \citenamefont {Ashikawa}, \citenamefont {Zhang}, \citenamefont {Tateno},\ and\ \citenamefont {Sasaki}}]{Takase2017}%
  \BibitemOpen
  \bibfield  {author} {\bibinfo {author} {\bibfnamefont {K.}~\bibnamefont {Takase}}, \bibinfo {author} {\bibfnamefont {Y.}~\bibnamefont {Ashikawa}}, \bibinfo {author} {\bibfnamefont {G.}~\bibnamefont {Zhang}}, \bibinfo {author} {\bibfnamefont {K.}~\bibnamefont {Tateno}},\ and\ \bibinfo {author} {\bibfnamefont {S.}~\bibnamefont {Sasaki}},\ }\bibfield  {title} {\bibinfo {title} {Highly gate-tuneable {R}ashba spin-orbit interaction in a gate-all-around {InAs} nanowire metal-oxide-semiconductor field-effect transistor},\ }\href {https://doi.org/10.1038/s41598-017-01080-0} {\bibfield  {journal} {\bibinfo  {journal} {Sci. Rep.}\ }\textbf {\bibinfo {volume} {7}},\ \bibinfo {pages} {930} (\bibinfo {year} {2017})}\BibitemShut {NoStop}%
\bibitem [{\citenamefont {Liang}\ and\ \citenamefont {Gao}(2012)}]{Liang2012}%
  \BibitemOpen
  \bibfield  {author} {\bibinfo {author} {\bibfnamefont {D.}~\bibnamefont {Liang}}\ and\ \bibinfo {author} {\bibfnamefont {X.~P.~A.}\ \bibnamefont {Gao}},\ }\bibfield  {title} {\bibinfo {title} {Strong tuning of {R}ashba spin–orbit interaction in single {InAs} nanowires},\ }\href {https://doi.org/10.1021/nl301325h} {\bibfield  {journal} {\bibinfo  {journal} {Nano Lett.}\ }\textbf {\bibinfo {volume} {12}},\ \bibinfo {pages} {3263} (\bibinfo {year} {2012})}\BibitemShut {NoStop}%
\bibitem [{\citenamefont {Sasaki}\ \emph {et~al.}(2013)\citenamefont {Sasaki}, \citenamefont {Tateno}, \citenamefont {Zhang}, \citenamefont {Suominen}, \citenamefont {Harada}, \citenamefont {Saito}, \citenamefont {Fujiwara}, \citenamefont {Sogawa},\ and\ \citenamefont {Muraki}}]{Sasaki2013}%
  \BibitemOpen
  \bibfield  {author} {\bibinfo {author} {\bibfnamefont {S.}~\bibnamefont {Sasaki}}, \bibinfo {author} {\bibfnamefont {K.}~\bibnamefont {Tateno}}, \bibinfo {author} {\bibfnamefont {G.}~\bibnamefont {Zhang}}, \bibinfo {author} {\bibfnamefont {H.}~\bibnamefont {Suominen}}, \bibinfo {author} {\bibfnamefont {Y.}~\bibnamefont {Harada}}, \bibinfo {author} {\bibfnamefont {S.}~\bibnamefont {Saito}}, \bibinfo {author} {\bibfnamefont {A.}~\bibnamefont {Fujiwara}}, \bibinfo {author} {\bibfnamefont {T.}~\bibnamefont {Sogawa}},\ and\ \bibinfo {author} {\bibfnamefont {K.}~\bibnamefont {Muraki}},\ }\bibfield  {title} {\bibinfo {title} {Encapsulated gate-all-around {InAs} nanowire field-effect transistors},\ }\href {https://doi.org/10.1063/1.4832058} {\bibfield  {journal} {\bibinfo  {journal} {Appl. Phys. Lett.}\ }\textbf {\bibinfo {volume} {103}},\ \bibinfo {pages} {213502} (\bibinfo {year} {2013})}\BibitemShut {NoStop}%
\bibitem [{\citenamefont {Burke}\ \emph {et~al.}(2015)\citenamefont {Burke}, \citenamefont {Carrad}, \citenamefont {Gluschke}, \citenamefont {Storm}, \citenamefont {Fahlvik~Svensson}, \citenamefont {Linke}, \citenamefont {Samuelson},\ and\ \citenamefont {Micolich}}]{micolich_2015}%
  \BibitemOpen
  \bibfield  {author} {\bibinfo {author} {\bibfnamefont {A.~M.}\ \bibnamefont {Burke}}, \bibinfo {author} {\bibfnamefont {D.~J.}\ \bibnamefont {Carrad}}, \bibinfo {author} {\bibfnamefont {J.~G.}\ \bibnamefont {Gluschke}}, \bibinfo {author} {\bibfnamefont {K.}~\bibnamefont {Storm}}, \bibinfo {author} {\bibfnamefont {S.}~\bibnamefont {Fahlvik~Svensson}}, \bibinfo {author} {\bibfnamefont {H.}~\bibnamefont {Linke}}, \bibinfo {author} {\bibfnamefont {L.}~\bibnamefont {Samuelson}},\ and\ \bibinfo {author} {\bibfnamefont {A.~P.}\ \bibnamefont {Micolich}},\ }\bibfield  {title} {\bibinfo {title} {{InAs} nanowire transistors with multiple, independent wrap-gate segments},\ }\href {https://doi.org/10.1021/nl5043243} {\bibfield  {journal} {\bibinfo  {journal} {Nano Lett.}\ }\textbf {\bibinfo {volume} {15}},\ \bibinfo {pages} {2836} (\bibinfo {year} {2015})}\BibitemShut {NoStop}%
\bibitem [{\citenamefont {Bindel}\ \emph {et~al.}(2016)\citenamefont {Bindel}, \citenamefont {Pezzotta}, \citenamefont {Ulrich}, \citenamefont {Liebmann}, \citenamefont {Sherman},\ and\ \citenamefont {Morgenstern}}]{Bindel2016}%
  \BibitemOpen
  \bibfield  {author} {\bibinfo {author} {\bibfnamefont {J.~R.}\ \bibnamefont {Bindel}}, \bibinfo {author} {\bibfnamefont {M.}~\bibnamefont {Pezzotta}}, \bibinfo {author} {\bibfnamefont {J.}~\bibnamefont {Ulrich}}, \bibinfo {author} {\bibfnamefont {M.}~\bibnamefont {Liebmann}}, \bibinfo {author} {\bibfnamefont {E.~Y.}\ \bibnamefont {Sherman}},\ and\ \bibinfo {author} {\bibfnamefont {M.}~\bibnamefont {Morgenstern}},\ }\bibfield  {title} {\bibinfo {title} {Probing variations of the {R}ashba spin--orbit coupling at the nanometre scale},\ }\href {https://doi.org/10.1038/nphys3774} {\bibfield  {journal} {\bibinfo  {journal} {Nat. Phys.}\ }\textbf {\bibinfo {volume} {12}},\ \bibinfo {pages} {920} (\bibinfo {year} {2016})}\BibitemShut {NoStop}%
\bibitem [{\citenamefont {Takase}\ \emph {et~al.}(2021)\citenamefont {Takase}, \citenamefont {Tateno},\ and\ \citenamefont {Sasaki}}]{Takase2021}%
  \BibitemOpen
  \bibfield  {author} {\bibinfo {author} {\bibfnamefont {K.}~\bibnamefont {Takase}}, \bibinfo {author} {\bibfnamefont {K.}~\bibnamefont {Tateno}},\ and\ \bibinfo {author} {\bibfnamefont {S.}~\bibnamefont {Sasaki}},\ }\bibfield  {title} {\bibinfo {title} {Electrical tuning of the spin–orbit interaction in nanowire by transparent {ZnO} gate grown by atomic layer deposition},\ }\href {https://doi.org/10.1063/5.0051281} {\bibfield  {journal} {\bibinfo  {journal} {Appl. Phys. Lett.}\ }\textbf {\bibinfo {volume} {119}},\ \bibinfo {pages} {013102} (\bibinfo {year} {2021})}\BibitemShut {NoStop}%
\bibitem [{\citenamefont {Pan}\ \emph {et~al.}(2025)\citenamefont {Pan}, \citenamefont {Taylor}, \citenamefont {Sau},\ and\ \citenamefont {Sarma}}]{Pan2025}%
  \BibitemOpen
  \bibfield  {author} {\bibinfo {author} {\bibfnamefont {H.}~\bibnamefont {Pan}}, \bibinfo {author} {\bibfnamefont {J.~R.}\ \bibnamefont {Taylor}}, \bibinfo {author} {\bibfnamefont {J.~D.}\ \bibnamefont {Sau}},\ and\ \bibinfo {author} {\bibfnamefont {S.~D.}\ \bibnamefont {Sarma}},\ }\href@noop {} {\bibinfo {title} {Ballistic transport in {1D} rashba systems in the context of majorana nanowires}} (\bibinfo {year} {2025}),\ \Eprint {https://arxiv.org/abs/2510.23961} {arXiv:2510.23961 [cond-mat.mes-hall]} \BibitemShut {NoStop}%
\bibitem [{\citenamefont {Datta}\ and\ \citenamefont {Das}(1990)}]{DattaDas1990}%
  \BibitemOpen
  \bibfield  {author} {\bibinfo {author} {\bibfnamefont {S.}~\bibnamefont {Datta}}\ and\ \bibinfo {author} {\bibfnamefont {B.}~\bibnamefont {Das}},\ }\bibfield  {title} {\bibinfo {title} {Electronic analog of the electro-optic modulator},\ }\href {https://doi.org/10.1063/1.102730} {\bibfield  {journal} {\bibinfo  {journal} {Appl. Phys. Lett.}\ }\textbf {\bibinfo {volume} {56}},\ \bibinfo {pages} {665} (\bibinfo {year} {1990})}\BibitemShut {NoStop}%
\bibitem [{\citenamefont {St\ifmmode~\check{r}\else \v{r}\fi{}eda}\ and\ \citenamefont {\ifmmode~\check{S}\else \v{S}\fi{}eba}(2003)}]{Streda2003}%
  \BibitemOpen
  \bibfield  {author} {\bibinfo {author} {\bibfnamefont {P.}~\bibnamefont {St\ifmmode~\check{r}\else \v{r}\fi{}eda}}\ and\ \bibinfo {author} {\bibfnamefont {P.}~\bibnamefont {\ifmmode~\check{S}\else \v{S}\fi{}eba}},\ }\bibfield  {title} {\bibinfo {title} {Antisymmetric spin filtering in one-dimensional electron systems with uniform spin-orbit coupling},\ }\href {https://doi.org/10.1103/PhysRevLett.90.256601} {\bibfield  {journal} {\bibinfo  {journal} {Phys. Rev. Lett.}\ }\textbf {\bibinfo {volume} {90}},\ \bibinfo {pages} {256601} (\bibinfo {year} {2003})}\BibitemShut {NoStop}%
\bibitem [{\citenamefont {Oreg}\ \emph {et~al.}(2010)\citenamefont {Oreg}, \citenamefont {Refael},\ and\ \citenamefont {von Oppen}}]{Yuval2010}%
  \BibitemOpen
  \bibfield  {author} {\bibinfo {author} {\bibfnamefont {Y.}~\bibnamefont {Oreg}}, \bibinfo {author} {\bibfnamefont {G.}~\bibnamefont {Refael}},\ and\ \bibinfo {author} {\bibfnamefont {F.}~\bibnamefont {von Oppen}},\ }\bibfield  {title} {\bibinfo {title} {Helical liquids and {M}ajorana bound states in quantum wires},\ }\href {https://doi.org/10.1103/PhysRevLett.105.177002} {\bibfield  {journal} {\bibinfo  {journal} {Phys. Rev. Lett.}\ }\textbf {\bibinfo {volume} {105}},\ \bibinfo {pages} {177002} (\bibinfo {year} {2010})}\BibitemShut {NoStop}%
\bibitem [{\citenamefont {Lutchyn}\ \emph {et~al.}(2010)\citenamefont {Lutchyn}, \citenamefont {Sau},\ and\ \citenamefont {Das~Sarma}}]{Roman2010}%
  \BibitemOpen
  \bibfield  {author} {\bibinfo {author} {\bibfnamefont {R.~M.}\ \bibnamefont {Lutchyn}}, \bibinfo {author} {\bibfnamefont {J.~D.}\ \bibnamefont {Sau}},\ and\ \bibinfo {author} {\bibfnamefont {S.}~\bibnamefont {Das~Sarma}},\ }\bibfield  {title} {\bibinfo {title} {{M}ajorana fermions and a topological phase transition in semiconductor-superconductor heterostructures},\ }\href {https://doi.org/10.1103/PhysRevLett.105.077001} {\bibfield  {journal} {\bibinfo  {journal} {Phys. Rev. Lett.}\ }\textbf {\bibinfo {volume} {105}},\ \bibinfo {pages} {077001} (\bibinfo {year} {2010})}\BibitemShut {NoStop}%
\bibitem [{\citenamefont {Kloeffel}\ \emph {et~al.}(2011)\citenamefont {Kloeffel}, \citenamefont {Trif},\ and\ \citenamefont {Loss}}]{Kloeffel2011}%
  \BibitemOpen
  \bibfield  {author} {\bibinfo {author} {\bibfnamefont {C.}~\bibnamefont {Kloeffel}}, \bibinfo {author} {\bibfnamefont {M.}~\bibnamefont {Trif}},\ and\ \bibinfo {author} {\bibfnamefont {D.}~\bibnamefont {Loss}},\ }\bibfield  {title} {\bibinfo {title} {Strong spin-orbit interaction and helical hole states in {Ge/Si} nanowires},\ }\href {https://doi.org/10.1103/PhysRevB.84.195314} {\bibfield  {journal} {\bibinfo  {journal} {Phys. Rev. B}\ }\textbf {\bibinfo {volume} {84}},\ \bibinfo {pages} {195314} (\bibinfo {year} {2011})}\BibitemShut {NoStop}%
\bibitem [{\citenamefont {Kammhuber}\ \emph {et~al.}(2017)\citenamefont {Kammhuber}, \citenamefont {Cassidy}, \citenamefont {Pei}, \citenamefont {Nowak}, \citenamefont {Vuik}, \citenamefont {G\"{u}l}, \citenamefont {Car}, \citenamefont {Plissard}, \citenamefont {Bakkers}, \citenamefont {Wimmer},\ and\ \citenamefont {Kouwenhoven}}]{Kammhuber2017}%
  \BibitemOpen
  \bibfield  {author} {\bibinfo {author} {\bibfnamefont {J.}~\bibnamefont {Kammhuber}}, \bibinfo {author} {\bibfnamefont {M.~C.}\ \bibnamefont {Cassidy}}, \bibinfo {author} {\bibfnamefont {F.}~\bibnamefont {Pei}}, \bibinfo {author} {\bibfnamefont {M.~P.}\ \bibnamefont {Nowak}}, \bibinfo {author} {\bibfnamefont {A.}~\bibnamefont {Vuik}}, \bibinfo {author} {\bibfnamefont {O.}~\bibnamefont {G\"{u}l}}, \bibinfo {author} {\bibfnamefont {D.}~\bibnamefont {Car}}, \bibinfo {author} {\bibfnamefont {S.~R.}\ \bibnamefont {Plissard}}, \bibinfo {author} {\bibfnamefont {E.~P. A.~M.}\ \bibnamefont {Bakkers}}, \bibinfo {author} {\bibfnamefont {M.}~\bibnamefont {Wimmer}},\ and\ \bibinfo {author} {\bibfnamefont {L.~P.}\ \bibnamefont {Kouwenhoven}},\ }\bibfield  {title} {\bibinfo {title} {Conductance through a helical state in an indium antimonide nanowire},\ }\href {https://doi.org/10.1038/s41467-017-00315-y} {\bibfield  {journal} {\bibinfo  {journal} {Nat. Commun.}\ }\textbf {\bibinfo {volume} {8}},\ \bibinfo {pages} {478}
  (\bibinfo {year} {2017})}\BibitemShut {NoStop}%
\bibitem [{\citenamefont {van Weperen}\ \emph {et~al.}(2013)\citenamefont {van Weperen}, \citenamefont {Plissard}, \citenamefont {Bakkers}, \citenamefont {Frolov},\ and\ \citenamefont {Kouwenhoven}}]{vanWeperen2013}%
  \BibitemOpen
  \bibfield  {author} {\bibinfo {author} {\bibfnamefont {I.}~\bibnamefont {van Weperen}}, \bibinfo {author} {\bibfnamefont {S.~R.}\ \bibnamefont {Plissard}}, \bibinfo {author} {\bibfnamefont {E.~P. A.~M.}\ \bibnamefont {Bakkers}}, \bibinfo {author} {\bibfnamefont {S.~M.}\ \bibnamefont {Frolov}},\ and\ \bibinfo {author} {\bibfnamefont {L.~P.}\ \bibnamefont {Kouwenhoven}},\ }\bibfield  {title} {\bibinfo {title} {Quantized conductance in an {InSb} nanowire},\ }\href {https://doi.org/10.1021/nl3035256} {\bibfield  {journal} {\bibinfo  {journal} {Nano Lett.}\ }\textbf {\bibinfo {volume} {13}},\ \bibinfo {pages} {387} (\bibinfo {year} {2013})}\BibitemShut {NoStop}%
\bibitem [{\citenamefont {Quay}\ \emph {et~al.}(2010)\citenamefont {Quay}, \citenamefont {Hughes}, \citenamefont {Sulpizio}, \citenamefont {Pfeiffer}, \citenamefont {Baldwin}, \citenamefont {West}, \citenamefont {Goldhaber-Gordon},\ and\ \citenamefont {de~Picciotto}}]{Quay2010}%
  \BibitemOpen
  \bibfield  {author} {\bibinfo {author} {\bibfnamefont {C.~H.~L.}\ \bibnamefont {Quay}}, \bibinfo {author} {\bibfnamefont {T.~L.}\ \bibnamefont {Hughes}}, \bibinfo {author} {\bibfnamefont {J.~A.}\ \bibnamefont {Sulpizio}}, \bibinfo {author} {\bibfnamefont {L.~N.}\ \bibnamefont {Pfeiffer}}, \bibinfo {author} {\bibfnamefont {K.~W.}\ \bibnamefont {Baldwin}}, \bibinfo {author} {\bibfnamefont {K.~W.}\ \bibnamefont {West}}, \bibinfo {author} {\bibfnamefont {D.}~\bibnamefont {Goldhaber-Gordon}},\ and\ \bibinfo {author} {\bibfnamefont {R.}~\bibnamefont {de~Picciotto}},\ }\bibfield  {title} {\bibinfo {title} {Observation of a one-dimensional spin--orbit gap in a quantum wire},\ }\href {https://doi.org/10.1038/nphys1626} {\bibfield  {journal} {\bibinfo  {journal} {Nat. Phys.}\ }\textbf {\bibinfo {volume} {6}},\ \bibinfo {pages} {336} (\bibinfo {year} {2010})}\BibitemShut {NoStop}%
\bibitem [{\citenamefont {Xu}\ \emph {et~al.}(2014)\citenamefont {Xu}, \citenamefont {Gan},\ and\ \citenamefont {Zhang}}]{Yong2014}%
  \BibitemOpen
  \bibfield  {author} {\bibinfo {author} {\bibfnamefont {Y.}~\bibnamefont {Xu}}, \bibinfo {author} {\bibfnamefont {Z.}~\bibnamefont {Gan}},\ and\ \bibinfo {author} {\bibfnamefont {S.-C.}\ \bibnamefont {Zhang}},\ }\bibfield  {title} {\bibinfo {title} {Enhanced thermoelectric performance and anomalous {S}eebeck effects in topological insulators},\ }\href {https://doi.org/10.1103/PhysRevLett.112.226801} {\bibfield  {journal} {\bibinfo  {journal} {Phys. Rev. Lett.}\ }\textbf {\bibinfo {volume} {112}},\ \bibinfo {pages} {226801} (\bibinfo {year} {2014})}\BibitemShut {NoStop}%
\bibitem [{\citenamefont {Szumniak}\ \emph {et~al.}(2024)\citenamefont {Szumniak}, \citenamefont {Loss},\ and\ \citenamefont {Klinovaja}}]{Szumniak2024}%
  \BibitemOpen
  \bibfield  {author} {\bibinfo {author} {\bibfnamefont {P.}~\bibnamefont {Szumniak}}, \bibinfo {author} {\bibfnamefont {D.}~\bibnamefont {Loss}},\ and\ \bibinfo {author} {\bibfnamefont {J.}~\bibnamefont {Klinovaja}},\ }\bibfield  {title} {\bibinfo {title} {Spin-resolved nonlocal transport in proximitized {Rashba} nanowires},\ }\href {https://doi.org/10.1103/PhysRevB.110.115413} {\bibfield  {journal} {\bibinfo  {journal} {Phys. Rev. B}\ }\textbf {\bibinfo {volume} {110}},\ \bibinfo {pages} {115413} (\bibinfo {year} {2024})}\BibitemShut {NoStop}%
\bibitem [{\citenamefont {Klinovaja}\ and\ \citenamefont {Loss}(2015)}]{Klinovaja2015}%
  \BibitemOpen
  \bibfield  {author} {\bibinfo {author} {\bibfnamefont {J.}~\bibnamefont {Klinovaja}}\ and\ \bibinfo {author} {\bibfnamefont {D.}~\bibnamefont {Loss}},\ }\bibfield  {title} {\bibinfo {title} {Fermionic and majorana bound states in hybrid nanowires with non-uniform spin-orbit interaction},\ }\href {http://dx.doi.org/10.1140/epjb/e2015-50882-2} {\bibfield  {journal} {\bibinfo  {journal} {Eur. Phys. J. B}\ }\textbf {\bibinfo {volume} {88}} (\bibinfo {year} {2015})}\BibitemShut {NoStop}%
\bibitem [{\citenamefont {Reeg}\ \emph {et~al.}(2018)\citenamefont {Reeg}, \citenamefont {Dmytruk}, \citenamefont {Chevallier}, \citenamefont {Loss},\ and\ \citenamefont {Klinovaja}}]{Klinovaja2018}%
  \BibitemOpen
  \bibfield  {author} {\bibinfo {author} {\bibfnamefont {C.}~\bibnamefont {Reeg}}, \bibinfo {author} {\bibfnamefont {O.}~\bibnamefont {Dmytruk}}, \bibinfo {author} {\bibfnamefont {D.}~\bibnamefont {Chevallier}}, \bibinfo {author} {\bibfnamefont {D.}~\bibnamefont {Loss}},\ and\ \bibinfo {author} {\bibfnamefont {J.}~\bibnamefont {Klinovaja}},\ }\bibfield  {title} {\bibinfo {title} {Zero-energy andreev bound states from quantum dots in proximitized {Rashba} nanowires},\ }\href {https://doi.org/10.1103/PhysRevB.98.245407} {\bibfield  {journal} {\bibinfo  {journal} {Phys. Rev. B}\ }\textbf {\bibinfo {volume} {98}},\ \bibinfo {pages} {245407} (\bibinfo {year} {2018})}\BibitemShut {NoStop}%
\bibitem [{\citenamefont {Dolcini}\ and\ \citenamefont {Rossi}(2018)}]{Dolcini2018}%
  \BibitemOpen
  \bibfield  {author} {\bibinfo {author} {\bibfnamefont {F.}~\bibnamefont {Dolcini}}\ and\ \bibinfo {author} {\bibfnamefont {F.}~\bibnamefont {Rossi}},\ }\bibfield  {title} {\bibinfo {title} {Magnetic field effects on a nanowire with inhomogeneous {Rashba} spin-orbit coupling: Spin properties at equilibrium},\ }\href {https://doi.org/10.1103/PhysRevB.98.045436} {\bibfield  {journal} {\bibinfo  {journal} {Phys. Rev. B}\ }\textbf {\bibinfo {volume} {98}},\ \bibinfo {pages} {045436} (\bibinfo {year} {2018})}\BibitemShut {NoStop}%
\bibitem [{\citenamefont {Rossi}\ \emph {et~al.}(2020)\citenamefont {Rossi}, \citenamefont {Dolcini},\ and\ \citenamefont {Rossi}}]{Rossi2020}%
  \BibitemOpen
  \bibfield  {author} {\bibinfo {author} {\bibfnamefont {L.}~\bibnamefont {Rossi}}, \bibinfo {author} {\bibfnamefont {F.}~\bibnamefont {Dolcini}},\ and\ \bibinfo {author} {\bibfnamefont {F.}~\bibnamefont {Rossi}},\ }\bibfield  {title} {\bibinfo {title} {Majorana-like localized spin density without bound states in topologically trivial spin-orbit coupled nanowires},\ }\href {https://doi.org/10.1103/PhysRevB.101.195421} {\bibfield  {journal} {\bibinfo  {journal} {Phys. Rev. B}\ }\textbf {\bibinfo {volume} {101}},\ \bibinfo {pages} {195421} (\bibinfo {year} {2020})}\BibitemShut {NoStop}%
\bibitem [{\citenamefont {Roddaro}\ \emph {et~al.}(2011)\citenamefont {Roddaro}, \citenamefont {Pescaglini}, \citenamefont {Ercolani}, \citenamefont {Sorba},\ and\ \citenamefont {Beltram}}]{Roddaro2011}%
  \BibitemOpen
  \bibfield  {author} {\bibinfo {author} {\bibfnamefont {S.}~\bibnamefont {Roddaro}}, \bibinfo {author} {\bibfnamefont {A.}~\bibnamefont {Pescaglini}}, \bibinfo {author} {\bibfnamefont {D.}~\bibnamefont {Ercolani}}, \bibinfo {author} {\bibfnamefont {L.}~\bibnamefont {Sorba}},\ and\ \bibinfo {author} {\bibfnamefont {F.}~\bibnamefont {Beltram}},\ }\bibfield  {title} {\bibinfo {title} {Manipulation of electron orbitals in hard-wall inas/inp nanowire quantum dots},\ }\href {https://doi.org/10.1021/nl200209m} {\bibfield  {journal} {\bibinfo  {journal} {Nano Lett.}\ }\textbf {\bibinfo {volume} {11}},\ \bibinfo {pages} {1695} (\bibinfo {year} {2011})}\BibitemShut {NoStop}%
\bibitem [{\citenamefont {Aseev}\ \emph {et~al.}(2019)\citenamefont {Aseev}, \citenamefont {Wang}, \citenamefont {Binci}, \citenamefont {Singh}, \citenamefont {Martí-Sánchez}, \citenamefont {Botifoll}, \citenamefont {Stek}, \citenamefont {Bordin}, \citenamefont {Watson}, \citenamefont {Boekhout}, \citenamefont {Abel}, \citenamefont {Gamble}, \citenamefont {Van~Hoogdalem}, \citenamefont {Arbiol}, \citenamefont {Kouwenhoven}, \citenamefont {de~Lange},\ and\ \citenamefont {Caroff}}]{Aseev2019}%
  \BibitemOpen
  \bibfield  {author} {\bibinfo {author} {\bibfnamefont {P.}~\bibnamefont {Aseev}}, \bibinfo {author} {\bibfnamefont {G.}~\bibnamefont {Wang}}, \bibinfo {author} {\bibfnamefont {L.}~\bibnamefont {Binci}}, \bibinfo {author} {\bibfnamefont {A.}~\bibnamefont {Singh}}, \bibinfo {author} {\bibfnamefont {S.}~\bibnamefont {Martí-Sánchez}}, \bibinfo {author} {\bibfnamefont {M.}~\bibnamefont {Botifoll}}, \bibinfo {author} {\bibfnamefont {L.~J.}\ \bibnamefont {Stek}}, \bibinfo {author} {\bibfnamefont {A.}~\bibnamefont {Bordin}}, \bibinfo {author} {\bibfnamefont {J.~D.}\ \bibnamefont {Watson}}, \bibinfo {author} {\bibfnamefont {F.}~\bibnamefont {Boekhout}}, \bibinfo {author} {\bibfnamefont {D.}~\bibnamefont {Abel}}, \bibinfo {author} {\bibfnamefont {J.}~\bibnamefont {Gamble}}, \bibinfo {author} {\bibfnamefont {K.}~\bibnamefont {Van~Hoogdalem}}, \bibinfo {author} {\bibfnamefont {J.}~\bibnamefont {Arbiol}}, \bibinfo {author} {\bibfnamefont {L.~P.}\ \bibnamefont {Kouwenhoven}}, \bibinfo {author} {\bibfnamefont
  {G.}~\bibnamefont {de~Lange}},\ and\ \bibinfo {author} {\bibfnamefont {P.}~\bibnamefont {Caroff}},\ }\bibfield  {title} {\bibinfo {title} {Ballistic {InSb} nanowires and networks via metal-sown selective area growth},\ }\href {https://doi.org/10.1021/acs.nanolett.9b04265} {\bibfield  {journal} {\bibinfo  {journal} {Nano Lett.}\ }\textbf {\bibinfo {volume} {19}},\ \bibinfo {pages} {9102} (\bibinfo {year} {2019})}\BibitemShut {NoStop}%
\bibitem [{\citenamefont {Fadaly}\ \emph {et~al.}(2017)\citenamefont {Fadaly}, \citenamefont {Zhang}, \citenamefont {Conesa-Boj}, \citenamefont {Car}, \citenamefont {G\"ul}, \citenamefont {Plissard}, \citenamefont {Op~het Veld}, \citenamefont {K\"{o}lling}, \citenamefont {Kouwenhoven},\ and\ \citenamefont {Bakkers}}]{Fadaly2017}%
  \BibitemOpen
  \bibfield  {author} {\bibinfo {author} {\bibfnamefont {E.~M.~T.}\ \bibnamefont {Fadaly}}, \bibinfo {author} {\bibfnamefont {H.}~\bibnamefont {Zhang}}, \bibinfo {author} {\bibfnamefont {S.}~\bibnamefont {Conesa-Boj}}, \bibinfo {author} {\bibfnamefont {D.}~\bibnamefont {Car}}, \bibinfo {author} {\bibfnamefont {O.}~\bibnamefont {G\"ul}}, \bibinfo {author} {\bibfnamefont {S.~R.}\ \bibnamefont {Plissard}}, \bibinfo {author} {\bibfnamefont {R.~L.~M.}\ \bibnamefont {Op~het Veld}}, \bibinfo {author} {\bibfnamefont {S.}~\bibnamefont {K\"{o}lling}}, \bibinfo {author} {\bibfnamefont {L.~P.}\ \bibnamefont {Kouwenhoven}},\ and\ \bibinfo {author} {\bibfnamefont {E.~P. A.~M.}\ \bibnamefont {Bakkers}},\ }\bibfield  {title} {\bibinfo {title} {Observation of conductance quantization in {InSb} nanowire networks},\ }\href {https://doi.org/10.1021/acs.nanolett.7b00797} {\bibfield  {journal} {\bibinfo  {journal} {Nano Lett.}\ }\textbf {\bibinfo {volume} {17}},\ \bibinfo {pages} {6511} (\bibinfo {year} {2017})}\BibitemShut
  {NoStop}%
\bibitem [{\citenamefont {Mourik}\ \emph {et~al.}(2012)\citenamefont {Mourik}, \citenamefont {Zuo}, \citenamefont {Frolov}, \citenamefont {Plissard}, \citenamefont {Bakkers},\ and\ \citenamefont {Kouwenhoven}}]{Mourik_2012}%
  \BibitemOpen
  \bibfield  {author} {\bibinfo {author} {\bibfnamefont {V.}~\bibnamefont {Mourik}}, \bibinfo {author} {\bibfnamefont {K.}~\bibnamefont {Zuo}}, \bibinfo {author} {\bibfnamefont {S.~M.}\ \bibnamefont {Frolov}}, \bibinfo {author} {\bibfnamefont {S.~R.}\ \bibnamefont {Plissard}}, \bibinfo {author} {\bibfnamefont {E.~P. A.~M.}\ \bibnamefont {Bakkers}},\ and\ \bibinfo {author} {\bibfnamefont {L.~P.}\ \bibnamefont {Kouwenhoven}},\ }\bibfield  {title} {\bibinfo {title} {Signatures of majorana fermions in hybrid superconductor–semiconductor nanowire devices},\ }\href {https://doi.org/10.1126/science.1222360} {\bibfield  {journal} {\bibinfo  {journal} {Science}\ }\textbf {\bibinfo {volume} {336}},\ \bibinfo {pages} {1003} (\bibinfo {year} {2012})}\BibitemShut {NoStop}%
\bibitem [{\citenamefont {Das}\ \emph {et~al.}(2012)\citenamefont {Das}, \citenamefont {Ronen}, \citenamefont {Most}, \citenamefont {Oreg}, \citenamefont {Heiblum},\ and\ \citenamefont {Shtrikman}}]{Heiblum_2012}%
  \BibitemOpen
  \bibfield  {author} {\bibinfo {author} {\bibfnamefont {A.}~\bibnamefont {Das}}, \bibinfo {author} {\bibfnamefont {Y.}~\bibnamefont {Ronen}}, \bibinfo {author} {\bibfnamefont {Y.}~\bibnamefont {Most}}, \bibinfo {author} {\bibfnamefont {Y.}~\bibnamefont {Oreg}}, \bibinfo {author} {\bibfnamefont {M.}~\bibnamefont {Heiblum}},\ and\ \bibinfo {author} {\bibfnamefont {H.}~\bibnamefont {Shtrikman}},\ }\bibfield  {title} {\bibinfo {title} {Zero-bias peaks and splitting in an al-inas nanowire topological superconductor as a signature of majorana fermions},\ }\href {https://doi.org/10.1038/nphys2479} {\bibfield  {journal} {\bibinfo  {journal} {Nat. Phys.}\ }\textbf {\bibinfo {volume} {8}},\ \bibinfo {pages} {887 – 895} (\bibinfo {year} {2012})}\BibitemShut {NoStop}%
\bibitem [{\citenamefont {Deng}\ \emph {et~al.}(2016)\citenamefont {Deng}, \citenamefont {Vaitiekėnas}, \citenamefont {Hansen}, \citenamefont {Danon}, \citenamefont {Leijnse}, \citenamefont {Flensberg}, \citenamefont {Nygård}, \citenamefont {Krogstrup},\ and\ \citenamefont {Marcus}}]{Marcus_2016}%
  \BibitemOpen
  \bibfield  {author} {\bibinfo {author} {\bibfnamefont {M.~T.}\ \bibnamefont {Deng}}, \bibinfo {author} {\bibfnamefont {S.}~\bibnamefont {Vaitiekėnas}}, \bibinfo {author} {\bibfnamefont {E.~B.}\ \bibnamefont {Hansen}}, \bibinfo {author} {\bibfnamefont {J.}~\bibnamefont {Danon}}, \bibinfo {author} {\bibfnamefont {M.}~\bibnamefont {Leijnse}}, \bibinfo {author} {\bibfnamefont {K.}~\bibnamefont {Flensberg}}, \bibinfo {author} {\bibfnamefont {J.}~\bibnamefont {Nygård}}, \bibinfo {author} {\bibfnamefont {P.}~\bibnamefont {Krogstrup}},\ and\ \bibinfo {author} {\bibfnamefont {C.~M.}\ \bibnamefont {Marcus}},\ }\bibfield  {title} {\bibinfo {title} {Majorana bound state in a coupled quantum-dot hybrid-nanowire system},\ }\href {https://doi.org/10.1126/science.aaf3961} {\bibfield  {journal} {\bibinfo  {journal} {Science}\ }\textbf {\bibinfo {volume} {354}},\ \bibinfo {pages} {1557} (\bibinfo {year} {2016})}\BibitemShut {NoStop}%
\bibitem [{\citenamefont {Prada}\ \emph {et~al.}(2020)\citenamefont {Prada}, \citenamefont {San-Jose}, \citenamefont {de~Moor}, \citenamefont {Geresdi}, \citenamefont {Lee}, \citenamefont {Klinovaja}, \citenamefont {Loss}, \citenamefont {Nygård}, \citenamefont {Aguado},\ and\ \citenamefont {Kouwenhoven}}]{Prada_2020}%
  \BibitemOpen
  \bibfield  {author} {\bibinfo {author} {\bibfnamefont {E.}~\bibnamefont {Prada}}, \bibinfo {author} {\bibfnamefont {P.}~\bibnamefont {San-Jose}}, \bibinfo {author} {\bibfnamefont {M.~W.~A.}\ \bibnamefont {de~Moor}}, \bibinfo {author} {\bibfnamefont {A.}~\bibnamefont {Geresdi}}, \bibinfo {author} {\bibfnamefont {E.~J.~H.}\ \bibnamefont {Lee}}, \bibinfo {author} {\bibfnamefont {J.}~\bibnamefont {Klinovaja}}, \bibinfo {author} {\bibfnamefont {D.}~\bibnamefont {Loss}}, \bibinfo {author} {\bibfnamefont {J.}~\bibnamefont {Nygård}}, \bibinfo {author} {\bibfnamefont {R.}~\bibnamefont {Aguado}},\ and\ \bibinfo {author} {\bibfnamefont {L.~P.}\ \bibnamefont {Kouwenhoven}},\ }\bibfield  {title} {\bibinfo {title} {From andreev to majorana bound states in hybrid superconductor–semiconductor nanowires},\ }\href {https://doi.org/10.1038/s42254-020-0228-y} {\bibfield  {journal} {\bibinfo  {journal} {Nat. Rev. Phys.}\ }\textbf {\bibinfo {volume} {2}},\ \bibinfo {pages} {575} (\bibinfo {year} {2020})}\BibitemShut {NoStop}%
\bibitem [{\citenamefont {Groth}\ \emph {et~al.}(2014)\citenamefont {Groth}, \citenamefont {Wimmer}, \citenamefont {Akhmerov},\ and\ \citenamefont {Waintal}}]{Groth2014}%
  \BibitemOpen
  \bibfield  {author} {\bibinfo {author} {\bibfnamefont {C.~W.}\ \bibnamefont {Groth}}, \bibinfo {author} {\bibfnamefont {M.}~\bibnamefont {Wimmer}}, \bibinfo {author} {\bibfnamefont {A.~R.}\ \bibnamefont {Akhmerov}},\ and\ \bibinfo {author} {\bibfnamefont {X.}~\bibnamefont {Waintal}},\ }\bibfield  {title} {\bibinfo {title} {Kwant: a software package for quantum transport},\ }\href {https://doi.org/10.1088/1367-2630/16/6/063065} {\bibfield  {journal} {\bibinfo  {journal} {New J. Phys.}\ }\textbf {\bibinfo {volume} {16}},\ \bibinfo {pages} {063065} (\bibinfo {year} {2014})}\BibitemShut {NoStop}%
\bibitem [{Note1()}]{Note1}%
  \BibitemOpen
  \bibinfo {note} {The reader should be aware that the results, including the TE properties, can be generalized if other independent bands are also present~\cite {May2012,Agrawal2024}}\BibitemShut {NoStop}%
\bibitem [{\citenamefont {Gan}\ \emph {et~al.}(2019)\citenamefont {Gan}, \citenamefont {Zhang}, \citenamefont {Christensen}, \citenamefont {Pryds},\ and\ \citenamefont {Chen}}]{Gan2019}%
  \BibitemOpen
  \bibfield  {author} {\bibinfo {author} {\bibfnamefont {Y.}~\bibnamefont {Gan}}, \bibinfo {author} {\bibfnamefont {Y.}~\bibnamefont {Zhang}}, \bibinfo {author} {\bibfnamefont {D.~V.}\ \bibnamefont {Christensen}}, \bibinfo {author} {\bibfnamefont {N.}~\bibnamefont {Pryds}},\ and\ \bibinfo {author} {\bibfnamefont {Y.}~\bibnamefont {Chen}},\ }\bibfield  {title} {\bibinfo {title} {Gate-tunable {R}ashba spin-orbit coupling and spin polarization at diluted oxide interfaces},\ }\href {https://doi.org/10.1103/PhysRevB.100.125134} {\bibfield  {journal} {\bibinfo  {journal} {Phys. Rev. B}\ }\textbf {\bibinfo {volume} {100}},\ \bibinfo {pages} {125134} (\bibinfo {year} {2019})}\BibitemShut {NoStop}%
\bibitem [{Note2()}]{Note2}%
  \BibitemOpen
  \bibinfo {note} {It is easy to generalize the results also to the cases where the magnetic field is not oriented along $x$. However, especially in the inhomogeneous case, one needs to specify the two angles between the magnetic fields and the RSOC, making the analysis slightly more involved.}\BibitemShut {Stop}%
\bibitem [{\citenamefont {Mateos}\ \emph {et~al.}(2024)\citenamefont {Mateos}, \citenamefont {Tosi}, \citenamefont {Braggio}, \citenamefont {Taddei},\ and\ \citenamefont {Arrachea}}]{PhysRevB.110.075415}%
  \BibitemOpen
  \bibfield  {author} {\bibinfo {author} {\bibfnamefont {J.~H.}\ \bibnamefont {Mateos}}, \bibinfo {author} {\bibfnamefont {L.}~\bibnamefont {Tosi}}, \bibinfo {author} {\bibfnamefont {A.}~\bibnamefont {Braggio}}, \bibinfo {author} {\bibfnamefont {F.}~\bibnamefont {Taddei}},\ and\ \bibinfo {author} {\bibfnamefont {L.}~\bibnamefont {Arrachea}},\ }\bibfield  {title} {\bibinfo {title} {Nonlocal thermoelectricity in quantum wires as a signature of bogoliubov-fermi points},\ }\href {https://link.aps.org/doi/10.1103/PhysRevB.110.075415} {\bibfield  {journal} {\bibinfo  {journal} {Phys. Rev. B}\ }\textbf {\bibinfo {volume} {110}},\ \bibinfo {pages} {075415} (\bibinfo {year} {2024})}\BibitemShut {NoStop}%
\bibitem [{\citenamefont {Qi}\ and\ \citenamefont {Zhang}(2011)}]{Qi2011}%
  \BibitemOpen
  \bibfield  {author} {\bibinfo {author} {\bibfnamefont {X.-L.}\ \bibnamefont {Qi}}\ and\ \bibinfo {author} {\bibfnamefont {S.-C.}\ \bibnamefont {Zhang}},\ }\bibfield  {title} {\bibinfo {title} {Topological insulators and superconductors},\ }\href {https://doi.org/10.1103/RevModPhys.83.1057} {\bibfield  {journal} {\bibinfo  {journal} {Rev. Mod. Phys.}\ }\textbf {\bibinfo {volume} {83}},\ \bibinfo {pages} {1057} (\bibinfo {year} {2011})}\BibitemShut {NoStop}%
\bibitem [{\citenamefont {Br{\"u}ne}\ \emph {et~al.}(2012)\citenamefont {Br{\"u}ne}, \citenamefont {Roth}, \citenamefont {Buhmann}, \citenamefont {Hankiewicz}, \citenamefont {Molenkamp}, \citenamefont {Maciejko}, \citenamefont {Qi},\ and\ \citenamefont {Zhang}}]{Brune2012}%
  \BibitemOpen
  \bibfield  {author} {\bibinfo {author} {\bibfnamefont {C.}~\bibnamefont {Br{\"u}ne}}, \bibinfo {author} {\bibfnamefont {A.}~\bibnamefont {Roth}}, \bibinfo {author} {\bibfnamefont {H.}~\bibnamefont {Buhmann}}, \bibinfo {author} {\bibfnamefont {E.~M.}\ \bibnamefont {Hankiewicz}}, \bibinfo {author} {\bibfnamefont {L.~W.}\ \bibnamefont {Molenkamp}}, \bibinfo {author} {\bibfnamefont {J.}~\bibnamefont {Maciejko}}, \bibinfo {author} {\bibfnamefont {X.-L.}\ \bibnamefont {Qi}},\ and\ \bibinfo {author} {\bibfnamefont {S.-C.}\ \bibnamefont {Zhang}},\ }\bibfield  {title} {\bibinfo {title} {Spin polarization of the quantum spin hall edge states},\ }\href {https://doi.org/10.1038/nphys2322} {\bibfield  {journal} {\bibinfo  {journal} {Nat. Phys.}\ }\textbf {\bibinfo {volume} {8}},\ \bibinfo {pages} {485} (\bibinfo {year} {2012})}\BibitemShut {NoStop}%
\bibitem [{\citenamefont {K{\"o}nig}\ \emph {et~al.}(2007)\citenamefont {K{\"o}nig}, \citenamefont {Wiedmann}, \citenamefont {Br{\"u}ne}, \citenamefont {Roth}, \citenamefont {Buhmann}, \citenamefont {Molenkamp}, \citenamefont {Qi},\ and\ \citenamefont {Zhang}}]{Konig2007}%
  \BibitemOpen
  \bibfield  {author} {\bibinfo {author} {\bibfnamefont {M.}~\bibnamefont {K{\"o}nig}}, \bibinfo {author} {\bibfnamefont {S.}~\bibnamefont {Wiedmann}}, \bibinfo {author} {\bibfnamefont {C.}~\bibnamefont {Br{\"u}ne}}, \bibinfo {author} {\bibfnamefont {A.}~\bibnamefont {Roth}}, \bibinfo {author} {\bibfnamefont {H.}~\bibnamefont {Buhmann}}, \bibinfo {author} {\bibfnamefont {L.~W.}\ \bibnamefont {Molenkamp}}, \bibinfo {author} {\bibfnamefont {X.-L.}\ \bibnamefont {Qi}},\ and\ \bibinfo {author} {\bibfnamefont {S.-C.}\ \bibnamefont {Zhang}},\ }\bibfield  {title} {\bibinfo {title} {Quantum spin hall insulator state in hgte quantum wells},\ }\href {https://doi.org/10.1126/science.1148047} {\bibfield  {journal} {\bibinfo  {journal} {Science}\ }\textbf {\bibinfo {volume} {318}},\ \bibinfo {pages} {766} (\bibinfo {year} {2007})}\BibitemShut {NoStop}%
\bibitem [{\citenamefont {Blanter}\ and\ \citenamefont {Büttiker}(2000)}]{Blanter2000}%
  \BibitemOpen
  \bibfield  {author} {\bibinfo {author} {\bibfnamefont {Y.~M.}\ \bibnamefont {Blanter}}\ and\ \bibinfo {author} {\bibfnamefont {M.}~\bibnamefont {Büttiker}},\ }\bibfield  {title} {\bibinfo {title} {Shot noise in mesoscopic conductors},\ }\href {https://doi.org/10.1016/S0370-1573(99)00123-4} {\bibfield  {journal} {\bibinfo  {journal} {Phys. Rep.}\ }\textbf {\bibinfo {volume} {336}},\ \bibinfo {pages} {1} (\bibinfo {year} {2000})}\BibitemShut {NoStop}%
\bibitem [{\citenamefont {B\"uttiker}\ \emph {et~al.}(1985)\citenamefont {B\"uttiker}, \citenamefont {Imry}, \citenamefont {Landauer},\ and\ \citenamefont {Pinhas}}]{Buttiker1985}%
  \BibitemOpen
  \bibfield  {author} {\bibinfo {author} {\bibfnamefont {M.}~\bibnamefont {B\"uttiker}}, \bibinfo {author} {\bibfnamefont {Y.}~\bibnamefont {Imry}}, \bibinfo {author} {\bibfnamefont {R.}~\bibnamefont {Landauer}},\ and\ \bibinfo {author} {\bibfnamefont {S.}~\bibnamefont {Pinhas}},\ }\bibfield  {title} {\bibinfo {title} {Generalized many-channel conductance formula with application to small rings},\ }\href {https://doi.org/10.1103/PhysRevB.31.6207} {\bibfield  {journal} {\bibinfo  {journal} {Phys. Rev. B}\ }\textbf {\bibinfo {volume} {31}},\ \bibinfo {pages} {6207} (\bibinfo {year} {1985})}\BibitemShut {NoStop}%
\bibitem [{\citenamefont {Lv}\ and\ \citenamefont {Li}(2012)}]{Lv2012}%
  \BibitemOpen
  \bibfield  {author} {\bibinfo {author} {\bibfnamefont {S.-H.}\ \bibnamefont {Lv}}\ and\ \bibinfo {author} {\bibfnamefont {Y.-X.}\ \bibnamefont {Li}},\ }\bibfield  {title} {\bibinfo {title} {Effects of the edge states on conductance and thermopower for the bilayer graphene nanoribbons},\ }\href {https://doi.org/10.1063/1.4747927} {\bibfield  {journal} {\bibinfo  {journal} {J. Appl. Phys.}\ }\textbf {\bibinfo {volume} {112}},\ \bibinfo {pages} {053701} (\bibinfo {year} {2012})}\BibitemShut {NoStop}%
\bibitem [{\citenamefont {Oji}(1984)}]{Oji1984}%
  \BibitemOpen
  \bibfield  {author} {\bibinfo {author} {\bibfnamefont {H.}~\bibnamefont {Oji}},\ }\bibfield  {title} {\bibinfo {title} {Thermomagnetic effects in two-dimensional electron systems},\ }\href {https://doi.org/10.1088/0022-3719/17/17/014} {\bibfield  {journal} {\bibinfo  {journal} {J. Phys. C: Solid State Phys.}\ }\textbf {\bibinfo {volume} {17}},\ \bibinfo {pages} {3059} (\bibinfo {year} {1984})}\BibitemShut {NoStop}%
\bibitem [{\citenamefont {Benenti}\ \emph {et~al.}(2017)\citenamefont {Benenti}, \citenamefont {Casati}, \citenamefont {Saito},\ and\ \citenamefont {Whitney}}]{Benenti2017}%
  \BibitemOpen
  \bibfield  {author} {\bibinfo {author} {\bibfnamefont {G.}~\bibnamefont {Benenti}}, \bibinfo {author} {\bibfnamefont {G.}~\bibnamefont {Casati}}, \bibinfo {author} {\bibfnamefont {K.}~\bibnamefont {Saito}},\ and\ \bibinfo {author} {\bibfnamefont {R.~S.}\ \bibnamefont {Whitney}},\ }\bibfield  {title} {\bibinfo {title} {Fundamental aspects of steady-state conversion of heat to work at the nanoscale},\ }\href {http://dx.doi.org/10.1016/j.physrep.2017.05.008} {\bibfield  {journal} {\bibinfo  {journal} {Phys. Rep.}\ }\textbf {\bibinfo {volume} {694}} (\bibinfo {year} {2017})}\BibitemShut {NoStop}%
\bibitem [{\citenamefont {de~Groot}\ and\ \citenamefont {Mazur}(2013)}]{Groot2013}%
  \BibitemOpen
  \bibfield  {author} {\bibinfo {author} {\bibfnamefont {S.~R.}\ \bibnamefont {de~Groot}}\ and\ \bibinfo {author} {\bibfnamefont {P.}~\bibnamefont {Mazur}},\ }\href@noop {} {\emph {\bibinfo {title} {Non-Equilibrium Thermodynamics}}},\ Dover Books on Physics\ (\bibinfo  {publisher} {Dover Publications},\ \bibinfo {year} {2013})\BibitemShut {NoStop}%
\bibitem [{\citenamefont {Lekwongderm}\ \emph {et~al.}(2019)\citenamefont {Lekwongderm}, \citenamefont {Chumkaew}, \citenamefont {Thainoi}, \citenamefont {Kiravittaya}, \citenamefont {Tandaechanurat}, \citenamefont {Nuntawong}, \citenamefont {Sopitpan}, \citenamefont {Yordsri}, \citenamefont {Thanachayanont}, \citenamefont {Kanjanachuchai}, \citenamefont {Ratanathammaphan},\ and\ \citenamefont {Panyakeow}}]{Lekwongderm2019}%
  \BibitemOpen
  \bibfield  {author} {\bibinfo {author} {\bibfnamefont {P.}~\bibnamefont {Lekwongderm}}, \bibinfo {author} {\bibfnamefont {R.}~\bibnamefont {Chumkaew}}, \bibinfo {author} {\bibfnamefont {S.}~\bibnamefont {Thainoi}}, \bibinfo {author} {\bibfnamefont {S.}~\bibnamefont {Kiravittaya}}, \bibinfo {author} {\bibfnamefont {A.}~\bibnamefont {Tandaechanurat}}, \bibinfo {author} {\bibfnamefont {N.}~\bibnamefont {Nuntawong}}, \bibinfo {author} {\bibfnamefont {S.}~\bibnamefont {Sopitpan}}, \bibinfo {author} {\bibfnamefont {V.}~\bibnamefont {Yordsri}}, \bibinfo {author} {\bibfnamefont {C.}~\bibnamefont {Thanachayanont}}, \bibinfo {author} {\bibfnamefont {S.}~\bibnamefont {Kanjanachuchai}}, \bibinfo {author} {\bibfnamefont {S.}~\bibnamefont {Ratanathammaphan}},\ and\ \bibinfo {author} {\bibfnamefont {S.}~\bibnamefont {Panyakeow}},\ }\bibfield  {title} {\bibinfo {title} {Study on raman spectroscopy of {I}n{S}b nano-stripes grown on {G}a{S}b substrate by molecular beam epitaxy and their {R}aman peak shift with magnetic
  field},\ }\href {https://doi.org/10.1016/j.jcrysgro.2019.02.033} {\bibfield  {journal} {\bibinfo  {journal} {J. Cryst. Growth}\ }\textbf {\bibinfo {volume} {512}},\ \bibinfo {pages} {198} (\bibinfo {year} {2019})}\BibitemShut {NoStop}%
\bibitem [{\citenamefont {Fan}\ \emph {et~al.}(2015)\citenamefont {Fan}, \citenamefont {Li}, \citenamefont {Kang}, \citenamefont {Caroff}, \citenamefont {Wang}, \citenamefont {Huang}, \citenamefont {Deng}, \citenamefont {Yu},\ and\ \citenamefont {Xu}}]{Fan2015}%
  \BibitemOpen
  \bibfield  {author} {\bibinfo {author} {\bibfnamefont {D.}~\bibnamefont {Fan}}, \bibinfo {author} {\bibfnamefont {S.}~\bibnamefont {Li}}, \bibinfo {author} {\bibfnamefont {N.}~\bibnamefont {Kang}}, \bibinfo {author} {\bibfnamefont {P.}~\bibnamefont {Caroff}}, \bibinfo {author} {\bibfnamefont {L.~B.}\ \bibnamefont {Wang}}, \bibinfo {author} {\bibfnamefont {Y.~Q.}\ \bibnamefont {Huang}}, \bibinfo {author} {\bibfnamefont {M.~T.}\ \bibnamefont {Deng}}, \bibinfo {author} {\bibfnamefont {C.~L.}\ \bibnamefont {Yu}},\ and\ \bibinfo {author} {\bibfnamefont {H.~Q.}\ \bibnamefont {Xu}},\ }\bibfield  {title} {\bibinfo {title} {Formation of long single quantum dots in high quality {I}n{S}b nanowires grown by molecular beam epitaxy},\ }\href {https://doi.org/10.1039/C5NR04273A} {\bibfield  {journal} {\bibinfo  {journal} {Nanoscale}\ }\textbf {\bibinfo {volume} {7}},\ \bibinfo {pages} {14822} (\bibinfo {year} {2015})}\BibitemShut {NoStop}%
\bibitem [{Note3()}]{Note3}%
  \BibitemOpen
  \bibinfo {note} {The effect of inhomogeneities in the RSOC direction has been mainly considered on the Andreev bound levels and Josephson current in the case of NWs proximitized by superconductors~\cite {Szumniak2024, Klinovaja2015,Klinovaja2018,Cayao2015}.}\BibitemShut {Stop}%
\bibitem [{\citenamefont {Svensson}\ \emph {et~al.}(2012)\citenamefont {Svensson}, \citenamefont {Persson}, \citenamefont {Hoffmann}, \citenamefont {Nakpathomkun}, \citenamefont {Nilsson}, \citenamefont {Xu}, \citenamefont {Samuelson},\ and\ \citenamefont {Linke}}]{Svensson2012}%
  \BibitemOpen
  \bibfield  {author} {\bibinfo {author} {\bibfnamefont {S.~F.}\ \bibnamefont {Svensson}}, \bibinfo {author} {\bibfnamefont {A.~I.}\ \bibnamefont {Persson}}, \bibinfo {author} {\bibfnamefont {E.~A.}\ \bibnamefont {Hoffmann}}, \bibinfo {author} {\bibfnamefont {N.}~\bibnamefont {Nakpathomkun}}, \bibinfo {author} {\bibfnamefont {H.~A.}\ \bibnamefont {Nilsson}}, \bibinfo {author} {\bibfnamefont {H.~Q.}\ \bibnamefont {Xu}}, \bibinfo {author} {\bibfnamefont {L.}~\bibnamefont {Samuelson}},\ and\ \bibinfo {author} {\bibfnamefont {H.}~\bibnamefont {Linke}},\ }\bibfield  {title} {\bibinfo {title} {Lineshape of the thermopower of quantum dots},\ }\href {https://doi.org/10.1088/1367-2630/14/3/033041} {\bibfield  {journal} {\bibinfo  {journal} {New J. Phys.}\ }\textbf {\bibinfo {volume} {14}},\ \bibinfo {pages} {033041} (\bibinfo {year} {2012})}\BibitemShut {NoStop}%
\bibitem [{\citenamefont {Erdman}\ \emph {et~al.}(2017)\citenamefont {Erdman}, \citenamefont {Mazza}, \citenamefont {Bosisio}, \citenamefont {Benenti}, \citenamefont {Fazio},\ and\ \citenamefont {Taddei}}]{Erdman2017}%
  \BibitemOpen
  \bibfield  {author} {\bibinfo {author} {\bibfnamefont {P.~A.}\ \bibnamefont {Erdman}}, \bibinfo {author} {\bibfnamefont {F.}~\bibnamefont {Mazza}}, \bibinfo {author} {\bibfnamefont {R.}~\bibnamefont {Bosisio}}, \bibinfo {author} {\bibfnamefont {G.}~\bibnamefont {Benenti}}, \bibinfo {author} {\bibfnamefont {R.}~\bibnamefont {Fazio}},\ and\ \bibinfo {author} {\bibfnamefont {F.}~\bibnamefont {Taddei}},\ }\bibfield  {title} {\bibinfo {title} {Thermoelectric properties of an interacting quantum dot based heat engine},\ }\href {https://doi.org/10.1103/PhysRevB.95.245432} {\bibfield  {journal} {\bibinfo  {journal} {Phys. Rev. B}\ }\textbf {\bibinfo {volume} {95}},\ \bibinfo {pages} {245432} (\bibinfo {year} {2017})}\BibitemShut {NoStop}%
\bibitem [{\citenamefont {Agrawal}\ \emph {et~al.}(2024)\citenamefont {Agrawal}, \citenamefont {{de Boor}},\ and\ \citenamefont {Dasgupta}}]{Agrawal2024}%
  \BibitemOpen
  \bibfield  {author} {\bibinfo {author} {\bibfnamefont {B.}~\bibnamefont {Agrawal}}, \bibinfo {author} {\bibfnamefont {J.}~\bibnamefont {{de Boor}}},\ and\ \bibinfo {author} {\bibfnamefont {T.}~\bibnamefont {Dasgupta}},\ }\bibfield  {title} {\bibinfo {title} {A multi-band refinement technique for analyzing electronic band structure of thermoelectric materials},\ }\href {https://doi.org/10.1016/j.xcrp.2024.101781} {\bibfield  {journal} {\bibinfo  {journal} {Cell Rep. Phys. Sci.}\ }\textbf {\bibinfo {volume} {5}},\ \bibinfo {pages} {101781} (\bibinfo {year} {2024})}\BibitemShut {NoStop}%
\bibitem [{\citenamefont {May}\ and\ \citenamefont {Snyder}(2012)}]{May2012}%
  \BibitemOpen
  \bibfield  {author} {\bibinfo {author} {\bibfnamefont {A.}~\bibnamefont {May}}\ and\ \bibinfo {author} {\bibfnamefont {G.~J.}\ \bibnamefont {Snyder}},\ }\bibfield  {title} {\bibinfo {title} {Introduction to modeling thermoelectric transport at high temperatures},\ }in\ \href@noop {} {\emph {\bibinfo {booktitle} {Materials, Preparation, and Characterization in Thermoelectrics}}},\ \bibinfo {editor} {edited by\ \bibinfo {editor} {\bibfnamefont {T.~M.}\ \bibnamefont {Tritt}}}\ (\bibinfo  {publisher} {CRC Press},\ \bibinfo {address} {Boca Raton, FL},\ \bibinfo {year} {2012})\ pp.\ \bibinfo {pages} {1--18}\BibitemShut {NoStop}%
\bibitem [{\citenamefont {Fisher}\ and\ \citenamefont {Lee}(1981)}]{PhysRevB.23.6851}%
  \BibitemOpen
  \bibfield  {author} {\bibinfo {author} {\bibfnamefont {D.~S.}\ \bibnamefont {Fisher}}\ and\ \bibinfo {author} {\bibfnamefont {P.~A.}\ \bibnamefont {Lee}},\ }\bibfield  {title} {\bibinfo {title} {Relation between conductivity and transmission matrix},\ }\href {https://doi.org/10.1103/PhysRevB.23.6851} {\bibfield  {journal} {\bibinfo  {journal} {Phys. Rev. B}\ }\textbf {\bibinfo {volume} {23}},\ \bibinfo {pages} {6851} (\bibinfo {year} {1981})}\BibitemShut {NoStop}%
\bibitem [{\citenamefont {Arrachea}\ \emph {et~al.}(2025{\natexlab{b}})\citenamefont {Arrachea}, \citenamefont {Braggio}, \citenamefont {Burset}, \citenamefont {Lee}, \citenamefont {Yeyati},\ and\ \citenamefont {Sánchez}}]{Lili2025}%
  \BibitemOpen
  \bibfield  {author} {\bibinfo {author} {\bibfnamefont {L.}~\bibnamefont {Arrachea}}, \bibinfo {author} {\bibfnamefont {A.}~\bibnamefont {Braggio}}, \bibinfo {author} {\bibfnamefont {P.}~\bibnamefont {Burset}}, \bibinfo {author} {\bibfnamefont {E.~J.~H.}\ \bibnamefont {Lee}}, \bibinfo {author} {\bibfnamefont {A.~L.}\ \bibnamefont {Yeyati}},\ and\ \bibinfo {author} {\bibfnamefont {R.}~\bibnamefont {Sánchez}},\ }\bibfield  {title} {\bibinfo {title} {Thermoelectric processes of quantum normal–superconductor interfaces},\ }\href {https://doi.org/10.1002/andp.202500197} {\bibfield  {journal} {\bibinfo  {journal} {Ann. Phys. (Leipzig)}\ ,\ \bibinfo {pages} {e00197}} (\bibinfo {year} {2025}{\natexlab{b}})}\BibitemShut {NoStop}%
\bibitem [{\citenamefont {Bolens}\ \emph {et~al.}(2017)\citenamefont {Bolens}, \citenamefont {Katsura}, \citenamefont {Ogata},\ and\ \citenamefont {Miyashita}}]{PhysRevB.95.235115}%
  \BibitemOpen
  \bibfield  {author} {\bibinfo {author} {\bibfnamefont {A.}~\bibnamefont {Bolens}}, \bibinfo {author} {\bibfnamefont {H.}~\bibnamefont {Katsura}}, \bibinfo {author} {\bibfnamefont {M.}~\bibnamefont {Ogata}},\ and\ \bibinfo {author} {\bibfnamefont {S.}~\bibnamefont {Miyashita}},\ }\bibfield  {title} {\bibinfo {title} {Synergetic effect of spin-orbit coupling and zeeman splitting on the optical conductivity in the one-dimensional hubbard model},\ }\href {https://doi.org/10.1103/PhysRevB.95.235115} {\bibfield  {journal} {\bibinfo  {journal} {Phys. Rev. B}\ }\textbf {\bibinfo {volume} {95}},\ \bibinfo {pages} {235115} (\bibinfo {year} {2017})}\BibitemShut {NoStop}%
\bibitem [{\citenamefont {Datta}(1995)}]{Datta_1995}%
  \BibitemOpen
  \bibfield  {author} {\bibinfo {author} {\bibfnamefont {S.}~\bibnamefont {Datta}},\ }\href@noop {} {\emph {\bibinfo {title} {Electronic Transport in Mesoscopic Systems}}}\ (\bibinfo  {publisher} {Cambridge University Press},\ \bibinfo {year} {1995})\BibitemShut {NoStop}%
\bibitem [{\citenamefont {Ouisse}(2013)}]{Thierry_2013}%
  \BibitemOpen
  \bibfield  {author} {\bibinfo {author} {\bibfnamefont {T.}~\bibnamefont {Ouisse}},\ }\href@noop {} {\emph {\bibinfo {title} {Electron Transport in Nanostructures and Mesoscopic Devices: An Introduction}}}\ (\bibinfo  {publisher} {Wiley, London},\ \bibinfo {year} {2013})\BibitemShut {NoStop}%
\bibitem [{\citenamefont {Beenakker}(1997)}]{RevModPhys.69.731}%
  \BibitemOpen
  \bibfield  {author} {\bibinfo {author} {\bibfnamefont {C.~W.~J.}\ \bibnamefont {Beenakker}},\ }\bibfield  {title} {\bibinfo {title} {Random-matrix theory of quantum transport},\ }\href {https://doi.org/10.1103/RevModPhys.69.731} {\bibfield  {journal} {\bibinfo  {journal} {Rev. Mod. Phys.}\ }\textbf {\bibinfo {volume} {69}},\ \bibinfo {pages} {731} (\bibinfo {year} {1997})}\BibitemShut {NoStop}%
\end{thebibliography}%


%

\end{document}